\documentclass[12pt]{article}
\usepackage{amssymb,amsfonts,latexsym}

\newcommand{\sv}{\vec{\s}}
\newcommand{\e}{\eta}

\newcommand{\sgb}{{\mbox{\scriptsize{\gb}}}}
\newcommand{\sgbn}{{\mbox{\scriptsize{\gbn}}}}
\newcommand{\shb}{{\mbox{\scriptsize{\hb}}}}

\newcommand{\lr}{{\mathcal{L}}}
\newcommand{\fr}{{\mathcal{F}}}
\newcommand{\nc}{{\mathcal{N}}}
\newcommand{\mc}{{\mathcal{M}}}
\newcommand{\tn}{{\hat{n}}}
\renewcommand{\j}{{\mathcal{J}}}
\newcommand{\g}{{\mathcal{G}}}
\newcommand{\pr}{{\mathcal{P}}}

\newcommand{\hl}{\lfloor\frac{\lambda}{2}\rfloor}

\newcommand{\sgcd}{{\textup{\scriptsize{ gcd }}}}

\renewcommand{\theequation}{\arabic{section}.\arabic{equation}}
\renewcommand{\(}{\begin{equation}}

\newcounter{saveeqn}
\newcounter{savealpheqn}

\newcommand{\alpheqn}{\setcounter{saveeqn}{\value{equation}}%
 \stepcounter{saveeqn}\setcounter{equation}{0}%
 \renewcommand{\theequation}{\mbox{\arabic{section}.\arabic{saveeqn}\alph{equation}}}
 \renewcommand{\)}{\end{equation}}}
\def\group#1{\refstepcounter{equation}\setcounter{saveeqn}{\value{equation}}%
 \label{#1}\setcounter{equation}{0}%
 \renewcommand{\theequation}{\mbox{\arabic{section}.\arabic{saveeqn}\alph{equation}}}
 \renewcommand{\)}{\end{equation}}}
\newcommand{\reseteqn}{\setcounter{equation}{\value{saveeqn}}%
 \renewcommand{\theequation}{\arabic{section}.\arabic{equation}}%
 \renewcommand{\)}{\end{equation}}}
\newcounter{alphcount}
\def\getletter#1{\renewcommand{\theequation}{\alph{equation}}%
		 \setcounter{alphcount}{\value{equation}}
		 \refstepcounter{equation}%
                 \label{#1}%
		 \setcounter{equation}{\value{alphcount}}%
                 \renewcommand{\theequation}{\mbox{\arabic{section}.\arabic{saveeqn}\alph{equation}}}}
\def\agetletter#1{\renewcommand{\theequation}{\alph{equation}}%
		 \begin{eqnarray}%
                 \label{#1}%
		 \nonumber\end{eqnarray}\vspace{-.666in}%
                 \renewcommand{\theequation}{\mbox{\Alph{subsection}.\arabic{saveeqn}\alph{equation}}}}
\newcommand{\aalpheqn}{\setcounter{saveeqn}{\value{equation}}%
 \stepcounter{saveeqn}\setcounter{equation}{0}%
 \renewcommand{\theequation}{\mbox{\Alph{subsection}.\arabic{saveeqn}\alph{equation}}}
  \renewcommand{\)}{\end{equation}}}
\newcommand{\areseteqn}{\setcounter{equation}{\value{saveeqn}}%
 \renewcommand{\theequation}{\Alph{subsection}.\arabic{equation}}%
 \renewcommand{\)}{\end{equation}}}

\renewcommand{\=}{\hspace{-.03in}=\hspace{-.02in}}
\def\group#1{\refstepcounter{equation}\setcounter{saveeqn}{\value{equation}}%
 \label{#1}\setcounter{equation}{0}%
 \renewcommand{\theequation}{\mbox{\arabic{section}.\arabic{saveeqn}\alph{equation}}}
 \renewcommand{\)}{\end{equation}}}
\renewcommand{\thefootnote}{\alph{footnote}}
\renewcommand{\(}{\begin{equation}}
\renewcommand{\)}{\end{equation}}
\newcommand{\ba}{\begin{eqnarray}}
\newcommand{\ea}{\end{eqnarray}}
\renewcommand{\l}{\lambda}
\renewcommand{\a}{\alpha}
\renewcommand{\b}{\beta}
\renewcommand{\r}{\rho}
\newcommand{\stheta}{\rho}
\newcommand{\Ltheta}{{\mathcal{R}}}

\newcommand{\sa}{\mathop{\vtop{\ialign{##\crcr
  $\hfil\displaystyle{\longrightarrow}\hfil$\crcr\noalign{\kern-1pt\nointerlineskip}
  \hspace{.12in}$^\sigma$\hskip6pt\crcr\noalign{\kern3pt}}}}}
\newcommand{\slra}{\mathop{\vtop{\ialign{##\crcr
  $\hfil\displaystyle{\longleftrightarrow}\hfil$\crcr\noalign{\kern-1pt\nointerlineskip}
  \hspace{.12in}$^\sigma$\hskip6pt\crcr\noalign{\kern3pt}}}}}
\newcommand{\sat}{\mathop{\vtop{\ialign{##\crcr
  $\hfil\displaystyle{\longrightarrow}\hfil$\crcr\noalign{\kern-1pt\nointerlineskip}
  \hspace{.12in}$^\sigma$\hskip6pt\crcr\noalign{\kern3pt}}}}}
\newcommand{\pa}{\mathop{\vtop{\ialign{##\crcr
  $\hfil\displaystyle{\oplus}\hfil$\crcr\noalign{\kern+1pt\nointerlineskip}
  \hspace{.08in}$^{\alpha=0}$\hskip6pt\crcr\noalign{\kern3pt}}}}}
\newcommand{\pan}{\mathop{\vtop{ialgin{##\crcr
  $\hfil\displaystyle{\oplus}\hfil$\crcr\noaligan{\kern+2pt\nointerlinkeskip}
  \hspace{.03in} $^{\alpha}$\hskip6pt\crcr\noalign{\kern3pit}}}}}
\newcommand{\ka}{\mathop{\vtop{\ialign{##\crcr
  $\hfil\displaystyle{\longleftrightarrow}\hfil$\crcr\noalign{\kern-1pt\nointerlineskip}
  \hspace{.12in}$^K$\hskip6pt\crcr\noalign{\kern3pt}}}}}
\newcommand{\bp}{\mathop{\vtop{ialign{##\crcr
  $\hfil\displaystyle{}\hfil$\crcr\noalign{\kern-13pt\nointerlineskip}
  \big{(}\hskip0pt\crcr\noalign{\kern3pt}}}}}
\newcommand{\cbp}{\mathop{\vtop{ialign{##\crcr
  $\hfil\displaystyle{}\hfil$\crcr\noalign{\kern-13pt\nointerlineskip}
  \big{)}\hskip0pt\crcr\noalign{\kern3pt}}}}}

\newcommand{\+}{\hspace{-.03in}+\hspace{-.02in}}
\newcommand{\s}{\sigma}
\newcommand{\srange}{\sigma=0,...,N_c-1}

\renewcommand{\sp}{,\hspace{.3in}}
\newcommand{\newsection}{\setcounter{equation}{0}\section}
\newcommand{\p}{^\prime}

\newcommand{\w}{\omega}

\newcommand{\uds}{U^\dag(\s)}

\newcommand{\lra}{\leftrightarrow}
\newcommand{\mod}{{\textup{\scriptsize{ mod }}}}

\newcommand{\reg}{O((z-w)^0)}
\newcommand{\hi}{\hat{i}}
\newcommand{\hj}{\hat{j}}
\renewcommand{\hl}{\hat{l}}
\newcommand{\hm}{\hat{m}}

\newcommand{\jh}{\hat{J}}
\newcommand{\tp}{{2\pi i}}
\newcommand{\lieH}{H}
\newcommand{\alie}{A(\textup{Lie} \hsp{.025}h)}
\newcommand{\alieH}{A(\textup{Lie} \hsp{.025}h(H))}

\newcommand{\ahh}{\frac{A(H)}{H}}
\newcommand{\alieHH}{\frac{A(\textup{Lie} \hsp{.025}h(H))}{H}}
\newcommand{\tahh}{$A(H)/H$}
\newcommand{\talieH}{A(\textup{Lie} \hsp{.025}h(H))/H}
\newcommand{\taz}{$A(\zl)/\zl$}

\newcommand{\tadd}{$A(\dl)/\dl$}
\newcommand{\bu}{$\bullet$}

\newcommand{\appendixa}
 {\renewcommand{\theequation}{\Alph{subsection}.\arabic{equation}}%
  \renewcommand{\thesubsection}%
               {Appendix \Alph{subsection}.\setcounter{equation}{0}}%
  \renewcommand{\alpheqn}{\aalpheqn}%
  \renewcommand{\reseteqn}{\areseteqn}
  \newcounter{savesec}}
\newcommand{\appendices}{\appendix\appendixa}
\def\app#1#2{\renewcommand{\thesubsection}{\Alph{subsection}}%
	\refstepcounter{subsection}%
	\setcounter{subsection}{\value{savesec}}%
	\stepcounter{savesec}\label{#1}%
	\renewcommand{\thesubsection}%
               {Appendix \Alph{subsection}.\setcounter{equation}{0}}%
	\subsection{#2}}

\def\foot#1{\mbox{\footnotesize $#1$}}
\def\scrs#1{\mbox{\scriptsize $#1$}}
\def\tyny#1{\mbox{\tiny $#1$}}
\def\smal#1{\mbox{\small $#1$}}
\def\big#1{\mbox{\large $#1$}}
\def\Big#1{\mbox{\Large $#1$}}

\def\srac#1#2{\smal{\frac{#1}{#2}}}

\def\gb            {\mbox{$\hat{\mathfrak g}$}}
\def\hb            {\mbox{$\hat{\mathfrak h}$}}
\def\gbn           {\mbox{$\mathfrak g$}}
\def\sm#1	   {\mbox{\scriptsize $#1$}}

\def\sz		   {\mbox{\scriptsize $\mathbb  Z$}}
\def\z		   {\mbox{$\mathbb  Z$}}
\def\zl		   {\mbox{$\mathbb  Z_\l$}}

\def\dl		   {\mbox{$\mathbb D_\l$}}

\mathchardef\endbar="375
\font\fivesans=cmss10 at 4.61pt
\font\sevensans=cmss10 at 6.81pt
\font\tensans=cmss10 at 12pt %added ``at 12pt''
\newfam\sansfam
\textfont\sansfam=\tensans\scriptfont\sansfam=\sevensans\scriptscriptfont
\sansfam=\fivesans
\def\sans{\fam\sansfam\tensans}
\def\Z{{\mathchoice
{\hbox{$\sans\textstyle Z\kern-0.455em Z$}} %was .4
{\hbox{$\sans\textstyle Z\kern-0.455em Z$}} %was .4
{\hbox{$\sans\scriptstyle Z\kern-0.355em Z$}} %was .3
{\hbox{$\sans\scriptscriptstyle Z\kern-0.255em Z$}}}} %was .2
\def\D{{\mathchoice
{\hbox{$\sans\textstyle D\kern-0.755em I$}} %was .4
{\hbox{$\sans\textstyle D\kern-0.755em I$}} %was .4
{\hbox{$\sans\scriptstyle D\kern-0.555em I$}} %was .3
{\hbox{$\sans\scriptscriptstyle D\kern-0.455em I$}}}} %was .2
\font\tensans=cmss10 at 14pt 
\newfam\sansfamb
\textfont\sansfamb=\tensans\scriptfont\sansfamb=\sevensans\scriptscriptfont
\sansfamb=\fivesans
\def\sansb{\fam\sansfamb\tensans}
\def\medZ{{\mathchoice
{\hbox{$\sansb\textstyle Z\kern-0.5em Z$}}
{\hbox{$\sansb\textstyle Z\kern-0.5em Z$}}
{\hbox{$\sansb\scriptstyle Z\kern-0.4em Z$}}
{\hbox{$\sansb\scriptscriptstyle Z\kern-0.3em Z$}}}}
\font\tensans=cmss10 at 17pt
\newfam\sansfamc
\textfont\sansfamc=\tensans\scriptfont\sansfamc=\sevensans\scriptscriptfont
\sansfamc=\fivesans
\def\sansc{\fam\sansfamc\tensans}
\def\bigZ{{\mathchoice
{\hbox{$\sansc\textstyle Z\kern-0.585em Z$}}
{\hbox{$\sansc\textstyle Z\kern-0.585em Z$}}
{\hbox{$\sansc\scriptstyle Z\kern-0.485em Z$}}
{\hbox{$\sansc\scriptscriptstyle Z\kern-0.385em Z$}}}}
\font\tensans=cmss10 at 10pt
\def\contr#1#2{\mathop{\vtop{\ialign{##\crcr
  $\hfil\displaystyle{#2}\hfil$\crcr\noalign{\kern3pt\nointerlineskip}
  \hspace{.09in}\rule[0in]{.01in}{.1in}\rule[0in]{#1in}{.01in}\rule[0in]{.01in}{.1in}\hskip6pt\crcr\noalign{\kern3pt}}}}}
\def\contrb#1#2#3{\mathop{\vtop{\ialign{##\crcr
  $\hfil\displaystyle{#3}\hfil$\crcr\noalign{\kern3pt\nointerlineskip}
  \hspace{#1in}\rule[0in]{.01in}{.1in}\rule[0in]{#2in}{.01in}\rule[0in]{.01in}{.1in}\hskip6pt\crcr\noalign{\kern3pt}}}}}
\def\namegroup#1{\begin{eqnarray}\label{#1}\nonumber\end{eqnarray}\vspace{-.5in}}
\def\hsp#1{\hspace{#1in}}
\def\rf#1{\ref{ref#1}}
\def\comment#1{\hsp{.3}\textup{#1}}

\catcode`\@=11
\def\vereq#1#2{\lower3pt\vbox{\baselineskip1.5pt \lineskip1.5pt
\ialign{$\m@th#1\hfill##\hfil$\crcr#2\crcr\sim\crcr}}}
\catcode`\@=12

\begin{document}
\begin{titlepage}
\begin{center}

May 19, 2000           \hfill UCB-PTH-00/16   \\
                                \hfill LBNL-45798    \\
%                                \hfill hep-th/9904105    \\

\vskip .75in
\def\thefootnote{\fnsymbol{footnote}}
{\large \bf More About All Current-Algebraic Orbifolds \\}

\vskip 0.3in

M. B. Halpern and J. E. Wang\footnote{E-Mail: hllywd2@physics.berkeley.edu}

\vskip 0.15in

{\em Department of Physics,
     University of California\\
     Berkeley, California 94720}\\
and\\
{\em Theoretical Physics Group\\
     Ernest Orlando Lawrence Berkeley National Laboratory\\
     University of California,
     Berkeley, California 94720}
        
\end{center}

\vskip .3in

\vfill

\begin{abstract}
Recently a construction was given for the stress tensors of all sectors of the general current-algebraic orbifold \tahh, where $A(H)$ is any current-algebraic conformal field theory with a finite symmetry group $H$.  Here we extend and further analyze this construction to obtain the mode formulation of each sector of each orbifold \tahh, including the twisted current algebra, the Virasoro generators, the orbifold adjoint operation and the commutator of the Virasoro generators with the modes of the twisted currents.  As applications, general expressions are obtained for the twisted current-current correlator and ground state conformal weight of each twisted sector of any permutation orbifold \tahh, $H\subset S_N$.  Systematics are also outlined for the orbifolds $\talieH$ of the ($H$ and Lie $h$)-invariant conformal field theories, which include the general WZW orbifold and the general coset orbifold.  Finally, two new large examples are worked out in further detail:  the general $S_N$ permutation orbifold $A(S_N)/S_N$ and the general inner-automorphic orbifold $A(H(d))/H(d)$.
\end{abstract}

\vfill

\end{titlepage}
\setcounter{footnote}{0}
\renewcommand{\thefootnote}{\alph{footnote}}

%THIS PAGE (PAGE ii) CONTAINS THE LBL DISCLAIMER
%TEXT SHOULD BEGIN ON NEXT PAGE (PAGE 1)
%\renewcommand{\thepage}{\roman{page}}
%\setcounter{page}{1}
%\mbox{ }
%
%\vskip 1in
%
%\begin{center}
%{\bf Disclaimer}
%\end{center}
%
%\vskip .2in
%
%\begin{scriptsize}
%\begin{quotation}
%This document was prepared as an account of work sponsored by the
%United
%States Government. While this document is believed to contain
%correct
% information, neither the United States Government nor any agency
%thereof, nor The Regents of the University of California, nor any
%of their
%employees, makes any warranty, express or implied, or assumes any legal

%liability or responsibility for the accuracy, completeness,

%or usefulness

%of any information, apparatus, product, or process disclosed, or

%represents that its use would not infringe privately owned rights.

%Reference herein

%to any specific commercial products process, or service by

%its trade name,

%trademark, manufacturer, or otherwise, does not necessarily

%constitute or

%imply its endorsement, recommendation, or favoring by the

%United States

%Government or any agency thereof, or The Regents of the

%University of

%California.  The views and opinions of authors expressed herein

%do not

%necessarily state or reflect those of the United States

%Government or any

%agency thereof, or The Regents of the University of California.

%\end{quotation}

%\end{scriptsize}

%

%\vskip 2in

%

%\begin{center}

%\begin{small}

%{\it Lawrence Berkeley National Laboratory is an equal opportunity

%employer.}

%\end{small}

%\end{center}

%\pagebreak

%\renewcommand{\thepage}{\arabic{page}}

\newsection{Introduction}
Orbifold theory$^{\rf{Th}-\rf{us3}}$ has a long history, yet until recently, orbifolds have been studied primarily at the level of examples.  This situation has now changed due to a recent synthesis of the principles of orbifold theory with the principles of current-algebraic conformal field theory, and we may now view at a glance the panorama of all current-algebraic orbifolds.$^{\rf{us2}}$

In particular, Ref.~\rf{us2} gave a construction\footnote{This construction drew heavily on recent advances in the theory of cyclic permutation orbifolds, including the discovery of orbifold affine algebra$^{\rf{Chris}}$ and the orbifold Virasoro master equation.$^{\rf{us1}}$} of the twisted currents $\hat{J}(\s)$ and stress \linebreak tensors $\hat{T}_\s$
\group{introarrow}
\(
T=L_H^{ab} :J_a J_b: \ \ \sa \ \ \hat{T}_\s = \lr^{n(r) \mu; -n(r),\nu}(L_H;\s) : \hat{J}(\s)_{n(r) \mu} \hat{J}(\s)_{-n(r),\nu}: 
\)
\(
\srange
\)
\reseteqn
of all sectors $\s$ of any current-algebraic orbifold $A(H)/H$.  Here $A(H)$, described by the stress tensor $T$, is any current-algebraic conformal field theory with a finite symmetry group $H$.  Technically, $A(H)$ is a member of the class of $H$-invariant CFT's$^{\rf{Lieh}-\rf{rev},\rf{us2}}$ on $g$, which includes all the CFT's with a symmetry $H\subset Aut(g)$ in the general affine-Virasoro construction.$^{\rf{vme},\rf{russ},\rf{rev}}$  The number of sectors $N_c$ of the orbifold $A(H)/H$ is the number of conjugacy classes of $H$ and the construction (\ref{introarrow}) is shown schematically in Fig.~1.

\begin{picture}(350,160)(0,0)

\put(135,90){\oval(85,70)}
\put(120,111){CFT's}
\put(125,88){\scriptsize{A(H)}}
\put(135,90){\circle{25}}

\put(147.5,90){\vector(1,0){112}}
\put(280,90){\circle{30}}
\put(268.5,85){$\frac{A(H)}{H}$}

\put(280,106){\line(0,1){15}}
\put(280,133){\circle{25}}
\put(270,131){$\scrs{\s=0}$}

\put(291.5,101.5){\line(1,1){11.7}}
\put(312,121){\circle{25}}
\put(302,119){$\scrs{\s=1}$}

%\put(324,104){\vector(-1,2){5}}
%\put(323,107){\vector(1,-2){3}}

\put(296,90){\line(1,0){15}}
\put(323,90){\circle{25}}
\put(313,88){$\scrs{\s=2}$}

%\put(322,68){\vector(1,2){5}}
%\put(323,71){\vector(-1,-2){3}}
%\put(291,77){\line(4,-5){10}}
%\put(310,57){\circle{25}}
%\put(300,65){$\scrs{\s=2}$}

\put(300,60){\circle*{2}}
\put(305,65){\circle*{2}}
\put(310,70){\circle*{2}}

\put(280,74){\line(0,-1){15}}
\put(280,47){\circle{25}}
\put(273,51){$\tyny{\s=}$}
\put(270,44){$\tyny{N_c\hsp{-.03}-\hsp{-.02}1}$}

%\put(250,54){\circle*{2}}

%\qbezier(327,123)(370,90)(297,57)
%\put(329,122){\vector(-2,1){5}}
%\put(329,58){\vector(-2,-1){5}}

%\qbezier(255,42)(250,22)(270,42)
%\put(268,39){\vector(-1,1){5}}
%\put(297,39){\vector(1,1){5}}

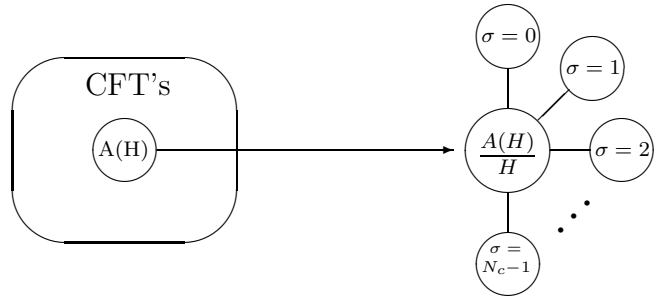
\put(50,10) {Fig. 1.  The $H$-invariant CFT's A(H) and their orbifolds $A(H)/H$.}
\end{picture}

The \textit{orbifold duality transformation} indicated in (\ref{introarrow})
\(
L_H \sa \lr(L_H;\s)
\)
gives the twisted inverse inertia tensor $\lr$ of each sector $\s$ in terms of the $H$-invariant inverse inertia tensor $L_H$ of the $H$-invariant CFT $A(H)$.  Other orbifold duality transformations exist for other twisted tensors of the orbifold, and the explicit form of the generic orbifold duality transformation is a discrete Fourier transform.

The central ingredients underlying the breadth and depth of this construction are

\bu \ local formulation of the theory in terms of currents, OPE's and OPE isomorphisms

\bu \ the $L_H^{ab}$ formulation of the $H$-invariant CFT's in the general affine-Virasoro construction.

\noindent The local formulation (as opposed to a mode formulation) is the key to the orbifold duality transformations, while the $L_H^{ab}$ formulation allows us to study all current-algebraic orbifolds at once, or any example.

The organization of this paper is best summarized by the names of its sections:

\vspace{.1in}
2 Local Formulation of Current-Algebraic Orbifolds 

\vspace{.03in}
3 The Mode Formulation of Orbifold Theory 

\vspace{.03in}
4 The Orbifold Adjoint Operation

\vspace{.03in}
5 The $\hat{T}\hat {J}$ OPE's of $A(H)/H$

\vspace{.03in}
6 The Orbifolds of the ($H$ and Lie $h$)-invariant CFT's

\vspace{.03in}
7 About Permutation Orbifolds

\vspace{.03in}
8 The Permutation Orbifolds $A(S_N)/S_N$

\vspace{.03in}
9 The Inner-Automorphic Orbifolds $A(H(d))/H(d)$.
\vspace{.1in} 
\\
In particular, Sec.~\ref{reviewsec} reviews and extends the local formulation$^{\rf{us2}}$ of current-algebraic orbifold theory in a form which exhibits the natural grading of the orbifold operator products.  In Secs.~\ref{modesec}--\ref{TJsec}, we work out the implied mode formulation and other consequences of the local formulation.  We mention in particular four results obtained in the mode formulation for all sectors $\s$ of all orbifolds $A(H)/H$: the general twisted current algebra in Eq.~(\ref{ghatgroup}), the orbifold Virasoro generators in Eq.~(\ref{Lmode}), the orbifold adjoint operation in Eqs.~(\ref{adjointgroup}) and (\ref{twistLdagger}) and the commutator (\ref{LJcommutator}) of the Virasoro generators with the modes of the twisted currents.

The ($H$ and Lie $h$)-invariant CFT's $\alieH$ are those ``doubly-invariant'' CFT's with both a finite symmetry and a Lie symmetry$^{\rf{Lieh}-\rf{rev},\rf{Jan}, \rf{us1}}$ although we mod out only by the finite symmetry to obtain (see Sec.~\ref{liehsec}) the orbifolds $\talieH$.  The general WZW orbifold and the general coset orbifold are discussed as special cases of $\talieH$ in Subsec.~\ref{Ksec}.  The twisted current-current correlators and ground state conformal weights of the general permutation orbifold are computed in Sec.~\ref{cwsec}.  We also work out two new large examples in further detail, including the general $S_N$ permutation orbifold in Sec.~\ref{SNsec} and the general inner-automorphic orbifold in Sec.~\ref{innersec}.  The story of the inner-automorphic orbifolds $A(H(d))/H(d)$ is particularly interesting, not least because of their overlap with the orbifolds $\talieH$.  Indeed, we will argue that this overlap contains almost all the inner-automorphic orbifolds which can be equivalently described by stress-tensor spectral flow$^{\rf{BH},\rf{FH},\rf{vme},\rf{rev}}$ whereas the generic inner-automorphic orbifold apparently can not be described in this way. 

The seminal case$^{\rf{us2}}$ of the cyclic permutation orbifolds \taz \ is used as an example in various sections, and the appendices include the setups for the permutation orbifolds $A(\dl)/\dl$ and the outer-automorphic orbifolds.

This is as far as we have worked out the consequences of the ``orbifold program'', but much remains to be done, including other large examples and analysis of the situation when $A(H)$ has a larger chiral algebra.

\newsection{Local Formulation of Current-Algebraic Orbifolds}
\label{reviewsec}
In this section we review and extend the local formulation of the general current-algebraic orbifold$^{\rf{us2}}$.  The extension includes a new periodic notation and new ``selection rules'', both of which are necessary to exhibit the natural grading of the orbifold operator products.  As an illustration, the seminal case of the general cyclic permutation orbifold$^{\rf{us2}}$ is recalled in Subsec.~\ref{cyclicsec}.

\subsection{The current-algebraic CFT's}
The study of current-algebraic CFT's begins with the general affine Lie algebra$^{\rf{Kac},\rf{Moody},\rf{BH},\rf{rev}}$ 
\group{affineliegroup}
\(
J_a(z) J_b (w) = \frac{G_{ab}}{(z-w)^2} + \frac{i f_{ab}^{\ \ c} J_{c}(w)}{z-w} + O((z-w)^0) \sp J_a(ze^{2 \pi i})=J_a(z) \label{firstJJOPE}
\)
\(
[J_a(m), J_b (n)] =  i f_{ab}^{\ \ c} J_{c}(m+n)+m G_{ab}\delta_{m+n,0} \label{untwistJalg}
\)
\(
g=\oplus_I \ g^I \sp G_{ab}=\oplus_I  \ k_I \eta_{ab}^{I} \sp a,b,c=1,...,\textup{dim} g \sp m,n \in \z
\)
\reseteqn
where $G_{ab}$ and $f_{ab}^{\ \ c}$ are the metric and structure constants of semisimple\footnote{Non-compact versions of $g$ are included and abelian components of $g$ are easily included$^{\rf{vme}}$ as well (see Subsec.~\ref{Virsec}).} Lie $g$.  The desired real form of the affine algebra is specified by the adjoint operation
\group{untwistdaggergroup}
\(
J_a(m)^\dagger =\stheta_a^{\ b} J_b(-m) \label{untwistJdagger} 
\sp \stheta_a^{\ c \ *} \stheta_c^{\ b}=\delta_a^{\ b} \sp \stheta^* \stheta=1  \label{rrstar}
\)
\(
G_{ab}^{\ \ *}=\stheta_a^{\ c} \stheta_b^{\ d} G_{cd} \label{gstar} \sp
f_{ab}^{\ \ c \ *}=\stheta_a^{\ d} \stheta_b^{\ e} f_{de}^{\ \ f} \stheta_f^{\ c \ *} \label{fstar} 
\)
\reseteqn
where the conjugation matrix $\r$ is unity in any Cartesian basis, and corresponding forms follow in other bases.  For integer level $x_I=2k_I/\psi_I^2$ of compact $g^I$ with highest root $\psi_I$, the adjoint operation in (\ref{rrstar}) guarantees unitarity of the affine Hilbert space.

The general affine-Virasoro construction$^{\rf{vme},\rf{russ},\rf{5ofus},\rf{Obers},\rf{rev}}$ is
\group{VMEgroup}
\(
T(z)T(w)=\frac{c/2}{(z-w)^4}+(\frac{2}{(z-w)^2}+\frac{\partial_w}{z-w})T(w)+ \reg
\)
\(
T(z)=L^{ab}:J_a(z) J_b(z): \sp L^{ab}=L^{ba}  \label{VMET}
\)
\(
L^{ab}= 2 L^{ac} G_{cd} L^{db} - L^{cd} L^{ef} f_{ce}^{\ \ a}f_{df}^{\ \ b}-L^{cd} f_{ce}^{\ \ f}f_{df}^{\ \ (a}L^{b) e} \sp c=2G_{ab}L^{ab} \label{VME}
\)
\reseteqn
where we have chosen\footnote{Because of the symmetry $L^{ab}=L^{ba}$, the stress tensor $T$ in (\ref{VMET}) can equivalently be described by various normal ordering prescriptions.  OPE normal ordering however comes equipped with an efficient calculus$^{\rf{Bouw}}$ for computing OPE's of products of ordered products and this prescription has been extended$^{\rf{us1},\rf{us2}}$ to compute corresponding OPE's in the twisted sectors of orbifolds (see Subsec.~\ref{introTsec}).} OPE normal ordering 
\(
: J_a(w)J_b(w): \  \ = \oint_w\frac{dz}{\tp}\frac{J_a(z) J_b(w)}{z-w} 
\)
and the inverse inertia tensor $L^{ab}$ satisfies the Virasoro master equation in (\ref{VME}).  The $L^*$ relation given below$^{\rf{rev}}$ 
\(
L^{ab \hsp{.02} *}= L^{cd} \stheta_c^{\ a \hsp{.02} *} \stheta_d^{\ b \hsp{.02} *} 
\ \lra \ L(m)^\dagger=L(-m) \label{Lstar}
\)
guarantees the indicated adjoint of the Virasoro generators, and this adjoint operation guarantees the unitarity of the CFT, given the unitarity of the affine Hilbert space.

\subsection{The automorphism group $H$ and the $H$-invariant CFT's}
\label{autosec}
We consider any $H$-covariant algebra $g$, that is, any algebra $g$ with a finite-order\footnote{Automorphism groups of infinite order (including Lie groups) are also included formally.  See our remarks in Subsecs.~\ref{Liehsec} and \ref{autoinnersec}.} automorphism group $H\subset Aut(g)$.  The automorphism group $H$ acts on the $g$ currents $J$ in matrix representation $\w$, 
\group{gfwrelationgroup}
\(
J_a (z)\p = \w (h_\s)_a^{\ b} J_b(z), \hspace{.3in} \w(h_\s) \in H\subset Aut(g) \label{Jprime}
\)
\(
\w (h_\s)_a^{\ c} \w^\dagger (h_\s)_c^{\ b}=\delta_a^{\ b} \label{wunitary}
\)
\getletter{fgautoslett}
\vspace{-.2in}
\(
\w (h_\s)_a^{\ c} \w (h_\s)_b^{\ d} G_{cd}=G_{ab} \label{GwwG} 
\sp \w (h_\s)_a^{\ d} \w (h_\s)_b^{\ e} f_{de}^{\ \ f} \w^\dagger (h_\s)_f^{\ c}= f_{ab}^{\ \ c} \label{fwwfw} 
\)
\getletter{guesslett}
\vspace{-.2in}
\(
\w(h_\s)^* \stheta \ \w^\dagger(h_\s)= \stheta  \ \lra \ 
J_a(m)\p \ ^{ \dagger} = J_a(m)^{\dagger}\ \p  \label{guess}
\)
\reseteqn
where the relations in (\ref{gfwrelationgroup}) hold for all $h_\s \in H$.  The relations (\ref{fwwfw}) -- which express the $H$-invariance of $G_{ab}$ and $f_{ab}^{\ \ c}$ -- guarantee that $J\p$ satisfies the same algebra as $J$, so that $H\subset Aut(g)$ is also an automorphism group of affine $g$.  Consequently, we often refer to the affine algebra as an $H$-covariant affine algebra on $g$.  The $H$-covariance of the conjugation matrix $\r$ in (\ref{guess}) is equivalent to the statement that the automorphism group preserves the real form of the affine algebra (\ref{untwistJalg}).

The $H$-invariant CFT's$^{\rf{Lieh}-\rf{rev},\rf{us2}}$ $A(H)$ are those CFT's for which the inverse inertia tensor $L_H$ is invariant under $H$
\alpheqn
\(
T = L_H^{ab}:J_a J_b: \sp L_H^{cd} \ \w (h_\s)_c^{\ a} \w (h_\s)_d^{\ b}=L_H^{ab} \sp \forall\  h_\s \in H  \subset Aut(g) \label{Hinv}
\)
\(
T\p \ = \ L_H^{ab}:J_a \ \p J_b \ \p: \ \  = \ L_H^{ab}:J_a J_b: \ \ = \ T \ .
\)
\reseteqn
These are the CFT's whose stress tensors also describe the untwisted sectors of the general current-algebraic orbifold \tahh.  For each symmetry group $H$, it is known$^{\rf{Lieh}}$ that the inverse inertia tensors of the $H$-invariant CFT's satisfy a consistent reduced Virasoro master equation (with an equal number of equations and unknowns).

\begin{sloppypar}The simplest $H$-invariant CFT's include $A_g(H)$, $H\subset Aut(g)$, whose stress tensor is the affine-Sugawara construction$^{\rf{BH},\rf{Halp}-\rf{Segal},\rf{rev}}$ on $g$, and the general $H$-invariant coset construction$^{\rf{us2},\rf{us3}}$ $\frac{g}{h}(H)$.  We will return to these popular examples as illustrations below (see Subsecs.~\ref{Liehsec}--\ref{Ksec} and \ref{innerVirsec}).\end{sloppypar}

In what follows all inverse inertia tensors $L_H$ are $H$-invariant, so we may drop the subscript $L_H^{ab} \ \rightarrow L^{ab}$ without confusion.

\subsection{The $H$-eigenvalue problem}
\label{Heigenvsec}
Choosing one representative $h_\s$ from each conjugacy class $\s$ of $H$, the $H$-eigenvalue problem is defined as 
\group{eigenprobgroup}
\(
\w (h_\s)_a^{\ b} \uds_b^{\ n(r) \mu} = \uds_a^{\ n(r) \mu}  E_{n(r)}(\s) 
\sp \srange \label{eigenprob}
\)
\(
U^\dagger (\s) U(\s)=1 \sp E_{n(r)}(\s)=e^{-\frac{2 \pi i n(r)}{\r(\s)}} \sp n(r)\in \z \ . \label{Edef}
\)
\reseteqn
Here $N_c$ is the number of conjugacy classes of $H$ and the integer $r$ runs over the distinct eigenvalues of the problem at each $\s$.  The eigenvalues $E_{n(r)}(\s)$ are labeled by the spectral index $n(r)\equiv n(r;\s)$ and the degeneracy index $\mu\equiv \mu(n(r;\s))$ labels the degenerate eigenstates with eigenvalue $E_{n(r)}(\s)$.  The quantity $\r(\s)\in \z^+$ is the order of the automorphism $h_\s$, defined as the smallest positive integer satisfying
\(
\w(h_\s)^{\r(\s)}=1 \ . \label{order}
\)
As a convention, we assign $\s=0$ to the trivial automorphism
\(
\w(h_0)_a^{\ b}=\delta_a^{\ b} \sp U(0)=U^\dagger(0)=1 \sp E_{n(r)}(0)=\r(0)=1 \ .
\)
Other useful forms of the eigenvalue problem include
\group{eigenprobU}
\(
\w =U^\dagger  E U \sp \w^\dagger=U^\dagger E^* U
\)
\(
\w U^\dagger=U^\dagger E ,\hsp{.1} U \w  =  E  U  
,\hsp{.1} U^* \w^* =  E^* U^*  ,\hsp{.1}  \w^\dagger U^\dagger 
=  U^\dagger E^*  ,\hsp{.1} U \w^\dagger = E^* U
\)
\reseteqn
where $\textup{diag}(E)=\{ E_{n(r)} \}$.  As seen here, we often suppress the label $\s$ for brevity.

The eigenvalues $E_{n(r)}$ are invariant under the transformation $n(r)\rightarrow n(r)\pm \r(\s)$, so we may require this periodicity for the eigenvectors as well,
\alpheqn
\(
E_{n(r)\pm \r(\s)}(\s)=E_{n(r)}(\s)
\)
\(
U^\dagger(\s)_a^{\ n(r)\pm \r(\s),\mu}=U^\dagger(\s)_a^{\ n(r)\mu} \sp U(\s)_{n(r)\pm \r(\s),\mu}^{\hsp{.65} a} =U(\s)_{n(r)\mu}^{\hsp{.3} a} \label{Uperiodicity}
\)
\(
\hsp{.8} \sum_{r,\mu} U^\dagger(\s)_a^{\ n(r) \mu} U(\s)_{n(r) \mu}^{\ \ \ \ \ \ b} = \delta_a^{\ b} \sp U(\s)_{n(r) \mu}^{\ \ \ \ \ \ a} U^\dagger(\s)_a^{\ n(s) \nu} = \delta_{\mu}^{\ \nu} \delta_{n(r)}^{\hsp{.2} n(s)}  \label{Uunitary}
\)
\vspace{-.2in}
\(
\delta_{n(r)}^{\hsp{.2} n(s)} \equiv \delta_{n(r)- n(s),0 \mod \r(\s)} \sp \sum_s f(n(s)) \ \delta_{n(s)}^{\hsp{.2} n(r)}=f(n(r)) 
\)
\reseteqn
where the last identity holds for all $f$ with period $\r(\s)$.  In what follows, relations such as the first relation in (\ref{Uunitary}) will generally be written as 
\(
U^\dagger(\s)_a^{\ n(r) \mu} U(\s)_{n(r) \mu}^{\ \ \ \ \ \ b} = \delta_a^{\ b}
\)
with an implied summation on repeated indices.

Using the $H$-covariance of $\r$ in (\ref{guess}), the eigenvalue problem can be recast in the form
\(
\w (\r U^\dagger)^* = (\r U^\dagger)^* E^* \ . \label{starevproblem}
\)
This tells us first that the eigenvalues $E^*$ are in the spectrum $E$
\(
E_{n(r)}(\s)^* = E_{\r(\s) -n(r)}(\s)=E_{-n(r)}(\s) \label{Estar}
\)
and, in particular, that $-n(r) \in \{ n(r) \}$ is a spectral index when $n(r)$ is a spectral index.  Moreover, (\ref{Estar}) tells us that the degeneracy of $E_{-n(r)}(\s)$ is the same as that of $E_{n(r)}(\s)$.  

Since $n(r)$ is only determined mod $\r(\s)$, it is useful to define the \textit{twist class} $\bar{n}(r)$ 
\(
\bar{n}(r) \equiv \overline{n(r)} \equiv n(r)- \r(\s) \lfloor \frac{n(r)}{\r(\s)} \rfloor \sp \bar{n}(r)\in\{ 0,...,\r(\s)-1 \}  \label{nnbar}
\)
where $\lfloor x \rfloor$ is the largest integer less than or equal to $x$.  The twist class is an integer which evaluates $n(r)$ in the fundamental range shown, and twist classes will control the monodromies of twisted fields in the orbifolds below.  Note that $\bar{n}(r)=n(r)$ when $n(r)$ is in the fundamental range.  We also compute the twist class corresponding to $-n(r)$ 
\ba
\overline{-n(r)}=\overline{-\bar{n}(r)}=\left\{
\begin{array}{cc}
\r(\s)-\bar{n}(r) \textup{ when } \bar{n}(r)\neq 0 \\
0 \comment{\hsp{.4} when $\bar{n}(r)= 0$}
\end{array} \right. 
\ea
where we have used $-\lfloor -x \rfloor = \lfloor x \rfloor +1$ for $ x $ not an integer.

In what follows, the description of each sector $\s$ of the general current-algebraic orbifold \tahh \ is given in terms of the solution $\{ U^\dagger(\s)$, $E(\s) \}$  of the $H$-eigenvalue problem (\ref{eigenprobgroup}).  It is unlikely that a solution of the $H$-eigenvalue problem can be found across all $H$, but solution of the $H$-eigenvalue problem is straightforward for any particular choice of $H$: The solutions for the cyclic permutation groups $H=\zl$
\group{zlUgroup}
\(
g=\oplus_{I=0}^{\l-1} \gbn^I \sp \gbn^I \cong \gbn
\)
\(
a \rightarrow a,I \sp a=1,...,\textup{dim} \gbn \sp I=0,...,\l-1
\)
\(
\w (h_\s)_a^{\ b} \rightarrow \w (h_\s)_{aI}^{\ \ \ bJ}=\delta_a^{\ b} \delta_{I+\s, J \mod \l} \sp \s=0,...,\l-1
\)
\(
n(r),\mu \rightarrow r, aj \sp \bar{r}=0,...,\r(\s)-1 \sp j=0,...,\frac{\l}{\r(\s)}-1
\)
\(
U^\dagger(\s)_a^{\ n(r)\mu} \rightarrow U^\dagger(\s)_{aI}^{\ \ \ rbj}=\frac{\delta_a^{\ b}}{\sqrt{\r(\s)}} e^{\frac{2 \pi i N(\s)r (j-I)}{\l}} \delta_{j,I \mod \frac{\l}{\r(\s)}} \sp E_r(\s) =e^{-\frac{2 \pi i r}{\r(\s)}} \label{zlU}
\)
\reseteqn
were given in Ref.~\rf{us2}, which also defines the integers $N(\s)$.  In this case, the labels $(a,j)$ are the degeneracy indices of the spectral indices $n(r)=r$, with $\bar{r}=r-\r(\s)\lfloor r/\r(\s) \rfloor$.  The $U^\dagger$ periodicity $r\rightarrow r\pm \r(\s)$ is a consequence of the support $((j-I)\r/\l) \in \z$ of the Kronecker factor.  More generally, the eigenvectors of each $H$-eigenvalue problem are the basis elements of discrete Fourier transforms with period $\r(\s)$ in the spectral indices $\{ n(r) \}$.

The solutions for the permutation group $H=S_N$ and the general group of inner automorphisms of simple $g$ will be given in Secs.~\ref{SNsec} and \ref{innersec} respectively.  The setups for the permutation group $H=\dl$ and the general group of outer automorphisms of simple $g$ are given in Apps.~\ref{dlapp} and \ref{outersapp} respectively.  Many other large examples remain to be worked out in further detail, including the other permutation subgroups of $S_N$ (see also Subsec.~\ref{twistalgsubsec}).

\subsection{The twisted currents of the orbifold \tahh}
We turn next to the twisted currents $\hat{J}$ of the current-algebraic orbifold
\(
\frac{A(H)}{H} \sp H \subset Aut(g)
\)
where $A(H)$ is any $H$-invariant CFT.

For each conjugacy class $\s$, the \textit{eigencurrent}$^{\rf{us2}}$ $\j (\s)$ is defined as the linear combination of the untwisted currents $J$
\alpheqn
\(
\j_{n(r) \mu}(z) \equiv {\foot{\chi(\s)_{n(r) \mu}}} U(\s)_{n(r) \mu}^{\ \ \ \ \ \ a} J_a(z), \hspace{.3in} J_a(z)= {\foot{\chi(\s)_{n(r) \mu}^{-1}}} U^\dagger(\s)_a^{\ n(r) \mu} \j_{n(r) \mu}(z) \label{geneigenj}
\)
\(
\j_{n(r)\mu}(z)\p=E_{n(r)}(\s) \hsp{.03}\j_{n(r) \mu}(z) 
\)
\reseteqn
which (according to (\ref{Jprime}) and (\ref{eigenprobU})) has a diagonal response $E_{n(r)}(\s)$ to the automorphism group.  The numbers ${\foot{\chi(\s)_{n(r)\pm \r(\s), \mu}}}={\foot{\chi(\s)_{n(r) \mu}}}$ with ${\foot{\chi(0)}}=1$ comprise an otherwise arbitrary set of normalization constants.

In general orbifold theory$^{\rf{DVVV},\rf{DMVV},\rf{Chris},\rf{Ban1},\rf{us2},\rf{us3}}$ the sectors of each orbifold \tahh \ are labeled by the conjugacy classes $\srange$ of $H$.  The twisted currents $\hat{J} (\s)$ of sector $\s$ are obtained from the eigencurrents of sector $\s$ by the \textbf{OPE isomorphism}$^{\rf{Chris},\rf{us2}}$ 
\group{Jarrowgroup}
\(
\j_{n(r)\mu}(z) \sa \hat{J}_{n(r) \mu}(z) \label{JJhatduality}
\)
\(
\hsp{-.14} \textup{automorphisms } E_{n(r)}(\s) \ \sa \ \textup{monodromies } E_{n(r)}(\s) \label{Earrow}
\)
\reseteqn
which includes  the automorphism/monodromy exchange in (\ref{Earrow}).  

The OPE isomorphism (\ref{Jarrowgroup}) gives the \textit{twisted current system}$^{\rf{us2}}$ of sector $\s$
\group{fulltwistedopegroup}
\(
\hat{J}_{n(r) \mu}(z) \hat{J}_{n(s) \nu}(w)=\frac{\g_{n(r) \mu;n(s) \nu}(\s)}{(z-w)^2}+\frac{i \fr_{n(r) \mu;n(s) \nu}^{\ \ \ \ \ \ \ \ \ \ \ n(t) \delta}(\s) \hat{J}_{n(t) \delta}(w)}{z-w} + O((z-w)^0) \label{hatJOPErecap}
\)
\(
\hat{J}_{n(r) \mu}(ze^{2 \pi i})= E_{n(r)}(\s) \hat{J}_{n(r) \mu}(z) \sp E_{n(r)}(\s)=e^{-\frac{2 \pi in(r)}{\r(\s)}} \label{hatJmonodromy}
\)
\(
\hat{J}_{n(r)_\pm \r(\s),\mu}(z)  =  \hat{J}_{n(r) \mu}(z) \label{Jperiod} 
\sp  \hat{J}_{\r(\s)-n(r), \mu}(z) = \hat{J}_{-n(r),\mu}(z) \label{Jnegperiod}
\)
\(
\g_{n(r) \mu; n(s) \nu}(\s) = {\foot{\chi(\s)_{n(r) \mu}}}{\foot{\chi(\s)_{n(s) \nu}}} U(\s)_{n(r) \mu}^{\ \ \ \ \ \ a} U(\s)_{n(s) \nu}^{\ \ \ \ \ \ b} G_{ab}  \label{twistg}
\)
\getletter{twistflett}
\vspace{-.2in}
\(
\fr_{n(r) \mu;n(s) \nu}^{\ \ \ \ \ \ \ \ \ \ \ \ n(t) \delta}(\s) = {\foot{\chi(\s)_{n(r) \mu}}}{\foot{\chi(\s)_{n(s) \nu}}}{\foot{\chi(\s)_{n(t) \delta}^{-1}}} U(\s)_{n(r) \mu}^{\ \ \ \ \ \ a} U(\s)_{n(s) \nu}^{\ \ \ \ \ \ b} f_{ab}^{\ \ c} \uds_c^{\ n(t) \delta} \label{twistf}
\)
\(
\# \{ \hat{J} (\s)\}=\# \{ J \}=\textup{dim}g \label{samedimensions}
\)
\reseteqn
and (\ref{hatJmonodromy}) tells us that the twisted current $\hat{J}_{n(r) \mu}$ has twist class $\bar{n}(r)$ in (\ref{nnbar}).  The relations (\ref{twistg}) and (\ref{twistf}) are the \textit{orbifold duality transformations}
\(
G \sa \g(G;\s)\sp f \sa \fr(f;\s)
\)
from the untwisted metric $G_{ab}$ and structure constants $f_{ab}^{\ \ c}$ to the twisted metric $\g(\s)$ and the twisted structure constants $\fr(\s)$ of sector $\s$.  

The twisted tensors $\g(\s)$ and $\fr(\s)$ inherit the spectral index periodicity 
\(
\g_{n(r) \pm \r(\s), \mu; n(s) \nu}(\s)=\g_{n(r) \mu; n(s) \nu}(\s)
\sp \fr_{n(r) \pm \r(\s), \mu;n(s) \nu}^{\hsp{.95} n(t) \delta}(\s) = \fr_{n(r) \mu;n(s) \nu}^{\ \ \ \ \ \ \ \ \ \ \ n(t) \delta}(\s) 
\)
from the periodicity (\ref{Uperiodicity}) of the eigenvectors $U(\s)$ and $U^\dagger(\s)$, and similarly for $n(s)$ and $n(t)$.  For the same reason, the periodicity $n(r)\rightarrow n(r)\pm \r(\s)$ holds for each spectral index of each of the twisted tensors introduced below.

For the untwisted sector $\s=0$ one has $U^\dagger(0)=1$, $\bar{n}(r)=0$, $\mu=a$ and
\(
\g_{ab}(0)=G_{ab} \sp \fr_{ab}^{\ \ c}(0)=f_{ab}^{\ \ c} \sp \hat{J}_a(\s=0)=\j_a(\s=0)=J_a
\)
so the system (\ref{fulltwistedopegroup}) reduces in this case to the OPE's (\ref{firstJJOPE}) of the affine Lie algebra of the $H$-covariant algebra $g$.  The number of twisted currents in (\ref{samedimensions}) is independent of $\s$ because $U^\dagger(\s)$ is an invertible square matrix.  The twisted current system (\ref{fulltwistedopegroup}) was called $\gbn(H\subset Aut(g);\s)$ in Ref.~\rf{us2}.   

The twisted metric of sector $\s$ also satisfies 
\group{twistedggroup}
\(
\g_{n(r) \mu; n(s) \nu}(\s)=\g_{n(s) \nu;n(r) \mu}(\s)
\)
\(
\g_{n(r) \mu; n(s) \nu}(\s)(1-E_{n(r)}(\s)E_{n(s)}(\s))=0 \label{dualg}
\)
\(
\g_{n(r) \mu; n(s) \nu}(\s) = \delta_{n(r)+n(s),0 \mod \r(\s)} \g_{n(r) \mu; -n(r), \nu}(\s)   \label{gwithdelta}
\) 
\reseteqn
where the \textit{selection rule} for $\g$ in (\ref{dualg}) is the dual\footnote{To prove the selection rule (\ref{dualg}), use (\ref{GwwG})  to replace $G$ by $\w^2 G$ in the orbifold duality transformation (\ref{twistg}), followed by the appropriate form of the $H$-eigenvalue problem in (\ref{eigenprobU}).  All the selection rules below follow similarly from the orbifold duality transformations, the $H$-eigenvalue problem and the corresponding $H$-invariances of the untwisted tensors.} in sector $\s$ of the $H$-invariance of $G$ in (\ref{GwwG}), and (\ref{gwithdelta}) is the solution\footnote{When $H$ is a direct product group, further Kronecker factors can occur in the reduced twisted tensor $\g_{n(r)\mu;-n(r),\nu}(\s)$.  Similarly, further Kronecker factors can occur in the reduced forms of each twisted tensor below.} of the selection rule.  Similarly, the twisted structure constants of sector $\s$ satisfy
\group{frpropertiesgroup}
\(
\fr_{n(r) \mu;n(s) \nu}^{\ \ \ \ \ \ \ \ \ \ \ n(t) \delta}(\s)=-\fr_{n(s) \nu;n(r) \mu}^{\ \ \ \ \ \ \ \ \ \ \ n(t) \delta}(\s)
\)
\[
\fr_{n(r) \mu; n(s) \nu}^{\ \ \ \ \ \ \ \ \ \ \ n(u) \mu \p}(\s) \fr_{n(t) \delta; n(u) \mu \p}^{\hsp{.62} n(v) \gamma}(\s)
+ \fr_{n(s) \nu;n(t) \delta}^{\ \ \ \ \ \ \ \ \ \ \ n(u) \mu \p}(\s) \fr_{n(r) \mu; n(u) \mu \p}^{\hsp{.65} n(v) \gamma}(\s)
\]
\(
\hsp{1.2} + \fr_{n(t) \delta; n(r) \mu }^{\hsp{.6} n(u) \mu \p}(\s) \fr_{n(s) \nu; n(u) \mu \p}^{\hsp{.65} n(v) \gamma}(\s) =0 \label{jacobi}
\)
\(
\fr_{n(r) \mu;n(s) \nu; n(t) \delta}(\s) \equiv \fr_{n(r) \mu;n(s) \nu}^{\ \ \ \ \ \ \ \ \ \ \ n(u) \epsilon}(\s) \g_{n(u) \epsilon; n(t) \delta}(\s)=-\fr_{n(r) \mu;n(t) \delta; n(s) \nu}(\s) \label{frantisymm}
\)
\(
\fr_{n(r) \mu;n(s) \nu}^{\ \ \ \ \ \ \ \ \ \ \ n(t) \delta}(\s) \ (1-E_{n(r)}(\s)E_{n(s)}(\s)E_{n(t)}(\s)^*)=0  \label{dualf}
\)
\(
\fr_{n(r) \mu;n(s) \nu}^{\ \ \ \ \ \ \ \ \ \ \ n(t) \delta}(\s) = \delta_{n(r)+n(s)-n(t),0 \mod \r(\s)} \fr_{n(r) \mu,n(s) \nu}^{\ \ \ \ \ \ \ \ \ \ \ n(r)+n(s), \delta}(\s)    \label{fwithdelta} 
\)
\getletter{friszerolett}
\vspace{-.2in}
\(
\fr_{n(r) \mu,n(s) \nu}^{\ \ \ \ \ \ \ \ \ \ \ n(r)+n(s), \delta}(\s) =0 \hsp{.3} \textup{unless } n(r)+n(s) \in \{ n(r) \} \ . 
\)
\reseteqn
The relation (\ref{jacobi}) is the Jacobi identity of the twisted structure constants, and the twisted structure constants in (\ref{frantisymm}) with all indices down are totally antisymmetric. The $\fr$-selection rule (\ref{dualf}) (and its solution in (\ref{fwithdelta},\ref{friszerolett})) is the dual in sector $\s$ of the $H$-invariance (\ref{fwwfw}) of the untwisted structure constants. 

The $H$-invariance conditions (\ref{fwwfw}) also imply\footnote{To prove that $\g$ is a class function, use the duality transformation (\ref{twistg}), the change of $U$ in (\ref{Uv}) and the $H$-invariance of $G$ in (\ref{GwwG}).  With the relevant duality transformation and $H$-invariance, each twisted tensor introduced below is similarly proved to be a class function.} that the twisted metric and twisted structure constants are class functions$^{\rf{us2}}$ under conjugation in $H$
\group{conjugationgroup}
\(
\w(h_\s) \rightarrow v^\dagger(\s) \w(h_\s) v(\s) \sp v(\s) v^\dagger(\s) =1 \sp v(\s)\in H \label{evunderconjugation}
\)
\(
U(\s) \rightarrow U(\s)v(\s) \sp U^\dagger(\s) \rightarrow v^\dagger(\s) U^\dagger(\s) , \hsp{.2}  \label{Uv}
\)
\(
\g (U(\s)v(\s);\s) = \g (U(\s);\s) \sp \fr (U(\s)v(\s);\s) = \fr (U(\s);\s)
\)
\reseteqn
where the $H$-eigenvalue problem (\ref{eigenprobU}) was used to obtain the change of $U(\s)$, $U^\dagger(\s)$ under conjugation.

A more informative presentation of the twisted current system (\ref{fulltwistedopegroup}) is 
\group{explicitJJOPEgroup}
\[
\hsp{-2} \hat{J}_{n(r) \mu}(z) \hat{J}_{n(s) \nu}(w)=\frac{\delta_{n(r)+n(s),0 \mod \r(\s)} \g_{n(r) \mu;-n(r), \nu}(\s)}{(z-w)^2}
\]
\(
\hsp{2} +\frac{i \fr_{n(r) \mu;n(s) \nu}^{\ \ \ \ \ \ \ \ \ \ \ n(r)+n(s),\delta}(\s) \hat{J}_{n(r)+n(s), \delta}(w)}{z-w} + \reg \label{twistJJOPEall}
\)
\(
\hat{J}_{n(r) \mu}(ze^{2 \pi i})= e^{-\frac{2 \pi in(r)}{\r(\s)}} \hat{J}_{n(r) \mu}(z) \sp \srange  \label{explcitJJmonodromies}
\)
\reseteqn
where the $\fr$ term is summed only on the repeated degeneracy index $\delta$.  In this form, we have incorporated the solutions of the $\g$- and $\fr$-selection rules to exhibit the grading of the orbifold system.  Looking back, one sees that this grading is a consequence of the $H$-covariance of the untwisted affine algebra $g$.

\subsection{The general orbifold affine-Virasoro construction} 
\label{introTsec}
The stress tensor $\hat{T}_\s$ of sector $\s$ of \tahh \ follows by first rewriting the stress tensor $T$ of the untwisted sector in terms of the eigencurrents and then using the OPE isomorphism (\ref{JJhatduality}) to obtain\footnote{Schematically, $T=L:JJ:=\lr:\j \j: \sa \hat{T}_\s=\lr:\hat{J} \hat{J}:$.} the \textbf{derived OPE isomorphism}$^{\rf{us2}}$
\(
T(z) \sa \hat{T}_\s (z) \label{TThatarrow} \ .
\)
The result is the \textit{general orbifold affine-Virasoro construction}$^{\rf{us2}}$
\group{twistTrecapgroup}
\(
\hat{T}_\s(z)\hat{T}_\s(w)=\frac{\hat{c}(\s)/2}{(z-w)^4}+(\frac{2}{(z-w)^2}+\frac{\partial_w}{z-w})\hat{T}_\s(w) + \reg \label{twistTTOPE}
\)
\(
\hat{T}_\s (z)=\lr^{n(r) \mu; n(s) \nu}(\s):\hat{J}_{n(r) \mu}(z) \hat{J}_{n(s) \nu}(z): \sp \hat{c}(\s)=c=2 G_{ab}L^{ab} \label{firsttwistT}
\)
\(
\lr^{n(r) \mu; n(s) \nu}(\s)= {\foot{\chi(\s)_{n(r) \mu}^{-1}}} {\foot{\chi(\s)_{n(s) \nu}^{-1}}}  L^{ab} U^\dagger(\s)_a^{\ \ n(r) \mu} U^\dagger(\s)_b^{\ \ n(s) \nu} \label{lr}
\)
\(
\lr^{n(r) \mu; n(s) \nu}(\s)=\lr^{n(s) \nu; n(r) \mu}(\s)
\)
\reseteqn
which reduces in the untwisted sector of each orbifold to the general $H$-invariant affine-Virasoro construction $\hat{T}_{\s=0}=T$ in (\ref{Hinv}).  Here (\ref{lr}) is the \textit{orbifold duality transformation}
\(
L \sa \lr(L;\s)
\)
from the $H$-invariant inverse inertia tensor $\lr(0)=L$ of the untwisted sector to the twisted inverse inertia tensor $\lr(\s)$ of sector $\s$.  

The symbol $:\hsp{-.02}(\cdot)\hsp{-.02}:$ in (\ref{firsttwistT}) denotes the orbifold extension of OPE normal ordering$^{\rf{Bouw},\rf{us1},\rf{us2}}$
\group{OPENOgroup}
\(
:\hat{J}_{n(r) \mu}(z) \hat{J}_{n(s) \nu}(w): \ \equiv \hat{J}_{n(r) \mu}(z) \hat{J}_{n(s) \nu}(w) - \frac{\g_{n(r) \mu;n(s) \nu}(\s)}{(z-w)^2}-\frac{i \fr_{n(r) \mu;n(s) \nu}^{\ \ \ \ \ \ \ \ \ \ \ n(t) \delta}(\s) \hat{J}_{n(t) \delta}(w)}{z-w} \label{HOPENO}
\)
\(
:\hat{J}_{n(r) \mu}(w) \hat{J}_{n(s) \nu}(w): \ =\oint_w\frac{dz}{\tp}\frac{\hat{J}_{n(r) \mu}(z) \hat{J}_{n(s) \nu}(w)}{z-w}  \label{JJintegralNO}
\)
\reseteqn
where the contour in (\ref{JJintegralNO}) does not encircle the origin.  We emphasize that the simplicity of the result (\ref{firsttwistT}) would be compromised should other ordering prescriptions be employed (see Eqs.~(\ref{twiststress}) and (\ref{TMp})). Following the usage in Refs.~\rf{us1}, \rf{us2}, and \rf{us3}, we will refer to the prescription (\ref{OPENOgroup}) as OPE normal ordering.\footnote{The OPE normal ordering in (\ref{OPENOgroup}) is the natural finite part of the orbifold operator product, but, except in the untwisted sectors,$^{\rf{Bouw}}$ this ordering is generally not a true normal ordering (with zero vev).  Indeed, this fact accounts for the non-zero ground state conformal weights of most orbifold sectors.}

The general orbifold affine-Virasoro construction (\ref{twistTrecapgroup}) is the center of the orbifold program.  Although the OPE isomorphisms (\ref{Jarrowgroup}) and (\ref{TThatarrow}) gave a simple derivation of this result, the Virasoro property (\ref{twistTTOPE}) of $\hat{T}_\s$ was checked against the earlier general Virasoro construction$^{\rf{us1}}$ for cyclic permutation orbifolds.  The Virasoro property (\ref{twistTTOPE}) was also checked by direct OPE computation in Ref.~\rf{us2} (see also Sec.~\ref{TJsec} and App.~\ref{TJapp}).  Because of the OPE normal ordering, one finds that every step of this computation follows by the \textit{duality algorithm}$^{\rf{us2}}$ 
\(
a \sat n(r)\mu  \sp G \sat \g \sp f \sat \fr  \sp L \sat \lr \label{fdalgorithm}
\)
from the corresponding step in the direct OPE verification of the Virasoro property of the general affine-Virasoro construction $T$ in (\ref{VMET}).  The final result of this computation is that the twisted inverse inertia tensor $\lr(\s)$ must satisfy the general orbifold Virasoro master equation$^{\rf{us2}}$
\vspace{.1in}
\(
\hsp{-2} \lr^{n(r) \mu; n(s) \nu}(\s) =2 \lr^{n(r) \mu; n(r') \mu'}(\s) \g_{n(r') \mu';n(s') \nu'}(\s) \lr^{n(s') \nu'; n(s) \nu}(\s) \label{HOVME}
\)
\vspace{-.25in}
\[
\hsp{.7}  - \lr^{n(r') \mu'; n(s') \nu'}(\s) \lr^{n(t) \delta; n(t') \delta'}(\s) \fr_{n(r') \mu'; n(t) \delta}^{\hsp{.6} n(r) \mu}(\s) \fr_{n(s') \nu'; n(t') \delta'}^{\hsp{.7} n(s) \nu} (\s)
\]
\[
\hsp{.8} - \lr^{n(r') \mu'; n(s') \nu'}(\s) \fr_{n(r') \mu'; n(t) \delta}^{\hsp{.65} n(t') \delta'}(\s) \fr_{n(s') \nu'; n(t') \delta'}^{\hsp{.7} (n(r) \mu} (\s) \lr^{n(s) \nu); n(t) \delta}(\s) 
\vspace{.05in} 
\]
which can indeed be obtained by the duality algorithm (\ref{fdalgorithm}) from the Virasoro master equation in (\ref{VME}).  Moreover, the general orbifold Virasoro master equation (\ref{HOVME}) is dual (by the orbifold duality transformation (\ref{lr})) to the Virasoro master equation.  A final connection is that (\ref{HOVME}) reduces to the Virasoro master equation (\ref{VME}) when the automorphism group $H$ is trivial.  The result (\ref{HOVME}) generalizes the earlier orbifold Virasoro master equation$^{\rf{us1}}$ for cyclic permutation orbifolds.

The twisted inverse inertia tensors $\lr(\s)$ of the various sectors of each orbifold are related by the orbifold duality transformation (\ref{lr}) and its inverse $L(\lr)$ according to
\[
\hsp{-5}\lr^{n(r)\p \mu \p; n(s)\p \nu \p}(\s\p)
\]
\(
=  \frac{ {\foot{\chi(\s)_{n(r) \mu}}} }{ {\foot{\chi(\s \p)_{n(r)\p \mu \p}}} } \frac{ {\foot{\chi(\s)_{n(s) \nu}}} }{ {\foot{\chi(\s \p)_{n(r)\p \mu \p}}} } \lr^{n(r) \mu ; n(s) \nu }(\s) (U(\s) U^\dagger(\s\p))_{n(r)\mu}^{\hsp{.3}  n(r)\p \mu \p} (U(\s) U^\dagger(\s\p))_{n(s)\nu}^{\hsp{.3}  n(s)\p \nu \p}  \ . \label{genlrduality}
\)
Similar relations are easily found among the sectors $\s$ of the twisted tensors $\g(\s)$, $\fr(\s)$ introduced above, as well as among the sectors of the other twisted tensors introduced below.

The twisted inverse inertia tensors $\lr(\s)$ also satisfy$^{\rf{us2}}$
\group{lrrecapgroup}
\(
\lr^{n(r) \mu; n(s) \nu}(\s) \ (1-E_{n(r)}(\s) E_{n(s)}(\s))=0 \label{leel}
\)
\(
\lr^{n(r) \mu; n(s) \nu}(\s) =\delta_{n(r)+n(s),0 \mod \r(\s)} \lr^{n(r) \mu;  -n(r), \nu}(\s)  \label{trivmon}
\)
\reseteqn
where the $\lr$-selection rule in (\ref{leel}) is the dual in sector $\s$ of the $H$-invariance (\ref{Hinv}) of the inverse inertia tensor $L$ in the untwisted sector.  Another consequence of the $H$-invariance (\ref{Hinv}) is that each $\lr(\s)$ is a class function$^{\rf{us2}}$
\(
\lr(U(\s) v(\s);\s)=\lr(U(\s);\s)
\)
under the $H$-conjugation in (\ref{conjugationgroup}).

Incorporation of the $\lr$-selection rule (\ref{lrrecapgroup}) gives a more informative presentation of the general orbifold affine-Virasoro construction
\group{bestTrecapgroup}
\(
\hat{T}_\s (z)= \lr^{n(r) \mu; -n(r), \nu}(\s):\hat{J}_{n(r) \mu}(z) \hat{J}_{-n(r), \nu}(z):  \sp \hat{T}_\s (ze^{2 \pi i})=\hat{T}_\s (z) \label{bestTrecap}
\)
\(
\hat{c}(\s)=2 \g_{n(r) \mu; -n(r), \nu}(\s) \lr^{n(r) \mu; -n(r), \nu}(\s) =2 G_{ab} L^{ab}=c \sp \srange
\)
\reseteqn
which shows that the orbifold stress tensors have trivial monodromy.  Moreover, the $\lr$-selection rule (\ref{lrrecapgroup}) is consistent with the general orbifold Virasoro master equation (\ref{HOVME}), and the two can be combined to obtain a reduced master equation for $\lr^{n(r) \mu;  -n(r), \nu}(\s)$.  Looking back, one sees that this consistency is a consequence of the underlying $H$-symmetry of the CFT $A(H)$.

\subsection{The cyclic permutation orbifolds \taz}
\label{cyclicsec}
As an example, we recall the seminal case$^{\rf{us2}}$ of the general cyclic permutation orbifold \taz,
\group{zlreviewgroup}
\(
a \rightarrow aI \sp L^{ab} \rightarrow L^{aI,bJ} =L^{ab}_{I-J} \sp I=0,...,\l-1
\)
\(
n(r), \mu \rightarrow r, aj \sp \hat{J}_{n(r) \mu} \rightarrow \hat{J}_{aj}^{(r)} \sp  {\foot{\chi(\s)_{n(r) \mu}}} \rightarrow {\foot{\chi(\s)_{raj}}}=\sqrt{\r(\s)} \label{zlindexarrow}
\)
\getletter{zlgflett}
\vspace{-.2in}
\(
\g_{raj;sbl}(\s)=\r(\s) k \eta_{ab} \delta_{jl}  \delta_{r+s,0 \mod \r(\s)} \sp \fr_{raj;sbl}^{\hsp{.35} tcm}(\s)=f_{ab}^{\ \ c} \delta_{jl} \delta_l^{\ m}  \delta_{r+s-t,0 \mod \r(\s)} \label{zlgf}
\)
\(
\jh_{aj}^{(r)}(z)\jh_{bl}^{(s)}(w) \hspace{-.03in} = \hspace{-.03in} \delta_{j l} \{ \frac{\r(\sigma) k\eta_{ab}\delta_{r+s,0\mod\r(\sigma)}}{(z-w)^2}+\frac{if_{ab}^{\ \ c}\jh_{cj}^{(r+s)}(w)}{(z-w)} \} +\reg \label{TCOPE}
\)
\getletter{zlJlett}
\vspace{-.1in}
\(
\hat{J}_{aj}^{(r)}(ze^{\tp}) = e^{-\frac{\tp r}{\r(\sigma)}} \hat{J}_{aj}^{(r)}
(z) \sp \hat{J}_{aj}^{(r\pm \r(\s))}(z)=\hat{J}_{aj}^{(r)}(z)
\)
\(
\hat{T}_\s (z)=\sum_{r=0}^{\r(\sigma)-1} \ \sum_{j,l=0}^{\frac{\l}{\r(\sigma)}-1} \lr^{raj;-r,bl}(\s):\jh_{aj}^{(r)}(z)\jh_{bl}^{(-r)}(z):  \label{twstress}
\)
\getletter{zlmaplett}
\vspace{-.1in}
\(
\lr^{raj;-r,bl}(\s)=\frac{1}{\r(\sigma)}\sum_{s=0}^{\r(\sigma)-1} e^{-\frac{\tp N(\s)rs}{\r(\sigma)}} L^{ab}_{\frac{\l}{\r(\sigma)} s+j-l} \label{themap}
\)
\(
\hat{\Delta}_0(\s)=\frac{\l k\eta_{ab}}{4\r^2(\s)} \{ \frac{\r^2(\s)-1}{3}L_0^{ab}-{\displaystyle{\sum_{r=1}^{\r(\s)-1}}} csc^2(\frac{\pi N(\s)r}{\r(\s)})\hsp{.02} L^{ab}_{\frac{\l}{\r(\s)}r} \}  \label{reviewzlcw} 
\)
\getletter{zlrangelett}
\vspace{-.1in}
\(
a,b=1,...,\textup{dim} \gbn \sp
\bar{r},\bar{s}=0,...,\r(\sigma)-1\sp
j,l=0,...,\frac{\l}{\r(\sigma)}-1 \sp \s=0,...,\l-1 
\)
\reseteqn
whose relation to the orbifold Virasoro master equation$^{\rf{us1}}$ provided a nontrivial check of the orbifold program.

The special case (\ref{zlreviewgroup}) follows from the discussion above and the $\zl$-eigendata in (\ref{zlUgroup}).  In this case the twisted current system (\ref{TCOPE},\ref{zlJlett}) describes a sector-dependent set of semisimple orbifold affine algebras.$^{\rf{Chris}}$  The orbifold duality transformation $\lr(L;\s)$ in (\ref{themap}) is a set of discrete Fourier transforms of $L$, with spectral index periodicity $r\rightarrow r+ \r(\s)$, where $L$ is the inverse inertia tensor of any $\zl$(permutation)-invariant CFT $A(\zl)$.  The twisted inverse inertia tensors $\lr^{raj;-r,bl}(\s)$ were called $\lr_r^{a(j)b(l)}(\s)$ in Ref.~\rf{us2} and the result in (\ref{reviewzlcw}) is the ground state conformal weight of sector $\s$.  

The orbifold duality transformations for $\g$ and $\fr$ in (\ref{zlgf}) can also be written as discrete Fourier transforms and, indeed, because of the periodicity of the eigenvectors of each $H$-eigenvalue problem, all the orbifold duality transformations of this paper
\(
\g(G;\s) \sp \fr(f;\s) \sp \lr(L;\s), \ ...
\)
are discrete Fourier transforms with period $\r(\s)$ in the spectral indices $\{ n(r) \}$.

\newsection{The Mode Formulation of Orbifold Theory}
\label{modesec}
As an application of the local formulation above, we turn now to the mode formulation of the general current-algebraic orbifold.

\subsection{The general twisted current algebra}
\label{twistalgsubsec}
The first step in the mode formulation is to find the twisted current algebra of each sector of each orbifold \tahh, beginning with the general twisted current system (\ref{explicitJJOPEgroup}).

The mode expansion of the twisted currents
\group{Jmodeexpansiongroup}
\(
\hat{J}_{n(r) \mu}(z)=\sum_{m \in \sz} \hat{J}_{n(r) \mu}(m+\srac{n(r)}{\r(\s)}) z^{-(m+\frac{n(r)}{\r(\s)})-1} \label{curenttomodetwistJ}
\)
\(
\hat{J}_{n(r) \mu}(z)=\hat{J}_{n(r) \pm \r(\s), \mu}(z)=\sum_{m \in \sz} \hat{J}_{n(r) \pm \r(\s), \mu}(m+\srac{n(r) \pm \r(\s)}{\r(\s)}) z^{-(m+\frac{n(r) \pm \r(\s)}{\r(\s)})-1} \label{Jmodeplus}
\)
\reseteqn
follows from the monodromies (\ref{explcitJJmonodromies}) and the periodicity (\ref{Jperiod}).  From (\ref{Jmodeplus}) we find the periodicity of the modes 
\group{Jmodeperiodg}
\(
\hat{J}_{n(r)_\pm \r(\s),\mu}(m \mp 1 +\srac{n(r) \pm \r(\s)}{\r(\s)})=
\hat{J}_{n(r)_\pm \r(\s),\mu}(m +\srac{n(r)}{\r(\s)})=
\hat{J}_{n(r)\mu}(m+\srac{n(r)}{\r(\s)}) \label{Jmodeperiod}
\)
\(
\hat{J}_{-n(r),\mu}(m - \srac{n(r)}{\r(\s)})=\hat{J}_{\r(\s) - n(r),\mu}(m -1 + \srac{\r(\s)-n(r)}{\r(\s)}) \label{Jmodeperioda}
\)
\(
\hat{J}_{n(r)_\pm \r(\s),\mu}(m+\srac{n(r) \pm \r(\s)}{\r(\s)}) \neq \hat{J}_{n(r)\mu}(m+\srac{n(r)}{\r(\s)}) \label{Jmodenotperiodic}
\)
\reseteqn
which includes and generalizes the mode periodicity of orbifold affine algebra.$^{\rf{Chris},\rf{us1},\rf{us2}}$  Note that the modes do \textit{not} satisfy the naive periodicity (\ref{Jmodenotperiodic}).

We introduce next the operators ${\hat{\hat{J}\hspace{.05in}}}$ with trivial monodromy 
\group{hathatJgroup}
\(
{\hat{\hat{J}\hspace{.05in}}}_{\hsp{-.05} n(r) \mu}(z) \equiv z^{\frac{n(r)}{\r(\s)}} \hat{J}_{n(r) \mu}(z)
\)
\[
%\hsp{-4.5}
 \hat{\hat{J}\hspace{.056in}}_{\hsp{-.05} n(r) \mu}(z) \hat{\hat{J}\hspace{.06in}}_{\hsp{-.05} n(s) \nu}(w)
%\]
%\[
= \delta_{n(r)+n(s),0 \mod \r(\s)} \w^{\frac{n(r)+n(s)}{\r(\s)}} (\frac{1}{(z-w)^2} + \frac{n(r)/ \r(\s)}{w(z-w)})\g_{n(r) \mu;-n(r), \nu}(\s)
\]
\(
\hsp{2} +\frac{i \fr_{n(r) \mu;n(s) \nu}^{\ \ \ \ \ \ \ \ \ \ \ n(r)+n(s),\delta}(\s) \hat{\hat{J}\hspace{.055in}}_{\hsp{-.05} n(r)+n(s), \delta}(w)}{z-w} + O((z-w)^0). \label{hathatJOPE}
\)
\reseteqn
The Kronecker delta in the first term of (\ref{hathatJOPE}) guarantees that these OPE's are free of branch cuts, which is a necessary condition for the monodromies (\ref{explcitJJmonodromies}) to be consistent with the OPE's (\ref{twistJJOPEall}).

Then standard analysis of the OPE's (\ref{hathatJOPE}) gives the \textit{general twisted current algebra} $\gb (\s) \equiv \gb(H\subset Aut(g);\s)$
\group{ghatgroup}
\[
[\hat{J}_{n(r) \mu}(m+\srac{n(r)}{\r(\s)}),\hat{J}_{n(s) \nu}(n+\srac{n(s)}{\r(\s)})] = i \fr_{n(r) \mu;n(s) \nu}^{\ \ \ \ \ \ \ \ \ \ \ n(r)+n(s),\delta}(\s) \hat{J}_{n(r)+n(s), \delta}(m+n+\srac{n(r)+n(s)}{\r(\s)}) 
\]
\(
\hsp{2.7} +(m+\srac{n(r)}{\r(\s)}) \delta_{m+n+\frac{n(r)+n(s)}{\r(\s)},0} \ \g_{n(r) \mu; -n(r), \nu}(\s)   \label{twistJmodealgebra} 
\)
\(
m,n\in \z \sp \srange
\)
\reseteqn
in sector $\s$ of each orbifold \tahh.  Here $N_c$ is the number of conjugacy classes of $H$, and the duality transformations for $\g$ and $\fr$ are given in Eq.~(\ref{fulltwistedopegroup}).  To obtain (\ref{ghatgroup}) we have used the identity 
\(
\delta_{n(r)+n(s),0 \mod \r(\s)} \delta_{m+n+\frac{n(r)+n(s)}{\r(\s)},0}=\delta_{m+n+\frac{n(r)+n(s)}{\r(\s)},0}
\)
to simplify the $\g$ term.

The general twisted current algebra (\ref{ghatgroup}) shows a natural grading across all orbifolds.  The general twisted current algebra also satisfies the Jacobi identity (see App.~\ref{selectionapp}), which is the ultimate check of the consistency of the principle (\ref{Jarrowgroup}).  

Looking back, one sees that the grading and the consistency of the general twisted current algebra are consequences of the $H$-covariance of the underlying affine algebra (\ref{untwistJalg}) on $g$.  The general twisted current algebra is also  consistent with the mode periodicity relations (\ref{Jmodeperiodg}) and, in the untwisted sector $\s=0$, the general twisted current algebra reduces to the $H$-covariant affine algebra on $g$.  

Another feature of the general twisted current algebra is the integral affine subalgebra 
\(
\gb^{(0)}(\s) \subset \gb (\s): \hsp{.1} [\hat{J}_{0 \mu}(m),\hat{J}_{0 \nu}(n)]=i
\fr_{0 \mu;0 \nu}^{\ \ \ \ \ 0 \delta}(\s) \hat{J}_{0 \delta}(m+n) 
+m \delta_{m+n,0} \ \g_{0 \mu;0 \nu}(\s)   \label{intaffinemodealgebra}
\)
which is an untwisted affine Lie algebra generated by the set of currents in twist class $\bar{n}(r)=0$.  The integral affine subalgebra $\gb^{(0)}(\s)$ is non-trivial for all known non-abelian $\gb (\s)$.

As an example of $\gb(\s)$, we mention the semisimple orbifold affine algebra at orbifold affine level $\hat{k}(\s)=\r(\s)k$
\group{zlalgebra}
\(
\gb(\s)=\gb(\zl \textup{(permutation)}\subset Aut(g);\s) \sp g=\oplus_{I=0}^{\l-1} \gbn^I \sp \gbn^I \cong \gbn
\)
\(
%\hsp{-3.7}
 [\hat{J}_{aj}^{(r)}(m+\srac{r}{\r(\s)}),\hat{J}_{bl}^{(s)}(n+\srac{s}{\r(\s)})]
%\]
%\(
=\delta_{jl} \{ i f_{ab}^{\ \ c} \hat{J}_{cj}^{(r+s)}(m+n+\srac{r+s}{\r(\s)}) + (m+\srac{r}{\r(\s)}) \delta_{m+n+\frac{r+s}{\r(\s)},0} \ \r(\s) k \eta_{ab} \}
\)
\(
\hat{J}_{aj}^{(r\pm \r(\s))}(m+\srac{r}{\r(\s)})=\hat{J}_{aj}^{(r)}(m+\srac{r}{\r(\s)}) \label{zlJprop}
\)
\(
a,b=1,...,\textup{dim}\gbn \sp \bar{r},\bar{s}=0,...,\r(\s)-1 \sp j,l=0,...,\frac{\l}{\r(\s)}-1 \sp \s=0,...,\l-1
\)
\reseteqn
which holds\footnote{For prime $\l$ one has $\r(\s)=\l$ and $j=0$ for all $\l-1$ twisted sectors, so each twisted sector has only the simple orbifold affine algebra at level $\hat{k}=\l k$.  This was first seen$^{\rf{Chris}}$ in the characters of cyclic copy orbifolds.} in sector $\s$ of each cyclic permutation orbifold \taz.  Using the data (\ref{zlindexarrow},\ref{zlgflett},\ref{zlrangelett}), the form (\ref{zlalgebra}) follows from the general twisted current algebra (\ref{ghatgroup}).  The algebra is also equivalent to the twisted current system in (\ref{zlreviewgroup}).

Here is a summary of what is known about (the $H$-eigenvalue problem and) the twisted current algebra $\gb (H \subset Aut(g);\s)$ :

\bu \ $H$ is any subgroup of $S_N$(permutation) $\rightarrow$ sets of commuting orbifold affine 

algebras$^{\rf{Chris}-\rf{Dong},\rf{us1},\rf{us2}-\rf{us3}}$ at various levels.

\bu \ $H$ is a group of inner or outer automorphisms of simple $g$ $\rightarrow$ inner-automorphically 

twisted$^{\rf{KP},\rf{GO},\rf{FH},\rf{Lerche},\rf{Birke}}$ and outer-automorphically twisted$^{\rf{book},\rf{GO},\rf{Birke}}$ affine Lie algebras.

\bu \  $H$ is a group of inner automorphisms of each copy $\gbn^I\cong \gbn$ times a permutation group 

which acts among the copies $\rightarrow$ doubly-twisted affine algebras.$^{\rf{us1},\rf{us2},\rf{us3}}$
\\
The results$^{\rf{us2}}$ for $H=\zl$(permutation) are collected in Eqs.~(\ref{zlUgroup}), (\ref{zlreviewgroup}) and (\ref{zlalgebra}).  The case $H=S_N$(permutation) is discussed in detail in Sec.~\ref{SNsec} and the case of the general group $H=H(d)$ of inner automorphisms of simple $g$ is discussed in Sec.~\ref{innersec}.  The setups for the case $H=\dl$(permutation) and the outer automorphism groups of simple $g$ are included in Apps.~\ref{dlapp} and \ref{outersapp} respectively.  

There are many other cases for which the twisted current algebras have not yet been worked out (e.g. $H$ is a group of outer automorphisms of each copy $\gbn^I\cong \gbn$ times a permutation group which acts among the copies).  Another basis for the general twisted current algebra $\gb (\s)$ is given in App.~\ref{constrainedbasisapp}.

\subsection{Operator products and mode ordering}
In addition to OPE normal ordering (\ref{OPENOgroup}), it is convenient to introduce a mode ordering\footnote{Many mode orderings are possible, the most effective ordering being the one that is closest to ``normal'' (zero vev), given the mode characterization of the ground state of a given sector.  The $M$-ordering in (\ref{MNOing}) is a normal ordering for the orbifold affine algebras of the general permutation orbifold (see Secs.~\ref{cwsec} and \ref{SNsec}).}
\[
:\hat{J}_{n(r) \mu}(m+\srac{n(r)}{\r(\s)}) \hat{J}_{n(s) \nu}(n+\srac{n(s)}{\r(\s)}):_M \ 
\equiv \theta(m+\srac{n(r)}{\r(\s)} \geq 0) \  \hat{J}_{n(s) \nu}(n+\srac{n(s)}{\r(\s)}) \hat{J}_{n(r) \mu}(m+\srac{n(r)}{\r(\s)})
\]
\(
+ \theta(m+\srac{n(r)}{\r(\s)} < 0) \ \hat{J}_{n(r) \mu}(m+\srac{n(r)}{\r(\s)}) \hat{J}_{n(s) \nu}(n+\srac{n(s)}{\r(\s)}) \label{MNOing}
\)
denoted by the subscript $M$.  Then one may use (\ref{curenttomodetwistJ}), (\ref{ghatgroup}) and (\ref{MNOing}) to compute the exact operator product of the twisted currents
\[
%\hsp{-1}
\hsp{-1.7} \hat{J}_{n(r) \mu}(z) \hat{J}_{n(s) \nu}(w)=(\frac{w}{z})^{\frac{\bar{n}(r)}{\r(\s)}} \{ [\frac{1}{(z-w)^2} + \frac{\bar{n}(r)/\r(\s)}{w(z-w)}] \g_{n(r) \mu;n(s) \nu}(\s)
\]
\(
%\hsp{1}
\hsp{1.3} +\frac{i \fr_{n(r) \mu;n(s) \nu}^{\hsp{.6} n(r)+n(s), \delta}(\s) \hat{J}_{n(r)+n(s), \delta}(w)}{z-w} \} + :\hat{J}_{n(r) \mu}(z) \hat{J}_{n(s) \nu}(w):_M \label{exactJJ}
\)
where $\bar{n}(r)$ is the twist class of $\hat{J}_{n(r) \mu}$ (see Eq.~(\ref{nnbar})).  The partial conversion $n(r) \rightarrow \bar{n}(r)$ seen here is detailed in App.~\ref{nnbarapp}, and it is easily checked that the result (\ref{exactJJ}) is consistent with the OPE's (\ref{explicitJJOPEgroup}).  We also note that every term in (\ref{exactJJ}) is periodic under $n(r)\rightarrow n(r)+\r(\s)$, $\bar{n}(r)\rightarrow \bar{n}(r)$.

Comparing (\ref{exactJJ}) with (\ref{OPENOgroup}), we may express the OPE normal ordered product
\[
:\hat{J}_{n(r) \mu}(z) \hat{J}_{n(s) \nu}(z): \  =  \ :\hat{J}_{n(r) \mu}(z) \hat{J}_{n(s) \nu}(z):_M - \frac{i \bar{n}(r)}{z\r(\s)} \fr_{n(r) \mu;n(s) \nu}^{\hsp{.6} n(r)+n(s), \delta}(\s) \hat{J}_{n(r)+n(s), \delta}(z)
\]
\(
+ \frac{1}{z^2}   \frac{\bar{n}(r)}{2 \r(\s)}(1-\frac{\bar{n}(r)}{\r(\s)}) \g_{n(r) \mu;n(s) \nu}(\s)  \label{OPEMNO}
\)
in terms of an $M$ ordered product.  

With the relation (\ref{OPEMNO}), the general orbifold stress tensor (\ref{bestTrecap}) can be expressed in terms of an M ordered product as 
\ba
\hat{T}_\s(z)&=&\sum_{r,\mu,\nu} \lr^{n(r) \mu;-n(r), \nu}(\s) \{ :\hat{J}_{n(r) \mu}(z) \hat{J}_{-n(r), \nu}(z):_M - \frac{i \bar{n}(r)}{\r(\s)} \fr_{n(r) \mu;-n(r), \nu}^{\hsp{.7} 0 \delta}(\s) \frac{\hat{J}_{0 \delta}(z)}{z} \nonumber \\
&&\hsp{1.3} + \frac{1}{z^2}   \frac{\bar{n}(r)}{2 \r(\s)}(1-\frac{\bar{n}(r)}{\r(\s)}) \g_{n(r) \mu;-n(r), \nu}(\s) \} \ . \label{twiststress}
\ea
In this form, we note the presence of the Freericks-Halpern$^{\rf{FH}}$ term $\hat{J}_{0 \delta}(z)/z$, where $\hat{J}_{0\delta}$ are the generators of the integral affine subalgebra.

\subsection{The Virasoro generators of sector $\s$}
\label{Virsec}
The OPE normal ordered product of the twisted current modes can also be expressed in terms of $M$ ordering 
\[
\hsp{-1.3} :\hat{J}_{n(r) \mu}(m+\srac{n(r)}{\r(\s)}) \hat{J}_{n(s) \nu}(n+\srac{n(s)}{\r(\s)}): \  =  \ :\hat{J}_{n(r) \mu}(m+\srac{n(r)}{\r(\s)}) \hat{J}_{n(s) \nu}(n+\srac{n(s)}{\r(\s)}):_M   
\]
\[
\hsp{2.4}-i\frac{\bar{n}(r)}{\r(\s)} \fr_{n(r) \mu; n(s) \nu}^{\hsp{.6} n(r)+n(s), \delta}(\s) \hat{J}_{n(r)+n(s), \delta}(m+n+\srac{n(r)+n(s)}{\r(\s)}) 
\]
\(
\hsp{2.4} +  \frac{\bar{n}(r)}{2 \r(\s)}(1-\frac{\bar{n}(r)}{\r(\s)}) \delta_{m+n+\frac{n(r)+n(s)}{\r(\s)},0} \g_{n(r) \mu;-n(r), \nu}(\s) \ .
\)
Then (using the mode expansion (\ref{curenttomodetwistJ}) of the twisted currents) the general orbifold stress tensor in (\ref{bestTrecapgroup}) or (\ref{twiststress}) gives the orbifold Virasoro generators 
\group{Lmode}
\(
\hat{T}_\s(z) =\sum_{m \in \sz} L_\s(m) z^{-m-2} \sp \srange
\)
\vspace{-.2in}
\ba
\hsp{-.5} L_\s (m)& =& \sum_{r,\mu,\nu} \lr^{n(r) \mu; -n(r), \nu}(\s) \ \sum_{p \in \sz} :\hat{J}_{n(r) \mu}(p+\srac{n(r)}{\r(\s)}) \hat{J}_{-n(r), \nu}(m-p-\srac{n(r)}{\r(\s)}):  \\
&=& \sum_{r,\mu,\nu} \lr^{n(r) \mu; -n(r), \nu}(\s) \ \{ \sum_{p \in \sz} :\hat{J}_{n(r) \mu}(p+\srac{n(r)}{\r(\s)}) \hat{J}_{-n(r), \nu}(m-p-\srac{n(r)}{\r(\s)}):_M   \nonumber \\
&&\hsp{1.7} -i\frac{\bar{n}(r)}{\r(\s)} \fr_{n(r) \mu; -n(r), \nu}^{\ \ \ \ \ \ \ \ \ \ \ \ \ \ 0 \delta}(\s) \hat{J}_{0 \delta}(m)  \nonumber \\
&&\hsp{1.7} + \delta_{m,0} \frac{\bar{n}(r)}{2 \r(\s)}(1-\frac{\bar{n}(r)}{\r(\s)}) \g_{n(r) \mu;-n(r), \nu}(\s) \} \label{Lmodeb} 
\ea
\reseteqn
for all sectors $\s$ of each orbifold \tahh.  This result is the mode form of the general orbifold affine-Virasoro construction.  For the untwisted sector $\s=0$, the general construction (\ref{Lmode}) reduces to the mode formulation of the general $H$-invariant affine-Virasoro construction$^{\rf{Lieh}}$ in (\ref{Hinv}).

The simplest example of this result is the orbifold $U(1)/\z_2$:
\alpheqn
\(
J(z)J(w)=\frac{1}{(z-w)^2}+\reg  \sp T(z)=\frac{1}{2}:J^2(z):
\)
\(
\s=1: \hsp{.3} J(z)\p =-J(w) \sp \r(1)=2 \sp U(1)={\foot{\chi(1)}}=1
\)
\(
\hat{J}(z)\hat{J}(w)=\frac{1}{(z-w)^2}+\reg \sp \hat{J}(ze^{2\pi i})=-\hat{J}(z)
\)
\(
\hsp{.75} \hat{J}(z)=\sum_{m\in \sz} \hat{J}(m+\srac{1}{2}) z^{-(m+\frac{1}{2})-1} , \hsp{.15} [\hat{J}(m+\srac{1}{2}), \hat{J}(n+\srac{1}{2})]=(m+\srac{1}{2}) \ \delta_{m+n+1,0} 
\)
\(
L_{\s=1}(m)=\frac{1}{2} \sum_{p\in \sz} :\hat{J}(p+\srac{1}{2})\hat{J}(m-p-\srac{1}{2}):_M +\delta_{m,0}\frac{1}{16} 
\)
\(
\hat{J}( \foot{(} m+\srac{1}{2} \foot{)} \geq 0)|0\rangle =(L_{\s=1}(m\geq 0)-\delta_{m,0}\frac{1}{16})|0\rangle=0 \ .
\)
\reseteqn
Here $\hat{J} \equiv \hat{J}_{n(r)=1}=\hat{J}_{n(r)=-1}=i \partial \hat{X}$ is the current of the standard $\frac{1}{2}$-integrally moded scalar field $\hat{X}$.  This twisted field was invented$^{\rf{Th}}$ by Halpern and Thorn, who used it to construct the first twisted sector of an orbifold.  

For the seminal non-abelian example,$^{\rf{us2}}$ the Virasoro generators of the permutation orbifolds \taz
\ba
\hsp{-.5} L_\s (m) & = & \sum_{r=0}^{\r(\s)-1} [\ \sum_{j, l=0}^{\frac{\l}{\r(\s)}-1} \lr^{r a j; -r, bl}(\s) \ \sum_{p \in \sz} :\hat{J}_{a j}^{(r)} (p+\srac{r}{\r(\s)}) \hat{J}_{bl}^{(-r)} (m-p-\srac{r}{\r(\s)}):_M   \nonumber \\
&& + \sum_{j=0}^{\frac{\l}{\r(\s)}-1} \lr^{r a j; -r, b j}(\s)  \{ \frac{-i r}{\r(\s)} f_{ab}^{\ \ c} \hat{J}_{cj}^{(0)} (m)  + \delta_{m,0}  \frac{r}{2 \r(\s)}(1-\frac{r}{\r(\s)}) k \r(\s) \eta_{ab} \} ] \label{zlL}
\ea
follow from (\ref{zlreviewgroup}) and (\ref{Lmode}).

\newsection{The Orbifold Adjoint Operation}
\subsection{The orbifold conjugation matrix}
To study the adjoint operation in the twisted sectors, we first introduce the orbifold conjugation matrix $\Ltheta$,
\group{Lthetawithdeltagroup}
\(
\r \ \sa \ \Ltheta(\r;\s) \label{rarrow}
\)
\(
\Ltheta_{n(r) \mu}^{\ \ \ \ \ \ n(s) \nu}(\s) \equiv {\foot{\chi(\s)_{n(r) \mu}}}^{*} {\foot{\chi(\s)_{n(s) \nu}^{-1}}} U(\s)_{n(r) \mu }^{\ \ \ \ \ a \ *} \stheta_a^{\ b} \uds_b^{\ n(s) \nu}  \label{Phi}
\)
\(
\Ltheta (\s)^*  \Ltheta (\s)= \Ltheta (\s) \Ltheta (\s)^*=1
\label{invphi}
\)
\(
\Ltheta_{n(r) \mu}^{\ \ \ \ \ \ n(s) \nu}(\s) \ (1-E_{n(r)}(\s) E_{n(s)}(\s))=0
\label{rdual} \)
\(
\Ltheta_{n(r) \mu}^{\ \ \ \ \ \ n(s) \nu}(\s) = \delta_{n(r)+n(s),0 \mod \r(\s)} \Ltheta_{n(r) \mu}^{\ \ \ \ \ \ -n(r), \nu}(\s)   \label{Lthetawithdelta}
\)
\reseteqn
which is the dual in sector $\s$ (see the following subsection) of the conjugation matrix $\r$ in Eq.~(\ref{untwistdaggergroup}).  Eq.~(\ref{rarrow}) and its realization in (\ref{Phi}) is another orbifold duality transformation on the same footing as the earlier ones for $\g, \fr$ and $\lr$.  From Eq.~(\ref{Phi}) and the $H$-eigenvalue problem (\ref{eigenprobU}), one sees that the $\Ltheta$-selection rule in (\ref{rdual}) (and its solution in (\ref{Lthetawithdelta})) is the dual in sector $\s$ of the $H$-covariance (\ref{guess}) of the conjugation matrix $\r$.  The orbifold conjugation matrix $\Ltheta$ also controls the complex conjugation of the other twisted tensors\footnote{To prove (\ref{lrstar}), for example, use the duality transformation $\lr(L)$ in (\ref{lr}) and the $L^*$ relation in (\ref{Lstar}), followed by the inverse $L(\lr)$ of the duality transformation.}
\group{stargroup}
\(
\g_{n(r) \mu;n(s) \nu}(\s)^*=\Ltheta_{n(r) \mu}^{\hsp{.3} n(t) \mu \p}(\s) \ \Ltheta_{n(s) \nu}^{\hsp{.3} n(u) \nu \p}(\s) \ \g_{n(t) \mu \p;n(u) \nu \p}(\s) \label{grstar}
\)
\(
\fr_{n(r) \mu; n(s) \nu}^{\ \ \ \ \ \ \ \ \ \ \ n(t) \delta}(\s)^*=\Ltheta_{n(r) \mu}^{\hsp{.3} n(u) \mu \p} (\s) \ \Ltheta_{n(s) \nu}^{\hsp{.3} n(v) \nu \p} (\s) \ \fr_{n(u) \mu \p; n(v) \nu \p}^{\hsp{.67} n(w) \delta \p}(\s) \ \Ltheta_{n(w) \delta \p}^{\hsp{.35} n(t) \delta}(\s)^* \label{frstar}
\)
\(
\lr^{n(r) \mu; n(s) \nu}(\s)^* = \lr^{n(t) \mu \p;n(u) \nu \p}(\s) \ \Ltheta_{n(t) \mu \p}^{\hsp{.32} n(r) \mu}(\s)^* \ \Ltheta_{n(u) \nu \p}^{\hsp{.32} n(s) \nu}(\s)^* \label{lrstar} 
\)
\reseteqn
in each sector $\s$ of each orbifold $A(H)/H$.  These results are dual to Eqs.~(\ref{gstar}) and (\ref{Lstar}), so that, in particular, the relation (\ref{lrstar}) holds when the CFT $A(H)$ is unitary.

Another consequence of the $H$-covariance of $\r$ in (\ref{guess}) is that each $\Ltheta (\s)$ is a class function
\(
\Ltheta(U(\s)v(\s);\s)=\Ltheta(U(\s);\s)
\)
under the $H$-conjugation in (\ref{conjugationgroup}).

\subsection{The adjoint of the twisted currents}
\label{realalgebra}
We can now give the adjoint of the twisted currents of sector $\s$
\group{adjointgroup}
\(
\hat{J}_{n(r) \mu}(m+\srac{n(r)}{\r(\s)})^\dagger=\sum_\nu \Ltheta_{n(r) \mu}^{\hsp{.3} -n(r), \nu}(\s) \hat{J}_{-n(r), \nu}(-m-\srac{n(r)}{\r(\s)}) \label{Jdagger}
\)
\(
\hat{J}_{n(r) \mu}(m+\srac{n(r)}{\r(\s)})^{\dagger \dagger}=\hat{J}_{n(r) \mu}(m+\srac{n(r)}{\r(\s)}) \label{hatJdoubledagger}
\)
\reseteqn
where $\Ltheta (\s)$ is the orbifold conjugation matrix defined above.  The \textit{orbifold adjoint operation} (\ref{Jdagger}) is the dual\footnote{The local form $\hat{J}(z)^\dagger=\Ltheta \hat{J}(z^{-1 \hsp{.02} *})z^{-2 \hsp{.02}*}$ of the adjoint operation (\ref{Jdagger}), including the form (\ref{Phi}) of $\Ltheta$, follows by duality transformation from the local form $J(z)^\dagger=\r J(z^{-1 \hsp{.02} *})z^{-2 \hsp{.02}*}$ of the adjoint operation (\ref{untwistdaggergroup}) in the untwisted sector.} in sector $\s$ of the adjoint operation (\ref{untwistJdagger}) on the untwisted currents.  The consistency relation (\ref{hatJdoubledagger}) is easily checked with Eqs.~(\ref{Jdagger}) and (\ref{partinvphi}).

The orbifold adjoint operation (\ref{Jdagger}) defines a real form of the general twisted current algebra $\gb (\s)$ in (\ref{ghatgroup}) because the adjoint of the algebra 
\[
[\hat{J}_{n(r) \mu}(m+\srac{n(r)}{\r(\s)})^\dagger,\hat{J}_{n(s) \nu}(n+\srac{n(s)}{\r(\s)})^\dagger]=
-(m+\srac{n(r)}{\r(\s)}) \g_{n(r) \mu;-n(r), \nu}(\s)^* \delta_{m+n+\frac{n(r)+n(s)}{\r(\s)},0}
\]
\(
+i \fr_{n(r) \mu;n(s) \nu}^{\hsp{.6} n(r)+n(s),\delta}(\s)^* \hat{J}_{n(r)+n(s), \delta}(m+n+\srac{n(r)+n(s)}{\r(\s)})^\dagger \label{consistentJdagger}
\)
is consistent with $\gb (\s)$.  To verify this statement, use (\ref{sgrstar}) and (\ref{sfrstar}).

In the case$^{\rf{us2}}$ of the permutation orbifolds \taz, the twisted current algebra is the semisimple orbifold affine algebra (\ref{zlalgebra}) and we obtain the orbifold adjoint operation
\alpheqn
\(
\r_a^{\ b} \rightarrow \r_{aI}^{\ \ \ bJ}=\r_a^{\ b} \delta_I^{\ J}  \sp I,J=0,...,\l-1
\)
\(
\Ltheta (\s) \rightarrow \Ltheta_{raj}^{\hsp{.2} sbl}(\s)=\r_a^{\ b} \delta_j^{\ l} \delta_{r+s,0 \mod \r(\s)} \sp \hat{J}_{aj}^{(r)}(m+\srac{r}{\r(\s)})^\dagger=\r_a^{\ b} \hat{J}_{bj}^{(-r)}(-m-\srac{r}{\r(\s)}) \label{zldagger}
\)
\reseteqn
from (\ref{zlUgroup}), (\ref{zlreviewgroup}) and (\ref{Jdagger}).  This adjoint operation is in agreement with the standard adjoint operation for orbifold affine algebra$^{\rf{Chris},\rf{us2}}$ and it is known$^{\rf{Chris}}$ that this adjoint operation guarantees unitarity of the twisted affine Hilbert space when the untwisted affine Hilbert space is unitary.

More generally, as we will note in Secs.~\ref{cwsec}, \ref{SNsec} and \ref{innersec}, the orbifold adjoint operation defines a unitary twisted affine Hilbert space for each sector of the orbifolds\footnote{In all these cases, unitarity of the twisted affine Hilbert spaces follows easily from known orbifold induction procedures.$^{\rf{KP},\rf{GO},\rf{FH},\rf{Lerche},\rf{Chris},\rf{us1}}$}

\bu \ $A(H)/H$, where $H$ is any subgroup of $S_N$(permutation)

\bu \ $A(H(d))/H(d)$, where $H(d)$ is any group of inner automorphisms of simple $g$

\bu \ $A(H)/H$, where $H$ is a product of any subgroup of $S_N$(permutation) times a group of 

inner automorphisms
\\
when the untwisted affine Hilbert space is unitary.  On the basis of this evidence, we conjecture that the orbifold adjoint operation implies unitarity of all the twisted affine Hilbert spaces of \tahh, given the unitarity of the untwisted affine Hilbert space.

\subsection{The adjoint of the orbifold Virasoro generators}
\label{Viradjoint}
The adjoint (\ref{Jdagger}) of the twisted currents gives the desired adjoint operation on the Virasoro generators
\(
L_\s (m)^\dagger =  L_\s (-m) \label{twistLdagger}
\)
in every sector $\s$ of each orbifold \tahh.  To see this, apply the orbifold adjoint operation (\ref{Jdagger}) to the mode form (\ref{Lmodeb}) of the Virasoro generators and use the $\lr^*$ relation (\ref{slrstar}).  In further detail, one finds that the $(\hat{J})^2$, $(\hat{J})^1$ and $(\hat{J})^0$ terms in (\ref{Lmodeb}) satisfy the adjoint relation (\ref{twistLdagger}) separately.

It follows that all the twisted sectors of the orbifolds bulleted in Subsec.~\ref{realalgebra} are unitary when the CFT's $A(H)$ are unitary,\footnote{This result was established for the permutation orbifolds \taz \ in Ref.~\rf{us2}.} and similarly for all $A(H)/H$ if our conjecture in that subsection can be proven.

\newsection{The $\hat{T} \hat{J}$ OPE's of \tahh}
\label{TJsec}
The OPE isomorphisms also allow us to compute the $\hat{T}_\s \hsp{.025} \hat{J}(\s)$ OPE's of orbifold theory. 

To begin this discussion, we remind the reader of the $TJ$ OPE of the general affine-Virasoro construction$^{\rf{vme},\rf{Yam},\rf{Giveon},\rf{Lieh},\rf{rev}}$ 
\group{MNgroup}
\(
T(z) J_a(w)=M(L)_a^{\ b} (\frac{1}{(z-w)^2}+\frac{\partial_w}{z-w})J_b(w)+ \frac{N(L)_a^{\ bc}:J_b(w)J_c(w):}{z-w} + \reg
\)
\(
[L(m),J_a(n)]=-nM(L)_a^{\ b}J_b(m+n) +N(L)_a^{\ bc} \sum_{p \in \sz}:J_b(p) J_c(m+n-p):
\)
\(
M(L)_a^{\ b}=2G_{ac}L^{cb}+f_{ad}^{\ \ e}L^{dc}f_{ce}^{\ \ b}
\sp N(L)_{a}^{\ bc}=-if_{ad}^{\ \ (b} L^{c)d}  \label{MNdef}
\)
\(
M(L)_{ab}=M(L)_a^{\ c}G_{cd}=M(L)_{ba} \sp N(L)_a^{\ bc}=N(L)_a^{\ cb}
\)
\(
M(L)^*=\stheta M(L) \stheta^* =M(L)^\dagger \sp N(L)_a^{\ bc \hsp{.02} *}=-\stheta_a^{\ d}N(L)_d^{\ ef}\stheta_c^{\ b \hsp{.02} *}\stheta_f^{\ c \hsp{.02} *} \ . \label{MNstar}
\)
\reseteqn
In the special case of the $H$-invariant CFT's $A(H)$, one finds that the $H$-invariances of the tensors $M(L)$ and $N(L)$ 
\(
[\w,M(L)]=0
\sp \w_a^{\ d}N(L)_d^{\ ef} (\w^\dagger)_e^{\ b}(\w^\dagger)_f^{\ c}=N(L)_a^{\ bc} \label{wMN} \sp \forall \ \w \in H \subset Aut(g) \label{MNwgroup}
\)
follow from (\ref{MNdef}), (\ref{GwwG}) and (\ref{Hinv}).

For the general orbifold \tahh, the OPE isomorphisms for $\hat{J}(\s)$ and $\hat{T}_\s$ in (\ref{JJhatduality}) and (\ref{TThatarrow}) combine to give the derived OPE isomorphism 
\(
T(z) \j_{n(r) \mu}(w) \ \sa \ \hat{T}_\s(z)\hat{J}_{n(r) \mu}(w) \label{TJarrow}
\)
and this isomorphism gives the general $\hat{T}_\s \hsp{.025} \hat{J}(\s)$ OPE in sector $\s$ of \tahh
\group{mcncdualitygroup}
\ba
\hat{T}_\s(z)\hat{J}_{n(r) \mu}(w) &=& 
 \mc_{n(r) \mu}^{\hsp{.3} n(s) \nu}(\s) [\frac{1}{(z-w)^2}  + \frac{ \partial_w }{(z-w)}] \hat{J}_{n(s) \nu}(w)  \label{hatsTJOPE} \\
& & + \frac{\nc_{n(r) \mu}^{\hsp{.3} n(s) \nu ; n(t) \delta}(\s) :\hat{J}_{n(s) \nu}(w) \hat{J}_{n(t) \delta}(w):}{z-w} + \reg  \nonumber
\ea
\(
\mc_{n(r)\mu}^{\hsp{.3} n(s)\nu}(\s) = {\foot{\chi(\s)_{n(r) \mu}}} {\foot{\chi(\s)_{n(s) \nu}^{-1}}}  U(\s)_{n(r) \mu}^{\hsp{.3} a} M(L)_a^{\ b} U^\dagger(\s)_{b}^{\ n(s) \nu} \label{mcdef}
\)
\getletter{ncdeflett}
\vspace{-.1in}
\(
\nc_{n(r)\mu}^{\hsp{.3} n(s)\nu;n(t)\delta}(\s) = \frac{{\foot{\chi (\s)_{n(r) \mu}}}}{{\foot{ \chi (\s)_{n(s) \nu} \chi (\s)_{n(t) \delta}}}} \ U(\s)_{n(r) \mu}^{\hsp{.3} a} N(L)_{a}^{\ bc} U^\dagger(\s)_{b}^{\ n(s) \nu} U^\dagger(\s)_{c}^{\ n(t) \delta} \ . \label{ncdef}
\) 
\reseteqn
\\
The twisted tensors $\mc (\s) $ and $\nc (\s)$ in (\ref{mcdef},\ref{ncdeflett}) are the dual in sector $\s$ of the untwisted tensors $M(L)$ and $N(L)$
\(
M(L) \sa \mc(M(L);\s) \sp N(L) \sa \nc(N(L);\s)  \label{mnarrow}
\)
and the same result (\ref{mcncdualitygroup}) can be obtained by direct OPE calculation (see App.~\ref{TJapp}) in the twisted sectors of the orbifolds.  The orbifold duality transformations (\ref{mnarrow}) (and their realization in (\ref{mcdef},\ref{ncdeflett})) are on the same footing as the earlier ones for $\g, \fr, \lr$ and $\Ltheta$.

One also finds the $\mc$- and $\nc$-selection rules and their solutions
\group{MNEgroup}
\(
\mc_{n(r) \mu}^{\hsp{.3} n(s) \nu}(\s)(1-E_{n(r)}(\s) E_{n(s)}(\s)^*)=0
\)
\(
\mc_{n(r) \mu}^{\hsp{.3} n(s) \nu}(\s)=\delta_{n(r)-n(s),0 \mod \r(\s)} \mc_{n(r) \mu}^{\hsp{.3} n(r) \nu}(\s) 
\)
\(
\nc_{n(r) \mu}^{\hsp{.3} n(s) \nu; n(t)\delta}(\s) (1-E_{n(r)}(\s) E_{n(s)}(\s)^* E_{n(t)}(\s)^*)=0 
\)
\(
\nc_{n(r) \mu}^{\hsp{.3} n(s) \nu; n(t)\delta}(\s)=\delta_{n(r)-n(s)-n(t),0 \mod \r(\s)} \nc_{n(r) \mu}^{\hsp{.3} n(s) \nu; n(r)-n(s),\delta}(\s)  
\)
\(
\nc_{n(r) \mu}^{\hsp{.3} n(s) \nu; n(r)-n(s),\delta}(\s) = 0 \hsp{.3} \textup{unless } n(r)-n(s) \in \{ n(r) \}
\)
\reseteqn
which are the dual in sector $\s$ of the $H$-invariances (\ref{MNwgroup}) of $M(L)$ and $N(L)$ in the untwisted sector.

Another consequence of the $H$-invariances in (\ref{MNwgroup}) is that the twisted tensors $\mc(\s)$ and $\nc(\s)$ are class functions
\(
\mc(U(\s)v(\s);\s)=\mc(U(\s);\s) \sp \nc(U(\s)v(\s);\s)=\nc(U(\s);\s)
\)
under the $H$-conjugation in (\ref{conjugationgroup}).

Using the $\mc$- and $\nc$-selection rules in (\ref{MNEgroup}), one obtains the $\hat{T}_\s \hsp{.025} \hat{J}(\s)$ OPE's in the more informative presentation
\ba
\hat{T}_\s(z)\hat{J}_{n(r) \mu}(w) &=& 
 \mc_{n(r) \mu}^{\hsp{.3} n(r) \nu}(\s) [\frac{1}{(z-w)^2}  + \frac{ \partial_w }{(z-w)}] \hat{J}_{n(r) \nu}(w)  \label{TJOPEb} \\
&  & \ + \frac{\nc_{n(r) \mu}^{\hsp{.3} n(s) \nu; \ n(r)-n(s), \delta}(\s) :\hat{J}_{n(s) \nu}(w) \hat{J}_{n(r)-n(s), \delta}(w):}{z-w} + \reg  \nonumber
\ea
which shows the consistency of the monodromies on both sides of the relation.  The associated commutator\footnote{The complex conjugates $\mc^*$ and $\nc^*$ can be computed with (\ref{MNstar}), and the results follow the pattern seen in Eq.~(\ref{stargroup}): a factor of $\Ltheta^*$ or $\Ltheta$ for an up or down index respectively.  Then it is easily checked that (\ref{LJcommutator}) is consistent with the adjoint operations in (\ref{Jdagger}) and (\ref{twistLdagger}).}
\(
\hsp{-.5} [L_\s(m),\hat{J}_{n(r) \mu}(n+\srac{n(r)}{\r(\s)})]=-(n+\srac{n(r)}{\r(\s)}) \mc_{n(r) \mu}^{\hsp{.3} n(r) \nu}(\s) \hat{J}_{n(r) \nu}(m+n+\srac{n(r)}{\r(\s)}) \label{LJcommutator}
\)
\[
\hsp{.5} + \nc_{n(r) \mu}^{\hsp{.3} n(s) \nu; n(r)-n(s), \delta}(\s) \sum_{p \in \sz}:\hat{J}_{n(s) \nu}(p+\srac{n(s)}{\r(\s)}) \hat{J}_{n(r)-n(s),\delta}(m+n-p+\srac{n(r)-n(s)}{\r(\s)}):
\]
follows from (\ref{TJOPEb}).

\newsection{The Orbifolds of the ($H$ and Lie $h$)-invariant CFT's}
\label{liehsec}
Our presentation of these ``doubly-invariant'' CFT's and their orbifolds will follow the progression
\(
\alie \ \ \rightarrow \ \ \alieH \ \ \rightarrow \ \ \alieHH
\)
starting with a review of Lie symmetry in current-algebraic conformal field theory.

\subsection{The Lie $h$-invariant CFT's}
\label{Liehsec}
The Lie $h$-invariant CFT's$^{\rf{Lieh}-\rf{rev},\rf{Jan},\rf{us1}}$ on $g$, called collectively $\alie$, are those CFT's with a Lie symmetry\footnote{The Lie $h$-invariant CFT's $\alie$ are also invariant (at least) under the connected part $(\textup{Lie} \hsp{.025} H)_c$ of the corresponding Lie group Lie $H$.  Moreover $(\textup{Lie} \hsp{.025} H)_c \subset \textup{Lie} \hsp{.025} G \subset Aut(g)$ and $(\textup{Lie} \hsp{.025} H)_c \subset Aut(h)$, so one may consider the orbifolds by a Lie group $\alie / (\textup{Lie} \hsp{.025} H)_c$ for any Lie $h$-invariant CFT.  We shall not do so here.} $h\subset g$, which may be realized globally or locally.  Large numbers of Lie $h$-invariant CFT's are known,$^{\rf{Lieh}}$ and simple examples of $\alie$ include the WZW model $A_g(\textup{Lie} \hsp{.025} g)$, described by the affine-Sugawara construction$^{\rf{BH},\rf{Halp}-\rf{Segal},\rf{rev}}$ on $g$, and the $g/h$ coset constructions.$^{\rf{BH},\rf{Halp},\rf{GKO},\rf{rev}}$

To describe this class of CFT's more precisely, we begin with the index decomposition 
\vspace{-.1in}
\alpheqn
\(
a=(A,I) , \hsp{.1} A \textup{ in } h , \hsp{.1} I \textup{ in } g/h : \hsp{.2}
G_{AI}=f_{AB}^{\hsp{.2} I}=f_{AI}^{\hsp{.2} B}=0 
\)
\vspace{-.25in}
\(
J_A(z) J_B (w) = \frac{G_{AB}}{(z-w)^2} + \frac{i f_{AB}^{\hsp{.2} C} J_{C}(w)}{z-w} + O((z-w)^0) \label{JJhOPE}
\)
\reseteqn
where $h$ is any reductive subalgebra of the ambient algebra $g$ and $J_h=\{ J_A \}$ are the $h$ currents.  The Lie $h$-invariant CFT's $\alie$ are those CFT's on $g$ whose inverse inertia tensor is invariant under infinitesimal transformations generated by Lie $h$
\vspace{-.05in}
\(
T=L^{ab}:J_a J_b: \sp \delta_AL^{ab}=L^{c (a}f_{cA}^{\ \ \ b)}= -i N(L)_A^{\hsp{.1} ab}=0  \sp A=1,...,\textup{dim} h \label{LiehLcondtion}
\vspace{-.05in}
\)
where $N(L)$ is defined in (\ref{MNgroup}).  The inverse inertia tensors of $\alie$ satisfy a consistent$^{\rf{Lieh}}$ reduced Virasoro master equation.  

At least for all unitary CFT's in $\alie$, it follows that$^{\rf{Giveon},\rf{Lieh}}$ 
\vspace{-.1in}
\group{hblockMgroup}
\(
M(L)_A^{\hsp{.1} I}=0
\sp M(L)_A^{\hsp{.1} C} M(L)_C^{\hsp{.1} B}= M(L)_A^{\hsp{.1} B} \label{Msquared}
\)
\vspace{-.2in}
\(
T(z) J_A(w)=M(L)_A^{\hsp{.1} B} (\frac{1}{(z-w)^2}+\frac{\partial_w}{z-w})J_B(w) + \reg \label{TJOPELieh} 
\)
\reseteqn
where $M(L)_A^{\hsp{.1} B}$ is the $h$ block of $M(L)$ (see Eq.~(\ref{MNgroup})).  As seen in (\ref{hblockMgroup}), the essential simplifications of the Lie $h$-invariant CFT's are that a) $M(L)_A^{\hsp{.1} B}$ is a projector and b) there is no two-current term on the right side of (\ref{TJOPELieh}).  It then follows$^{\rf{Giveon},\rf{Lieh}}$ in a left eigenbasis of $M(L)_A^{\hsp{.1} B}$ that the $h$ currents are either $(1,0)$ or $(0,0)$ operators of $T(z)$.  For the affine-Sugawara constructions, one has $h=g$ and all the $h$  currents are $(1,0)$ operators, so that Lie $h$ is realized globally.  For the $g/h$ coset constructions the $h$ currents are $(0,0)$ operators, so that Lie $h$ is realized locally.  For general Lie $h$-invariant CFT's, the affine $h$ subalgebras (\ref{JJhOPE}) decompose as$^{\rf{Lieh}}$
\vspace{-.05in}
\(
h=h_0 \oplus h_1 \label{01untwistedcommute}
\vspace{-.05in}
\)
where $h_0$ and $h_1$ are the closed affine subalgebras of the $(0,0)$ and the $(1,0)$ operators respectively.\footnote{In what follows, we will generally assume unitarity of the Lie $h$-invariant CFT's and hence the Lie $h$ relations (\ref{hblockMgroup}).  For the affine-Sugawara and coset constructions, however, it is clear that (\ref{hblockMgroup}) holds without assumption of unitarity.  Section~\ref{innersec} gives further evidence that the relations (\ref{hblockMgroup}) may follow directly from the Virasoro master equation.}

\subsection{The ($H$ and Lie $h$)-invariant CFT's}
\label{Hliehcftsec}
We consider next the ($H$ and Lie $h$)-invariant CFT's, called collectively $\alieH$, 
\alpheqn
\(
\alieH \subset A(\textup{Lie} \hsp{.025} h) \sp \alieH \subset A(H) 
\)
\(
h\subset g \sp \lieH \subset Aut(g) \sp \lieH \subset Aut(h) \ .
\)
\reseteqn
These are the CFT's which are simultaneously invariant under some Lie $h$ and \textit{also} a finite group $\lieH$, and the CFT's of $\alieH$ can be used to from the orbifolds $\talieH$.  Because $g$, $h$ and $H$ are not fixed, these \textit{doubly-invariant} CFT's provide a different slice of orbifold theory than the large examples in Subsec.~\ref{cyclicsec} and Secs.~\ref{SNsec} and \ref{innersec}.  

We will not attempt to classify the ($H$ and Lie $h$)-invariant CFT's, but the set is large, including the general WZW model 
\(
A_g(H)=A_g(\textup{Lie} \hsp{.025} g)=A(\textup{Lie} \hsp{.025} g (H)) \sp H \subset Aut(g)
\)
which is described by the affine-Sugawara construction$^{\rf{BH},\rf{Halp}-\rf{Segal},\rf{rev}}$ on $g$, and the general $H$-invariant coset construction$^{\rf{us2},\rf{us3}}$
\(
\frac{g}{h}(H) \subset \alieH \ .
\)
Many other examples in $\alieH$ are known including $a)$ the level families that live on the Lie $h$-invariant graphs with a graph symmetry$^{\rf{Obers},\rf{Lieh},\rf{rev}}$ and $b)$ the Lie $(h=g_{diag})$-invariant CFT's$^{\rf{us1},\rf{us2}}$ in $A(\zl(\textup{permutation}))$, whose stress tensors describe the untwisted sectors of the $G_{diag(\s)}$-invariant cyclic orbifolds.  

Another large class of doubly-invariant CFT's $A(\textup{Cartan} \hsp{.025} g(H(d)))$ is discussed in Sec.~\ref{innersec}, where $H=H(d)$ is any group of inner automorphisms.  These doubly-invariant CFT's are particularly important because, as we shall see, they underlie the connection between inner-automorphic orbifolds and stress-tensor spectral flow.

For CFT's in $\alieH$, the subalgebra $h\subset g$ is an $H$-covariant subalgebra$^{\rf{us3}}$
\(
\w(h_\s)_A^{\hsp{.1} I}=0 \sp J_A \ \p=\w(h_\s)_A^{\hsp{.1} B}J_B \sp \forall \ \w(h_\s) \in \lieH \sp A=1,...,\textup{dim}h
\)
of the $H$-covariant ambient algebra $g$.  Here $\w(h_\s)_A^{\ B}$ is the $h$ block of $\w(h_\s)$ and $J_A \ \p$ satisfies the same OPE as $J_A$, given in (\ref{JJhOPE}).  Then, using (\ref{wunitary},\ref{fgautoslett}) and (\ref{JJhOPE}), the properties
\alpheqn
\(
\w(h_\s)_A^{\hsp{.1} C} \w^\dagger(h_\s)_C^{\hsp{.1} B}=\delta_A^{\hsp{.1} B} \sp \w(h_\s)_I^{\ A}=0 
\)
\(
\w(h_\s)_A^{\hsp{.1} C} \w(h_\s)_B^{\hsp{.1} D}G_{CD}=G_{AB} \sp \w(h_\s)_A^{\hsp{.1} D} \w(h_\s)_B^{\hsp{.1} E}f_{DE}^{\hsp{.2} F} \w^\dagger(h_\s)_F^{\hsp{.1} C}=f_{AB}^{\hsp{.2} C} \label{gfhautos}
\)
\reseteqn
are obtained for all $h_\s$.  The relations in (\ref{gfhautos}) express the $\lieH$-invariance of the $h$ tensors $G_{AB}$ and $f_{AB}^{\hsp{.2} C}$.  It follows that each matrix action $\w(h_\s)$ is block diagonal
\(
\w(h_\s)=\pmatrix{\w(h_\s)_A^{\hsp{.1} B} & 0 \cr 0 & \w(h_\s)_I^{\hsp{.1} J} \cr} \label{wblockdiagonal}
\)
and that the $h$ and $g/h$ blocks $\w(h_\s)_A^{\hsp{.1} B}$ and $\w(h_\s)_I^{\hsp{.1} J}$ of $\w(h_\s)$ are separately unitary.  Then the unitary eigenvector matrices $U(\s)$ and $U^\dagger (\s)$ may also be taken block diagonal with each block separately unitary.  Finally, Eqs.~(\ref{MNwgroup}) and (\ref{wblockdiagonal}) imply the vanishing commutator
\(
\w(h_\s)_A^{\hsp{.1} C} M(L)_C^{\hsp{.1} B}=M(L)_A^{\hsp{.1} C}\w(h_\s)_C^{\hsp{.1} B}  \label{wMMw}
\)
among the $h$ blocks of $\w(h_\s)$ and $M(L)$.

\subsection{The orbifolds $\talieH$}
\label{Liehorbifoldssec}
We turn now to the orbifolds by $H$ of the doubly-invariant CFT's $\alieH$
\(
\alieHH \subset \ahh \sp h \subset g \sp \lieH\subset Aut(g) \sp \lieH \subset Aut(h) \ .
\)
The twisted $g$ currents $\hat{J}_{n(r)\mu}$ of these orbifolds satisfy the twisted current algebra 
\(
\gb (H\subset Aut(g), \ H \subset Aut(h);\s) \label{ghdoubleH}
\)
whose form is included in (\ref{ghatgroup}).  In this case, the ambient algebra (\ref{ghdoubleH}) has a twisted $h$ subalgebra$^{\rf{us3}}$ generated by the twisted $h$ currents $\hat{J}_{\shb (\s)}$, which we will discuss below.  When Lie $h$ is realized locally, one must learn to gauge the twisted $h$ currents in an action formulation of these orbifolds.  

To study the twisted $h$ currents in a ``conformal weight'' basis (see Eq.~(\ref{heigenJgroup})), we begin with the simultaneous left-eigenvalue problem
\group{simevprob}
\(
U(\s)_{\tn(r) \theta_i \mu}^{\hsp{.4} B} \w(h_\s)_B^{\ \ A} = E_{\tn(r)}(\s) U(\s)_{\tn(r) \theta_i \mu}^{\hsp{.4} A}   \label{hblockwev}
\)
\(
U(\s)_{\tn(r) \theta_i \mu}^{\hsp{.4} B} M(L)_B^{\hsp{.1} A} =\theta_i U(\s)_{\tn(r) \theta_i \mu}^{\hsp{.4} A}  \label{hblockMev}
\)
\(
E_{\tn(r)}(\s)=e^{-\frac{2 \pi i \tn(r)}{\r(\s)}} \sp \tn(r) \in \{ n(r) \} \sp \theta_i \in \{ 0,1 \} \sp \srange \label{thetarange}
\)
\reseteqn
defined on the (unitary) $h$ block of $U$.  Eq.~(\ref{hblockwev}) is the induced $H$-eigenvalue problem$^{\rf{us3}}$ for the $H$-covariant subalgebra $h$, and the consistency of the simultaneous eigenvalue problem (\ref{simevprob}) is guaranteed by the vanishing commutator in Eq.~(\ref{wMMw}).  The allowed values of $\theta_i$ in (\ref{thetarange}) follow from $M^2=M$ in (\ref{Msquared}), and we emphasize that the matrix $M(L)$ and its eigenvalues $\theta_i(\s)=\theta_i(0)=\theta_i$ are \textit{independent} of $\s$.

The analysis$^{\rf{us2}}$ reviewed in Sec.~\ref{reviewsec} for the ambient algebra $g$ can now be applied, mutatis mutandis, to the $H$-covariant subalgebra $h$.  With (\ref{hblockMgroup}) and (\ref{simevprob}), we find that the $h$-eigencurrents $\j_h(\s) =\{ \j_{\tn(r) \theta_i \mu} \}$ of sector $\s$
\group{heigenJgroup}
\(
\j_{\tn(r) \theta_i \mu}(z) \equiv {\foot{\chi(\s)_{\tn(r) \theta_i \mu}}} U(\s)_{\tn(r) \theta_i \mu}^{\hsp{.4} A} J_A(z)  \sp \j_{\tn(r) \theta_i \mu} (z)\p = E_{\tn(r)}(\s) \j_{\tn(r) \theta_i \mu} (z) \label{heigenJdef}
\)
\(
T(z) \j_{\tn(r) \theta_i \mu} (w)=\theta_i \hsp{.03} (\frac{1}{(z-w)^2}+\frac{\partial_w}{z-w}) \j_{\tn(r) \theta_i \mu} (w) +\reg \label{teigenhj}
\)
\reseteqn
have conformal weight $\theta_i \in \{ 0,1 \}$ and simultaneously a diagonal response $E_{\tn(r)}(\s)$ to the automorphism group $\lieH$.  

Then using the OPE's (\ref{JJhOPE}) of the $h$ current algebra, the $h$-eigencurrents (\ref{heigenJdef}) and the OPE isomorphism 
\group{hatHharrowgroup}
\vspace{-.1in}
\(
\j_h(\s) \ \sa \  \hat{J}_{\shb (\s)} \sp \hat{J}_{\shb (\s)} = \{ \hat{J}_{\tn(r) \theta_i \mu} \}
\)
\(
\textup{automorphisms } E_{\tn(r)}(\s) \ \sa \ \textup{monodromies } E_{\tn(r)}(\s) 
\)
\reseteqn
one finds the twisted $h$ current system (read Eq.~(\ref{hatJOPErecap}) with $n(r)\mu \rightarrow \tn(r) \theta_i \mu$).  The twisted $h$ tensors of this system are
\alpheqn
\(
\g_{\tn(r) \theta_i \mu; \tn(s) \theta_j \nu}(\s) = {\foot{\chi(\s)_{\tn(r) \theta_i \mu}}} {\foot{\chi(\s)_{\tn(s) \theta_j \nu}}} U(\s)_{\tn(r) \theta_i \mu}^{\hsp{.4} A} U(\s)_{\tn(s) \theta_j \nu}^{\hsp{.4} B} G_{AB} 
\)
\(
\fr_{\tn(r) \theta_i \mu;\tn(s) \theta_j \nu}^{\hsp{.8} \tn(t) \theta_k \delta}(\s) = \frac{ {\foot{\chi (\s)_{\tn(r) \theta_i \mu}}} {\foot{\chi (\s)_{\tn(s) \theta_j \nu}}} }{ {\foot{\chi(\s)_{\tn(t) \theta_k \delta}}}} \ U(\s)_{\tn(r) \theta_i \mu}^{\hsp{.4} A} U(\s)_{\tn(s) \theta_j \nu}^{\hsp{.4} B} f_{AB}^{\ \ \ C} \uds_C^{\ \ \tn(t) \theta_k \delta} 
\)
\[
\smal{\fr_{\tn(r) \theta_i \mu; \tn(s) \theta_j \nu}^{\hsp{.8} \tn(u) \theta_l \epsilon}(\s) \fr_{\tn(t) \theta_k \delta; \tn(u) \theta_l \epsilon}^{\hsp{.8} \tn(v) \theta_m \gamma}(\s)
+ \fr_{\tn(s) \theta_j \nu;\tn(t) \theta_k \delta}^{\hsp{.8} \tn(u) \theta_l \epsilon}(\s) \fr_{\tn(r) \theta_i \mu; \tn(u) \theta_l \epsilon}^{\hsp{.8} \tn(v) \theta_m \gamma}(\s)}
\]
\(
+ \fr_{\tn(t) \theta_k \delta; \tn(r) \theta_i \mu }^{\hsp{.8} \tn(u) \theta_l \epsilon}(\s) \fr_{\tn(s) \theta_j \nu; \tn(u) \theta_l \epsilon}^{\hsp{.8} \tn(v) \theta_m \gamma}(\s) =0 \label{hjacobi}
\)
\getletter{hfrantisymmlett}
\vspace{-.2in}
\(
\fr_{\tn(r) \theta_i \mu;\tn(s) \theta_j \nu; \tn(t) \theta_k \delta}(\s) \equiv \fr_{\tn(r) \theta_i \mu;\tn(s) \theta_j \nu}^{\hsp{.8} \tn(u) \theta_l \epsilon}(\s) \g_{\tn(u) \theta_l \epsilon; \tn(t) \theta_k \delta}(\s)=-\fr_{\tn(r) \theta_i \mu;\tn(t) \theta_k \delta; \tn(s) \theta_j \nu}(\s) \label{hfrantisymm}
\)
\(
\g_{\tn(r) \theta_i \mu; \tn(s) \theta_j \nu}(\s)(1-E_{\tn(r)}(\s) E_{\tn(s)}(\s))=0
\)
\(
\fr_{\tn(r) \theta_i \mu;\tn(s) \theta_j \nu}^{\hsp{.82} \tn(t) \theta_k \delta}(\s)(1-E_{\tn(r)}(\s) E_{\tn(s)}(\s) E_{\tn(t)}(\s)^*)=0
\)
\reseteqn
and the corresponding twisted $h$ subalgebra is
\group{JJOPELiehgroup}
\(
\hb(\s) \ \equiv \ \hb (H\subset Aut(h); \s) \ \ \subset \ \ \gb (H\subset Aut(g), \ H \subset Aut(h);\s)
\)
\[
[\hat{J}_{\tn(r) \theta_i \mu}(m+\srac{\tn(r)}{\r(\s)}), \hat{J}_{\tn(s) \theta_j \nu} (n+\srac{\tn(s)}{\r(\s)})] = (m+\srac{\tn(r)}{\r(\s)}) \delta_{m+n+\frac{\tn(r)+\tn(s)}{\r(\s)},0} \ \g_{\tn(r) \theta_i \mu;-\tn(r), \theta_j \nu}(\s)
\]
\(
\hsp{1} + i \fr_{\tn(r) \theta_i \mu;\tn(s) \theta_j \nu}^{\hsp{.82} \tn(r)+\tn(s), \theta_k \delta}(\s) \hat{J}_{\tn(r)+\tn(s), \theta_k \delta}(m+n+\srac{\tn(r)+\tn(s)}{\r(\s)})
\)
\(
\# \{ \hat{J}_{\shb (\s)} \} = \# \{ J_h \} = \textup{dim}h \ .
\)
\reseteqn
The subalgebra $\hb(\s)$ is the dual in sector $\s$ of the untwisted affine $h$ subalgebra in (\ref{JJhOPE}), and the Jacobi identity of $\hb(\s)$ follows with (\ref{hjacobi},\ref{hfrantisymmlett}).

We turn next to the twisted inverse inertia tensor $\lr(\s)$ in the stress tensor $\hat{T}_\s$ of each orbifold in $\talieH$.  In addition to the selection rules (\ref{lrrecapgroup}) -- which are dual to the $H$-symmetry of $\alieH$ -- we find that $\lr(\s)$ also satisfies the condition\footnote{A special case of this condition was first seen for cyclic permutation orbifolds in Ref.~\rf{us1}.}
\(
\lr^{n(u) \gamma; (n(s)\nu } (\s) \fr_{n(u) \gamma; \tn(r) \theta_i \mu;}^{\hsp{.7}  n(t) \delta) } (\s) =-i \nc_{\tn(r) \theta_i \mu}^{\hsp{.4} n(s) \nu; n(t) \delta}(\s) =0
\)
which is dual to the Lie $h$ condition (\ref{LiehLcondtion}).  Moreover, the $\hat{T}_\s \hat{J}_{\shb(\s)}$ relations 
\group{TJOPELiehgroup}
\(
\mc_{\tn(r) \theta_i \mu}^{\hsp{.4} \tn(s) \theta_j \nu}(\s)= \theta_i \delta_{\tn(r) \theta_i \mu}^{\hsp{.4} \tn(s) \theta_j \nu} 
\)
\(
\hat{T}_\s(z) \hat{J}_{\tn(r) \theta_i \mu}(w)=\theta_i \hsp{.03} (\frac{1}{(z-w)^2}+\frac{\partial_w}{z-w})\hat{J}_{\tn(r) \theta_i \mu}(w) + \reg \label{twistTJhOPE}
\)
\(
[L_\s(m), \hat{J}_{\tn(r) \theta_i \mu} (n+\srac{\tn(r)}{\r(\s)})]=-\theta_i \hsp{.03} (n+ \srac{\tn(r)}{\r(\s)})\hat{J}_{\tn(r) \theta_i \mu}(m+n+ \srac{\tn(r)}{\r(\s)})  \label{LJLieh}
\)
\reseteqn
can be obtained from (\ref{teigenhj}), (\ref{hatHharrowgroup}) and the derived isomorphism (\ref{TThatarrow}), or as a special case of (\ref{TJOPEb}) and (\ref{LJcommutator}).  These results show that each twisted $h$ current of any orbifold $\talieH$ is either$^{\rf{us3}}$ a twisted $(1,0)$ operator (defined by $\theta_i=1$) or a twisted $(0,0)$ operator (defined by $\theta_i=0$).  Looking back, we see that the properties (\ref{TJOPELiehgroup}) are orbifold reflections of the Lie $h$ properties in (\ref{hblockMgroup}) and (\ref{teigenhj}).

\subsection{WZW orbifolds, coset orbifolds and orbifold $K$-conjugation}
\label{Ksec}
We work out here the general WZW orbifold$^{\rf{us2}}$ and the general coset orbifold,$^{\rf{us3}}$ which provide simple sets of examples in $\talieH$.  In this discussion, we emphasize the role of the WZW orbifolds in orbifold $K$-conjugation,$^{\rf{us2}}$ and the role of orbifold $K$-conjugation in the construction of the coset orbifolds.

The affine-Sugawara construction$^{\rf{BH},\rf{Halp}-\rf{Segal},\rf{rev}}$ on $g$
\group{ASgroup}
\(
A_g(H)=A(\textup{Lie} \hsp{.025} g(H)) \sp H \subset Aut(h=g) 
\) 
\(
T_g=L_g^{ab}:J_a J_b: \sp L_g^{ab}=\oplus_I \frac{\eta_I^{ab}}{2k_I+Q_I}
\)
\(
L_g^{cd} \w_c^{\ a} \w_c^{\ b}=L_g^{ab} \sp \forall \ \w \in H \subset Aut(h=g)
\)
\(
N(L_g)_a^{\ bc}=0  \sp M(L_g)_a^{\ b}=\delta_a^{\ b}
\)
\(
[L_g(m), J_a(n)]= -n J_a(m+n) 
\)
\reseteqn
is always a (global Lie$\hsp{.025}h=g$)-invariant CFT which is also $H$-invariant under any $H\subset Aut(g)$.  Then the description of the general WZW orbifold$^{\rf{us2}}$ 
\group{OASgroup}
\(
\frac{A_g(H)}{H} = \frac{A(\textup{Lie} \hsp{.025} g(H))}{H} \sp \hb(\s)=\gb(\s)
\)
\getletter{OASgroupletta}
\vspace{-.15in}
\(
\hat{T}_{\sgb (\s)} \ = \ \lr_{\sgb (\s)}^{n(r) \mu; -n(r),\nu}(\s):\hat{J}_{n(r) \mu} \hat{J}_{-n(r), \nu}: \sp \srange \label{OASstresstensor}
\)
\getletter{OASgrouplettb}
\vspace{-.2in}
\( 
\lr_{\sgb (\s)}^{n(r) \mu; -n(r), \nu}(\s) \ = \ {\foot{\chi(\s)_{n(r) \mu}^{-1}}} {\foot{\chi(\s)_{-n(r), \nu}^{-1}}} L_g^{ab} U^\dagger(\s)_{a}^{\ n(r) \mu} U^\dagger(\s)_{b}^{\ -n(r), \nu}
\)
\(
 \nc_{n(r) \mu}^{\hsp{.3} n(s)\nu; n(t) \delta}(\s)=0 \sp \mc_{n(r)\mu}^{\hsp{.3} n(s)\nu}(\s)=\delta_{n(r)\mu}^{\hsp{.3} n(s)\nu} 
\)
\(
[L_\s^{\sgb(\s)}(m), \hat{J}_{n(r) \mu} (n+\srac{n(r)}{\r(\s)})]= -(n+\srac{n(r)}{\r(\s)}) \hat{J}_{n(r) \mu} (m+n+\srac{n(r)}{\r(\s)})  \label{WZWcommutator}
\)
\reseteqn
is obtained from the results of the previous subsection.  In the WZW orbifolds, the twisted $g$ currents $\{ \hat{J}(\s) \}$ satisfy the general twisted current algebra $\gb(\s)$ in (\ref{ghatgroup}) and each of the twisted $g$ currents is a twisted (1,0) operator (with $\theta_i(\s)=\theta_i=1$) under the orbifold affine-Sugawara construction $\hat{T}_{\sgb (\s)}$.  The case of the WZW cyclic permutation orbifolds was discussed in Ref.~\rf{us2} and further discussion of WZW orbifolds is found in Secs.~\ref{cwsec} and \ref{innersec}.

Returning for a moment to the general affine-Virasoro construction (\ref{VMEgroup}), we remind the reader that the affine-Sugawara construction $T_g$ also plays a central role in the operation known as $K$-conjugation$^{\rf{BH},\rf{Halp},\rf{GKO},\rf{Kir},\rf{vme},\rf{rev}}$ 
\group{genKuntwist}
\(
A \ka \tilde{A}
\)
\getletter{genKuntwista}
\vspace{-.2in}
\(
T_{g} (z)=T (z)+\tilde{T} (z)
\sp \tilde{T} (z) T (w)=\reg
\)
\(
L_g^{ab}= L^{ab}+\tilde{L}^{ab} \sp c_g=c+\tilde{c} \label{genKconjuntwist}
\)
\reseteqn
which relates $K$-conjugate pairs $A$, $\tilde{A}$ of current-algebraic CFT's on $g$.  The simplest example of $K$-conjugation is the $g/h$ coset construction$^{\rf{BH},\rf{Halp},\rf{GKO},\rf{rev}}$ 
\(
T_{g/h}=T_g-T_h \sp c_{g/h}=c_g-c_h \sp [L_{g/h}(m), J_A(n)]=0 \sp A=1,...,\textup{dim}h
\)
which is a (local Lie $h$)-invariant CFT (with $\theta_i=0$ for each $h$ current).

We emphasize that $K$-conjugation is closed on the space of $H$-invariant CFT's
\(
A(H) \ka \tilde{A}(H) \sp H\subset Aut(g) \label{KHinv}
\)
because the affine-Sugawara construction $T_g$ describes the $H$-invariant CFT $A_g(H)$.  

For the corresponding orbifolds by $H$, one finds that $K$-conjugation (\ref{KHinv}) and the duality transformation (\ref{lr}) combine to give orbifold $K$-conjugation$^{\rf{us2}}$
\group{genK}
\(
\frac{A(H)}{H} \ka \frac{\tilde{A}(H)}{H} \label{Kconjugateorbifolds}
\)
\getletter{genKa}
\vspace{-.1in}
\(
\hat{T}_{\sgb (\s)} (z)=\hat{T}_\s (z)+\hat{\tilde{T}\hspace{.035in}}_{\hspace{-.05in}\s} (z)
\sp \hat{\tilde{T}\hspace{.035in}}_{\hspace{-.05in}\s} (z) \hat{T}_\s (w)=\reg
\)
\vspace{-.2in}
\getletter{genKb}
\(
\lr_{\sgb (\s)}^{n(r) \mu; n(s) \nu}(\s)=\lr^{n(r) \mu; n(s) \nu}(\s)+\tilde{\lr}^{n(r) \mu; n(s) \nu}(\s) \label{genKconjtwist}
\)
\vspace{-.3in}
\getletter{genKc}
\(
\hat{c}_{\sgb (\s)}=c_g \sp \hat{c}(\s)=c \sp \hat{\tilde{c}\hspace{.035in}}_{\hspace{-.05in}}(\s)=\tilde{c}  \label{genKctwist}
\)
\reseteqn
which relates $K$-conjugate pairs of stress tensors in \tahh \ through the orbifold affine-Sugawara construction $\hat{T}_{\sgb (\s)}$. 

The simplest example of orbifold $K$-conjugation, namely the general coset orbifold 
\group{cosetorbifoldgroup}
\(
\frac{\frac{g}{h}(H)}{H} \subset \alieHH \sp h\subset g \sp H \subset Aut(h) \sp H \subset Aut(g) \label{cosetname}
\)
\getletter{cosetname2lett}
\vspace{-.2in}
\(
T_{g/h}(z)=T_g(z)-T_h(z) \ \sa \ \hat{T}_{\frac{g/h}{H}}(z)_\s=\hat{T}_{\sgb(\s)}-\hat{T}_{\shb(\s)} 
\)
\(
T_h=L_h^{AB}:J_A J_B: \ \ \sa \ \ \hat{T}_{\shb(\s)}=\lr_{\shb(\s)}^{\tn(r) 0 \mu; -\tn(r), 0 \nu}(\s):\hat{J}_{\tn(r)0 \mu} \hat{J}_{-\tn(r), 0 \nu}:
\)
\(
\lr_{\shb(\s)}^{\tn(r) 0 \mu; \tn(s) 0  \nu}(\s)= {\foot{\chi(\s)^{-1}_{\tn(r) 0 \mu}}} {\foot{\chi(\s)^{-1}_{\tn(s) 0 \nu}}} L_h^{AB} U^\dagger(\s)_A^{\ \tn(r) 0 \mu} U^\dagger(\s)_B^{\ \tn(s) 0 \nu} 
\)
\reseteqn
is also found in $\talieH$.  The untwisted sectors of these orbifolds are formed from the $H$-invariant coset constructions $\frac{g}{h}(H)$, and the twisted $h$ currents $\hat{J}_{\shb (\s)}=\{ \hat{J}_{\tn(r)0 \mu}\}$ have $\theta_i(\s)=\theta_i=0$ in this case because the untwisted $h$ currents $J_A$ are $(0,0)$ operators of $T_{g/h}$.  

In further detail, the twisted $h$ currents are twisted $(1,0)$ operators under $\hat{T}_{\sgb(\s)}$ and $\hat{T}_{\shb(\s)}$ and hence twisted $(0,0)$ operators under $\hat{T}_{(g/h)/H}$:
\alpheqn
\(
[L_\s^{\sgb(\s)}(m), \hat{J}_{\tn(r) 0 \mu} (n+\srac{\tn(r)}{\r(\s)})]
=-(n+\srac{\tn(r)}{\r(\s)}) \hat{J}_{\tn(r)0 \mu} (m+n+\srac{\tn(r)}{\r(\s)})
\)
\(
[L_\s^{\shb(\s)}(m), \hat{J}_{\tn(r) 0 \mu} (n+\srac{\tn(r)}{\r(\s)})]
=-(n+\srac{\tn(r)}{\r(\s)}) \hat{J}_{\tn(r)0 \mu} (m+n+\srac{\tn(r)}{\r(\s)})
\)
\(
[L_\s^{\frac{g/h}{H}}(m), \hat{J}_{\tn(r) 0 \mu} (n+\srac{\tn(r)}{\r(\s)})]=0 \ \ . \label{LhatJ00commutator} 
\)
\reseteqn
The eigenvalue $\theta_i=0$ of the twisted $h$ currents controls the vanishing final commutator with the coset orbifold Virasoro generators.  The results (\ref{cosetname},\ref{cosetname2lett}) and (\ref{LhatJ00commutator}) are equivalent to those given in Ref.~\rf{us3}, which also discusses the orbifolds of the $\zl$-invariant coset constructions in further detail.  Finally, the twisted $h$ currents satisfy the twisted $h$ subalgebra
\[
[\hat{J}_{\tn(r) 0 \mu}(m+\srac{\tn(r)}{\r(\s)}), \hat{J}_{\tn(s) 0 \nu} (n+\srac{\tn(s)}{\r(\s)})] = (m+\srac{\tn(r)}{\r(\s)}) \delta_{m+n+\frac{\tn(r)+\tn(s)}{\r(\s)},0} \ \g_{\tn(r) 0 \mu;-\tn(r), 0 \nu}(\s)
\]
\(
\hsp{1} + i \fr_{\tn(r) 0 \mu;\tn(s) 0 \nu}^{\hsp{.73} \tn(r)+\tn(s), 0 \delta}(\s) \hat{J}_{\tn(r)+\tn(s), 0 \delta}(m+n+\srac{\tn(r)+\tn(s)}{\r(\s)})
\)
which is dual to the untwisted $h$ algebra whose OPE's are given in (\ref{JJhOPE}).

Other applications of orbifold $K$-conjugation are found in the following subsection and Subsec.~\ref{flowsec}.

\subsection{Decomposition of the twisted $h$ subalgebras}
\label{decompostionsubsec}
With the help of orbifold $K$-conjugation, we will show in this subsection that the twisted affine $h$ subalgebra $\hb(\s)$ of each sector of each orbifold $\talieH$ decomposes as
\(
\hb (\s)=\hb_0(\s) \oplus \hb_1(\s) \label{twist10commute}
\)
where $\hb_0(\s)$ and $\hb_1(\s)$ are the closed affine subalgebras of the twisted (0,0) and (1,0) operators respectively.  The algebra $\hb(\s) \subset \gb (\s)$ is given in (\ref{JJOPELiehgroup}) and the orbifold argument given below parallels the argument in Ref.~\rf{Lieh} for the decomposition (\ref{01untwistedcommute}) of the $h$ subalgebra of any Lie $h$-invariant CFT. 

For the orbifolds $\talieH$, we may combine orbifold $K$-conjugation (\ref{genK}) with the $L_\s$, $\hat{J}_{\shb (\s)}$ commutator in (\ref{TJOPELiehgroup}) to find the relation
\(
\alieHH \ka \frac{\tilde{A}(\textup{Lie} \hsp{.025}h(H))}{\lieH} \ : \hsp{.3}
\theta_i^{\sgb (\s)}=1=\theta_i + \tilde{\theta}_i  \label{thetaKcong}
\)
among the ``conformal weights'' $\theta$ of the twisted $h$ currents of each $K$-conjugate orbifold pair in $\talieH$.

In what follows, we consider various pairs of generators $\hat{J}_{\shb (\s)}$ of the twisted $h$ subalgebra $\hb(\s)$ of sector $\s$ in any particular orbifold $\talieH$, starting with an arbitrary pair of twisted $(0,0)$ operators.  According to (\ref{TJOPELiehgroup}), the commutator of this pair of twisted currents commutes with the stress tensor $\hat{T}_\s$ of this orbifold
\(
[L_\s(m),[\hat{J}_{\tn(r) 0 \mu} (n+\srac{\tn(r)}{\r(\s)}), \hat{J}_{\tn(s) 0 \nu} (p+\srac{\tn(s)}{\r(\s)})]]=0
\)
so the set of twisted $(0,0)$ operators of this orbifold is closed under commutation.  Similarly the set of twisted $(1,0)$ operators of this orbifold is closed under commutation because, according to (\ref{thetaKcong}), these currents and their commutators are twisted $(0,0)$ operators of the $K$-conjugate stress tensors $\hat{\tilde{T}\hspace{.035in}}_{\hspace{-.05in}\s}$.  To prove that the twisted $(0,0)$ and twisted $(1,0)$ operators of the orbifold commute, use the chain rule and the commutator (\ref{TJOPELiehgroup}) to see that
\[
\hsp{-2} [L_\s(m),[\hat{J}_{\tn(r) 0 \mu} (n+\srac{\tn(r)}{\r(\s)}), \hat{J}_{\tn(s) 1 \nu} (p+\srac{\tn(s)}{\r(\s)})]]
\]
\(
\hsp{2} =-(p+\srac{\tn(s)}{\r(\s)})[\hat{J}_{\tn(r) 0 \mu} (n+\srac{\tn(r)}{\r(\s)}), \hat{J}_{\tn(s) 1 \nu} (p+m+\srac{\tn(s)}{\r(\s)})] \ .
\)
Then use the twisted $h$ algebra (\ref{JJOPELiehgroup}) on both sides, followed by the commutator (\ref{TJOPELiehgroup}) again on the left side. The result of this computation 
\(
\g_{n(r) 0 \mu, n(s) 1 \nu}=\fr_{n(r) 0 \mu, n(s) 1 \nu}^{\hsp{.72} n(r)+n(s), \theta_k \delta}=0 
\)
establishes the decomposition (\ref{twist10commute}) of the twisted $h$ subalgebra $\hb (\s)$ of sector $\s$.  As discussed in Subsec.~\ref{Ksec}, the general WZW orbifold and the general coset orbifold provide simple examples of this decomposition.

\newsection{About Permutation Orbifolds}
\label{cwsec}
In this section, we discuss some features common to all permutation orbifolds 
\(
\frac{A(H)}{H} \sp H \subset S_N (\textup{permutation}) \subset Aut(g) \sp g=\oplus_{I=0}^{K-1}\gbn^I \sp K \leq N  \label{permutationorbifolds}
\)
where the copies $\gbn^I \cong \gbn$ in the ($H \subset S_N$) permutation-invariant CFT's are taken at level $k$.  In this case, the general twisted current algebra $\gb(\s)$ of sector $\s$ consists of a set of commuting orbifold affine algebras$^{\rf{Chris},\rf{Ban1},\rf{us2},\rf{Ban2}}$ at various orders (see also Sec.~\ref{SNsec}).  We have also checked that the orbifold adjoint operation (\ref{Jdagger}) gives the standard$^{\rf{Chris},\rf{us2}}$ adjoint operation (see Eq.~(\ref{zldagger}) and Sec.~\ref{SNsec}) for each commuting orbifold affine algebra in each sector $\s$, so that unitarity of the twisted sectors of any permutation orbifold $A(H)/H$ follows from unitarity of the permutation-invariant CFT $A(H)$. 

From the orbifold induction procedure for orbifold affine algebras$^{\rf{Chris}}$ we know that the ground state $|0\rangle_\s$ of sector $\s$ (which corresponds to the twist field of sector $\s$) is a twisted affine primary state 
\(
\hat{J}_{n(r) \mu}( \foot{(} m+\srac{n(r)}{\r(\s)} \foot{)}  \geq 0)|0\rangle_\s =0 \sp  _\s\langle0| \hat{J}_{n(r) \mu}( \foot{(} m+\srac{n(r)}{\r(\s)} \foot{)} \leq 0)=0  \label{Jvac} 
\)
where the second relation follows from the first by the orbifold adjoint operation.  It follows from (\ref{Jvac}) that $M$ ordering, defined in (\ref{MNOing}), is a true normal ordering
\group{Permstatesrelations}
\(
\langle \hat{J}_{n(r) \mu}(z)\rangle_\s=\langle :  \hat{J}_{n(r) \mu}(z) \hat{J}_{n(s) \nu}(w):_M \rangle_\s=0 \label{JJvac}
\)
\(
\sum_{p \in \sz} : \hat{J}_{n(r) \mu}(p+\srac{n(r)}{\r(\s)}) \hat{J}_{-n(r), \nu}(m-p-\srac{n(r)}{\r(\s)}):_M |0\rangle_\s=0 , \hsp{.15} \textup{for} \ m\geq 0 , \hsp{.15} \forall\ n(r), \mu ,\nu \label{modeNOA} 
\)
\reseteqn
for all permutation orbifolds.  The relations (\ref{Jvac}) and (\ref{Permstatesrelations}) hold for all $n(r)$ and $n(s)$, not necessarily in the fundamental range.  As a consequence of (\ref{JJvac}), we obtain the twisted current-current correlators in each sector $\s$ of the general permutation orbifold
%\vspace{.1in}
\(
%\hsp{-4.5}
 \langle \hat{J}_{n(r) \mu}(z) \hat{J}_{n(s) \nu}(w) \rangle_\s
=\delta_{n(r)+n(s),0 \mod \r(\s)} (\frac{w}{z})^{\frac{\bar{n}(r)}{\r(\s)}} [\frac{1}{(z-w)^2} + \frac{\bar{n}(r)/\r(\s)}{w(z-w)}] \g_{n(r) \mu;-n(r), \nu}(\s)
\)
from the exact operator product (\ref{exactJJ}).  Correlators of more than two twisted currents can be computed from (\ref{curenttomodetwistJ}), (\ref{ghatgroup}) and (\ref{Jvac}).

Using (\ref{Lmodeb}), (\ref{Jvac}) and (\ref{modeNOA}), we may also compute the ground state conformal weight $\hat{\Delta}_0(\s)$ of sector $\s$
\alpheqn
\(
L_\s (m \geq 0)|0\rangle_\s = \delta_{m,0} \hat{\Delta}_0(\s)|0\rangle_\s
\)
\ba
\hat{\Delta}_0(\s)&=&\sum_{r,s,\mu,\nu} \frac{\bar{n}(r)}{2 \r(\s)}(1-\frac{\bar{n}(r)}{\r(\s)}) \lr^{n(r) \mu; n(s) \nu }(\s) \g_{n(r) \mu;n(s) \nu}(\s) \label{cwlrs} \\
&=&\sum_{r,\mu,\nu} \frac{\bar{n}(r)}{2 \r(\s)}(1-\frac{\bar{n}(r)}{\r(\s)}) \lr^{n(r) \mu; -n(r), \nu }(\s) \g_{n(r) \mu;-n(r), \nu}(\s) \label{cwlr} \ .
\ea
\reseteqn
The ground state conformal weights are class functions and the factor $\bar{n}(r)$ in (\ref{cwlr}) tells us that twist class $\bar{n}(r)=0$ (the integral affine subalgebra) does not contribute to $\hat{\Delta}_0(\s)$.  When the CFT $A(H)$ is unitary, the ground state conformal weights $\hat{\Delta}_0(\s)$ must be real
\(
\hat{\Delta}_0(\s)^*=\hat{\Delta}_0(\s) \label{cwreal} 
\)
because the twisted sectors of these orbifolds are also unitary in this case.  To see this explicitly, use the $\g^*$ relation (\ref{sgrstar}), the $\lr^*$ relation (\ref{slrstar}), the relation (\ref{partinvphi}) for $\Ltheta$ and the symmetry of $\g$ and $\lr$.

Using the duality transformations for $\g$ and $\lr$ in Eqs.~(\ref{twistg}) and (\ref{lr}), the ground state conformal weights (\ref{cwlrs}) can also be expressed in terms of the inverse inertia tensor $L^{ab}$ of the untwisted sector\footnote{The structure $\frac{\bar{n}}{\r}(1-\frac{\bar{n}}{\r})$ was first seen in the ground state conformal weights of the free-field examples discussed by Dixon, Harvey, Vafa and Witten in Ref.~\rf{DixO2}.}
\group{projectorgroup}
\(
\hat{\Delta}_0(\s)=L^{ac}G_{cb} \sum_{r} \frac{\bar{n}(r)}{2 \r(\s)}(1-\frac{\bar{n}(r)}{\r(\s)}) \pr(\bar{n}(r);\s)_a^{\ b} \label{cwL}
\)
\(
\pr(\bar{n}(r);\s)_a^{\ b} = {\textstyle {\Big {\bf \sum}}_{\mu}} \ \uds_a^{\ n(r) \mu} U(\s)_{n(r) \mu}^{\ \ \ \ \ \ b}  \sp \w(h_\s)_a^{\ b}= {\textstyle {\Big {\bf \sum}}_{r}}  \ E_{n(r)}(\s) \pr(\bar{n}(r);\s)_a^{\ b}
\)
\(
\pr(\bar{n}(r);\s)_a^{\ c} \ \pr(\bar{n}(s);\s)_c^{\ b} =\delta_{n(r)}^{\hsp{.2} n(s)} \pr(\bar{n}(r);\s)_a^{\ b}
\)
\reseteqn
where $\pr(\bar{n}(r);\s)$ is the projector onto the $\bar{n}(r)$ subspace of sector $\s$.  Other properties of these projectors are collected in App.~\ref{projectorsapp}.  

To go further for $H\subset S_N$, we need a more explicit notation for the semisimplicity of the Lie algebra $g$ in (\ref{permutationorbifolds}) and the degeneracy indices of the $H$-eigenvalue problem:
\alpheqn
\(
a\rightarrow a,I \sp n(r),\mu \rightarrow n(r),a j \sp a=1,...,\textup{dim}\gbn \sp I=0,...,K-1 \label{indexarrow}
\)
\vspace{-.30in}
\(
G_{ab} \rightarrow G_{aI,bJ}= k \eta_{ab} \delta_{IJ} \sp f_{ab}^{\ \ c} \rightarrow f_{aI,bJ}^{\hsp{.3} cM}= f_{ab}^{\ \ c} \delta_{IJ} \delta_{J}^{\ M} 
\)
\(
J_{aI}(z) J_{bJ}(w)=\delta_{IJ} \{ \frac{k \eta_{ab}}{(z-w)^2}+\frac{f_{ab}^{\ \ c} J_{c J}(w)}{z-w} \} +\reg
\)
\(
L^{ab} \rightarrow L^{aI,bJ} \sp \r_a^{\ b} \rightarrow \r_{aI}^{\ \ \ bJ}=\r_a^{\ b} \delta_I^{\ J}  \sp \w(h_\s)_a^{\ b}\rightarrow \w (h_\s)_{aI}^{\ \ \ bJ}=\delta_a^{\ b} \w (h_\s)_{I}^{\ J}
\)
\(
U^\dagger(\s)_a^{\ n(r) \mu} \rightarrow U^\dagger(\s)_{aJ}^{\ \ \ n(r) b j}=\delta_a^{\ b} U^\dagger(\s)_J^{\ n(r) j} ,\hsp{.1} \w(h_\s)_I^{\ J} U^\dagger(\s)_J^{\ n(r) j}=U^\dagger(\s)_I^{\ n(r) j}E_{n(r)}(\s) \label{reducedevprob}
\)
\(
\pr(\bar{n}(r);\s)_a^{\ b} \rightarrow \pr(\bar{n}(r);\s)_{aI}^{\ \ \ b J}=\delta_a^{\ b} \pr(\bar{n}(r);\s)_{I}^{\ J}
\)
\(
\pr(\bar{n}(r);\s)_{I}^{\ J}  = \sum_j \uds_I^{\ n(r) j} U(\s)_{n(r) j}^{\ \ \ \ \ J} 
\)
\vspace{-.13in}
\(
\g_{n(r)aj; n(s)bl}(\s) \propto k \eta_{ab} , \hsp{.15}  \fr_{n(r)aj; n(s)bl}^{\hsp{.7} n(t)cm}(\s) \propto f_{ab}^{\ \ c} , \hsp{.15} \Ltheta_{n(r)aj}^{\hsp{.35} n(s)bl}(\s) \propto \r_a^{\ b} \label{permgf}
\)
\(
T=L^{aI,bJ}:J_{aI}J_{bJ}: \ \ \sa \ \ \hat{T}_\s = \lr^{n(r) aj; n(s) bl}(\s):\hat{J}_{n(r)aj} \hat{J}_{n(s) b l}: \ .
\)
\reseteqn
Here the $a$ indices to the right of the arrows are the Lie algebra indices of the copies $\gbn^I$, with Killing metric $\eta_{ab}$, and $\mu=(a,j)$ labels the $\bar{n}(r)$ subspace.  The reduced matrix $U^\dagger(\s)_{J}^{\ n(r)j}$ (which solves the reduced eigenvalue problem in (\ref{reducedevprob})) is also unitary, and the reduced matrix $\pr(\bar{n}(r);\s)_{I}^{\ J}$ is also a projector onto the $\bar{n}(r)$ subspace.

Then the ground state conformal weight (\ref{cwL}) takes the form
\(
\hat{\Delta}_0(\s)=k L^{aI,bJ} \eta_{ab} \sum_r  \frac{\bar{n}(r)}{2 \r(\s)}(1-\frac{\bar{n}(r)}{\r(\s)}) \pr(\bar{n}(r);\s)_{I J} \label{finalpermcw}
\)
for all $H\subset S_N$, where $P_{IJ}=P_I^{\ K} \delta_{KJ}$.  For the case of the cyclic permutation orbifolds \taz, a more explicit form of the ground state conformal weights can be obtained
\alpheqn
\(
L^{aI,bJ}=L_{I-J}^{ab} \sp \pr(r;\s)_{I J}=\frac{1}{\r(\s)}e^{\frac{2 \pi i N(\s) (J-I)r}{\l}}\delta_{I,J \mod \frac{\l}{\r(\s)}}
\)
\(
\hat{\Delta}_0(\s)=\frac{\l k\eta_{ab}}{4\r^2(\s)} \{ \frac{\r^2(\s)-1}{3}L_0^{ab}-{\displaystyle{\sum_{r=1}^{\r(\s)-1}}} csc^2(\frac{\pi N(\s)r}{\r(\s)})\hsp{.02} L^{ab}_{\frac{\l}{\r(\s)}r} \} \sp \s=1,...,\l-1 \label{zlcw} 
\)
\reseteqn
from (\ref{zlUgroup}) and (\ref{finalpermcw}), in agreement with the result given in Ref.~\rf{us2}.

Returning to the general permutation orbifold, we find that the reduced matrices $\w$ and $U$ satisfy
\group{permsumrules}
\(
\sum_I \w(h_\s)_{I}^{\ J}=1 \label{wsumrule}
\)
\(
(\sum_I U^\dagger(\s)_I^{\ n(r)j})(1-E_{n(r)}(\s))= (\sum_I U(\s)_{n(r)j}^{\ \ \ \ \ I}) (1-E_{n(r)}(\s))=0 \label{SNUsumrule}
\)
\(
(\sum_I U^\dagger(\s)_I^{\ n(r)j}) ,\hsp{.2} (\sum_I U(\s)_{n(r)j}^{\ \ \ \ \ I}) \ \textup{  and  } \sum_I \pr(\bar{n}(r);\s)_{I}^{\ J} \propto \delta_{n(r),0 \mod \r(\s)} \label{permUidentity}
\)
\reseteqn
for all $H \subset S_N$.  Here (\ref{wsumrule}) follows because $\w(h_\s)_{IJ}$ is a permutation matrix, while the selection rules (\ref{SNUsumrule}) follow from (\ref{wsumrule}) and the reduced eigenvalue problem.

The relations (\ref{finalpermcw}) and (\ref{permUidentity}) give a simple form for the ground state conformal weights of the permutation orbifolds $A(S_N)/S_N$
\alpheqn
\(
L^{aI,bJ}=\l^{ab} \delta^{IJ}+l^{ab}
\sp \sum_I \pr(\bar{n}(r);\s)_I^{\ I}=\textup{dim} [\bar{n}(r)] \label{SNL}
\)
\(
\hat{\Delta}_0(\s)=k \eta_{ab} \lambda^{ab} \sum_r \frac{\bar{n}(r)}{2 \r(\s)}(1-\frac{\bar{n}(r)}{\r(\s)}) \ \textup{dim} [\bar{n}(r)] \label{genpermcw}
\)
\reseteqn
where $\textup{dim} [\bar{n}(r)]$ is the dimension of the $\bar{n}(r)$ subspace.  This simplification depends on the form of the inverse inertia tensor in (\ref{SNL}), which describes the general $S_N$-invariant CFT.  A more explicit form of these ground state conformal weights will be obtained in Sec.~\ref{SNsec}.

The relations (\ref{permsumrules}) also give a simple form for the ground state conformal weights of the general permutation copy orbifold\footnote{For the case $H=\zl$(permutation), results equivalent to (\ref{permcopyorbifold}) were given in Refs.~\rf{us2} and \rf{us3}.}
\group{permcopyorbifold}
\(
\frac{\times_{I=0}^{K-1}A_I}{H} \subset \ahh \sp H\subset S_N(\textup{permutation})
\)
\(
L^{aI,bJ}=\l^{ab} \delta^{IJ} 
\)
\(
\hat{\Delta}_0(\s)=\frac{c}{2} \sum_r \frac{\bar{n}(r)}{2 \r(\s)}(1-\frac{\bar{n}(r)}{\r(\s)}) \ \textup{dim} [\bar{n}(r)]
\sp c=2 k \eta_{ab}\l^{ab}
\)
\reseteqn
where $A_I \cong A$ are $K$ copies (permuted by $H\subset S_N$) of any affine-Virasoro construction $A$ with central charge $c$.  Included in this set of copy orbifolds is the general WZW permutation orbifold (see also Subsec.~\ref{Ksec}), whose twisted inverse inertia tensors are given by
\(
\lr_{\sgb(\s)}^{n(r)aj; n(s)bl}(\s)=\frac{x_{\sgbn}}{2(x_{\sgbn}+\tilde{h}_{\sgbn})} \g^{n(r)aj; n(s)bl}(\s)  \sp \g^{n(r)aj; n(s)bl}(\s) \propto \frac{\eta^{ab}}{k} \sp \srange \ .
\)
Here $N_c$ is the number of conjugacy classes of $H\subset S_N$, $\tilde{h}_{\sgbn}$ is the dual Coxeter number of each copy $\gbn$ and $x_{\sgbn}$ is the invariant level of affine $\gbn$.  The twisted tensor $\g$ with all indices up is the inverse of the twisted metric $\g_{n(r)aj; n(s)bl}(\s)$ in (\ref{permgf}).

For both $A(S_N)/S_N$ and the general permutation copy orbifold (\ref{permcopyorbifold}) we find an $a\lra b$ symmetry of $\lr$
\alpheqn
\(
\lr^{n(r)aj; n(s)bl}(\s)=\lr^{n(r)bj; n(s)al}(\s) 
\)
\(
\hat{T}_\s(z)=\lr^{n(r)aj; -n(r),bl}(\s):\hat{J}_{n(r)aj}(z) \hat{J}_{-n(r),bl}(z):_M +\frac{1}{z^2}\hat{\Delta}_0(\s)
\)
\reseteqn
so that no linear terms in the currents appear in the $M$-ordered form of the stress tensors.  For $\zl$-(permutation) copy orbifolds, many examples of this phenomenon were seen in Ref.~\rf{Chris}.

\newsection{The Permutation Orbifolds $A(S_N)/S_N$}
\label{SNsec}
\subsection{The permutation group $S_N$}
As a large example beyond \taz, we will work out the permutation orbifolds 
\(
\frac{A(S_N)}{S_N} \sp S_N(\textup{permutation})\subset Aut(g)
\)
in further detail, where $A(S_N)$ is any $S_N$(permutation)-invariant affine-Virasoro construction.  The case of $A(S_3)/S_3$ was worked out previously in Ref.~\rf{us2}.  In the $S_N$-invariant CFT's, we have $N$ copies $\gbn^I$ of an affine Lie algebra on simple $\gbn$ with currents $J_{aI}$
\group{SNalgebra}
\(
g=\oplus_{I=0}^{N-1}\gbn^I\sp
\gbn^{I}\cong\gbn
\)
\(
[J_{aI}(m),J_{bJ}(n)]=\delta_{IJ} \{ if_{ab}^{\ \ c}J_{cI}(m+n)+mk\e_{ab}\delta_{m+n,0} \}
\)
\(
J_{aI}\ \p=\w_I^{\ J}J_{aJ}\sp
\w \in S_N\textup{(permutation)} \label{SNauto}
\)
\(
a,b=1,...,\textup{dim}\gbn\sp I,J=0,...,N-1\sp m,n\in\z
\)
\reseteqn
and the action of $S_N$(permutation) on the currents is given in (\ref{SNauto}).

In the cyclic notation for the elements of $S_N$, the action of a cycle of length $l$
\(
(I_{\hj=0}...I_{\hj=l-1}): \hsp{.3}
I_{\hj}=0,...,N-1 \sp 
\hj=0,...,l-1 \sp
I_{\hj}\neq I_{\hi}\textup{ when }\hj\neq\hi
\)
is a cyclic permutation of the copies of $\gbn$ according to 
\(
\gbn^{I_{\hj}}\ \rightarrow\ \gbn^{I_{\hj+1}}\sp \hj=0,...,l-2 ; \hsp{.3}
\gbn^{I_{l-1}}\ \rightarrow\ \gbn^{I_{0}}.
\)
All elements of $S_N$ are products of disjoint cycles, and the general element can be written as a product of elements of disjoint cyclic groups $\z_{\s_j}$
\(
\prod_{j=0}^{n-1}(I_{\hj=0}^j...I_{\hj=\s_j-1}^j)\sp
I_{\hj}^j\neq I_{\hi}^{i}\textup{ when }j\neq i
\label{cycnot}
\)
where $\s_j$ is the length of the $j$th disjoint cycle, $\hj$ counts within a cycle and $n$ (the number of disjoint cycles) is some positive integer.  

A property of each element of $S_N$ is its cycle type $\sv$, which is the collection of lengths of cycles written in order of decreasing length
\(
\sv\equiv\{\s_0,...,\s_{n-1}\} \sp \s_{j+1}\leq\s_j \sp \sum_{j=0}^{n(\sv)-1}\s_j=N \label{cycletype} \ .\label{solveme}
\)
The cycle types $\sv$ are the $\s_{j+1}\leq\s_j$ partitions of $N$ for any $n=n(\sv)$, and $\sv$ also labels the conjugacy classes of $S_N$.  Conversely, the elements of conjugacy class $\sv$ are obtained by relaxing the restriction on the ordering of the lengths in $\sv$.  As an example, the table below 
\(
\begin{tabular}{r|l} 
Conj.Class $\sv$&Elements\\ \hline
$\{\s_0,\s_1,\s_2\}$=\{1,1,1\}&(0)(1)(2)\\ 
$\{\s_0,\s_1\}$=\{2,1\}&(01)(2),(02)(1),(12)(0)\\ 
$\{\s_0\}$=\{3\}&(012),(021)\\ 
\end{tabular} 
\)
\\
shows the conjugacy classes of $S_3$.

Modeling our choice on the cycle types, we choose one representative from each conjugacy class $\sv$ by determining $\{ I_{\hj}^j \}$ as follows
\group{rep}
\(
j,\hj\rightarrow \{ I_{\hj}^j \}: \hsp{.2}I_{\hj}^j\ \equiv \sum_{k=0}^{j-1}\s_k+\hj \sp \s_{j+1}\leq\s_j \sp \sum_{j=0}^{n(\sv)-1}\s_j=N
\)
\(
j=0,...,n(\sv)-1\sp\hj=0,...,\s_j-1 \label{jhj}
\)
\reseteqn
given $\sv$ and the ranges of $j,\hj$ in (\ref{jhj}).  Here $j$ labels the disjoint cycles and $\hj$ labels the integers in each disjoint cycle.  For example, the chosen representatives for the elements of $S_3$ and $S_4$ are
\(
\begin{tabular}{r|l} 
$S_3$: Conj. Class $\sv$&Representative\\ \hline
\{1,1,1\}&(0)(1)(2)=$(I_0^0)(I_0^1)(I_0^2)$\\ 
\{2,1\}&(01)(2)=$(I_0^0I_1^0)(I_0^1)$\\ 
\{3\}&(012)=$(I_0^0I_1^0I_2^0)$\\ 
\end{tabular}
\label{s3reps}
\hsp{.1}
\begin{tabular}{r|l} 
$S_4$: Conj. Class $\sv$&Representative\\ \hline
\{1,1,1,1\}&(0)(1)(2)(3)\\ 
\{2,1,1\}&(01)(2)(3)\\ 
\{2,2\}&(01)(23)\\ 
\{3,1\}&(012)(3)\\ 
\{4\}&(0123).\\ 
\end{tabular}
\label{s4}
\)
\\
We will refer to the representative of conjugacy class $\sv$ as $h_{\sv}\in S_N$.

\subsection{Automorphisms and the twisted currents}
\label{SNcurrentssec}
To construct the action $\w$ of the automorphism $h_{\sv}$ on the currents, we begin by relabeling the copies of $\gbn$
\alpheqn
\(
\gbn^I\rightarrow\gbn^{j\hj}\sp a,I\rightarrow a,j\hj\sp
J_{aI}\rightarrow J_{aj\hj}
\)
\(
\sum_{k=0}^{j-1}\s_k+\hj=I\sp
j\in\{0,...,n(\sv)-1\}\sp\hj\in\{0,...,\s_j-1\} \label{Idecomposition}
\)
\reseteqn
where $I$ is the semisimplicity index in (\ref{SNalgebra}), $j$ and $\hj$ are the unique solutions to (\ref{Idecomposition}) given $I$, and $\sv$ solves Eq.~(\ref{rep}).  This relabeling (which is the inverse of (\ref{rep})) corresponds to the pictorial representation in Fig.~2 of the action of $h_{\sv}$ on the copies 

\begin{picture}(350,105)(0,0)
\put(100,45){\line(0,1){30}}
\put(121,80){\sm{j=0} }
\put(160,45){\line(0,1){30}}
\put(175,80){\sm{j=1} }
\put(210,45){\line(0,1){30}}
\put(230,66){.    .     .    .    .}
\put(285,45){\line(0,1){30}}
\put(282,80){\sm{j=n(\sv)-1} }
\put(325,45){\line(0,1){30}}
\put(100,45){\line(1,0){225}}
\put(90,34){\sm{I=0} }
\put(157,34){\sm{\s_0} }
\put(196,34){\sm{\s_0+\s_1} }
\put(310,34){\sm{I=N-1} }
\put(105,55){\line(1,0){50}}
\put(105,50){\line(1,0){50}}
\put(105,55){\vector(1,0){28.2}}
%\put(155,50){\vector(-1,0){28.2}}
\put(105,50){\line(0,1){5}}
\put(155,50){\line(0,1){5}}

\put(165,55){\line(1,0){40}}
\put(165,50){\line(1,0){40}}
\put(165,55){\vector(1,0){23}}
%\put(205,50){\vector(-1,0){23}}
\put(165,50){\line(0,1){5}}
\put(205,50){\line(0,1){5}}

\put(290,55){\line(1,0){30}}
\put(290,50){\line(1,0){30}}
\put(290,55){\vector(1,0){18}}
%\put(320,50){\vector(-1,0){18}}
\put(290,50){\line(0,1){5}}
\put(320,50){\line(0,1){5}}

\put(113,10){Fig. 2.  The action of $h_{\sv}$ on the copies.}
\end{picture}\\
where $\hj$ labels the integers which are cyclically permuted $\hj\rightarrow\hj+1$ inside each box (disjoint cycle) $j$.  It follows that the action $\w(\sv)\equiv \w(h_{\sv})$ on the currents is
\alpheqn
\(
\w\ \rightarrow\ \w_{aI}^{\ \ \ bJ}(\sv) = \w_{aj\hj}^{\ \ \ bl\hl}(\sv)=\delta_a^{\ b}\w_{j\hj}^{\ \ l\hl}(\sv)
\)
\(
J_{aj\hj}\ \p=\sum_{l,\hl}\w_{j\hj}^{\ \ l\hl}(\sv)J_{al\hl}\sp
\w_{j\hj}^{\ \ l\hl}(\sv) = \delta_{jl}\delta_{\hj+1,\hl\mod\s_j}
\)
\reseteqn
where the block-diagonal matrix $\w_{j\hj}^{\ \ l\hl}(\sv)$ is orthogonal.  In what follows, all quantities have the spectral index periodicity $\hj \rightarrow \hj +\s_j$.

The next step is to solve the $S_N$-eigenvalue problem
\alpheqn
\(
n(r),\mu \ \ \rightarrow  \ \ \hj, aj \sp U^\dagger \rightarrow\ U^\dagger(\sv)_{\hj aj}^{\ \ \ \hl bl}=\delta_a^{\ b} U^\dagger(\sv)_{\hj j}^{\ \ \hl l}
\)
\(
\sum_{\hl,l}\w_{\hat{m} m}^{\hsp{.2} \hl l}(\sv) U^\dagger(\sv)_{\hl l}^{\ \ \hj j}
=U^\dagger (\sv)_{\hat{m} m}^{\hsp{.2} \hj j}E_{\hj}^j(\sv)
\)
\reseteqn
for the block-diagonal and unitary matrix $U^\dagger(\sv)$ and the eigenvalues $E(\sv)$:
\alpheqn
\(
U^\dagger (\sv)_{\hj j}^{\ \ \hl l}=\frac{\delta_{jl}}{\sqrt{\s_j}}e^{-\frac{\tp\hj\hl}{\s_j}}\sp 
U(\sv)_{\hj j}^{\ \ \hl l}=\frac{\delta_{jl}}{\sqrt{\s_j}}e^{\frac{\tp\hj\hl}{\s_j}}\sp 
E_{\hat{j}}^j(\sv)=e^{-\frac{\tp\hj}{\s_j}}
\)
\(
\sum_{\hm,m} U(\sv)_{\hj j}^{\ \ \hm m \hsp{.05} *} \ U^\dagger(\sv)_{\hm m}^{\hsp{.25} \hl l}= \delta_{jl}\delta_{\hj+\hl,0\mod\s_j} \sp \sum_{\hj,j} U^\dagger(\sv)_{\hj j}^{\ \ \hl l}=\sqrt{\s_l}\delta_{\hl,0\mod\s_l} \label{johnsident} \ .
\)
\reseteqn
The second identity in Eq.~(\ref{johnsident}) is the form taken by Eq.~(\ref{permUidentity}) in this case.

With the choice ${\foot{\chi_{\hj aj}(\sv)}}=\sqrt{\s_j}$ we find the twisted current system of sector $\sv$ of the orbifold $A(S_N)/S_N$
\alpheqn
\(
\hat{J}_{n(r)\mu} \ \rightarrow \ \hat{J}_{aj}^{(\hj)} \sp \g \ \rightarrow \ \g_{\hj aj;\hl bl}(\sv)= \delta_{jl}\s_j k\e_{ab} \delta_{\hj+\hl,0\mod\s_j}
\)
\(
\fr \ \rightarrow \ \fr_{\hj aj;\hl bl}^{\ \ \ \ \ \ \ \hm cm}(\sv)=f_{ab}^{\ \ c} \delta_{jl}\delta_l^{\ m}\delta_{\hj+\hl,\hm\mod\s_j}
\)
\(
\hat{J}_{aj}^{(\hj)}(z)\hat{J}_{bl}^{(\hl)}(w)=\delta_{jl} \{ \frac{(\s_j k) \e_{ab}\delta_{\hj+\hl,0\mod\s_j}}{(z-w)^2}+\frac{if_{ab}^{\ \ c}\hat{J}_{cj}^{(\hj+\hl)}(w)}{z-w} \} +\reg
\)
\(
\hat{J}_{aj}^{(\hj)}(ze^{\tp})=e^{-\frac{\tp\hj}{\s_j}}\hat{J}_{aj}^{(\hj)}(z) \sp \hat{J}_{aj}^{(\hj \pm \s_j)}(z)=\hat{J}_{aj}^{(\hj)}(z)   \label{SNmonodromies}
\)
\(
\# \{ \hat{J} \} = \textup{dim} \gbn (\sum_{j=0}^{n(\sv)-1}\s_j) =N \textup{dim} \gbn =\textup{dim}g= \# \{ J \} 
\)
\(
a,b=1,...,\textup{dim} \gbn \sp
j,l=0,...,n(\sv)-1 \sp
\bar{\hj}=0,...,\s_j-1\sp \bar{\hl}=0,...,\s_l-1 
\)
\reseteqn
\\
where $\bar{\hj}=\hj-\s_j \lfloor \hj/\s_j \rfloor$ evaluates $\hj$ in its fundamental range.  According to (\ref{SNmonodromies}), the fraction 
\(
\frac{n(r)}{\r(\s)}=\frac{\hj}{\s_j} \label{njrelation}
\)
controls the monodromies of the twisted current $\hat{J}_{aj}^{(\hj)}$.

The modes of the twisted currents follow from the monodromies 
\alpheqn
\(
\hat{J}_{aj}^{(\hj)}(z)=\sum_{m\in\sz}\hat{J}_{aj}^{(\hj)}(m+\srac{\hj}{\s_j})z^{-(m + \frac{\hj}{\s_j})-1}
\)
\(
\hat{J}_{aj}^{(\hj+\s_j)}(m-1+\srac{\hj+\s_j}{\s_j})=\hat{J}_{aj}^{(\hj)}(m+\srac{\hj}{\s_j}) \label{SnhatJperiodicity}
\)
\(
\hat{J}_{aj}^{(\hj)}(m+\srac{\hj}{\s_j})|0\rangle_{\sv}=0\textup{\ \ when\ \ }(m+\srac{\hj}{\s_j})\geq 0
\)
\reseteqn
where the periodicity and ground state condition are special cases of Eqs.~(\ref{Jmodeperiod}) and (\ref{Jvac}) respectively.

This leads to the twisted current algebra $\gb(\sv)$ of sector $\sv$
\alpheqn
\(
\gb(\sv)=\gb(S_N(\textup{permutation})\subset Aut(g);\sv)=\oplus_{j=0}^{n(\sv)-1}\gb_{\s_j}
\)
\(
[\hat{J}_{aj}^{(\hj)}(m\+\srac{\hj}{\s_j}),\hat{J}_{bl}^{(\hl)}(n\+\srac{\hl}{\s_l})] = \delta_{jl} \{ if_{ab}^{\ \ c}\hat{J}_{cj}^{(\hj+\hl)}(m+n+\srac{\hj+\hl}{\s_j})+ (\s_jk) \e_{ab}(m\+\srac{\hj}{\s_j})\delta_{m+n+\frac{\hj+\hl}{\s_j},0} \}
\)
\(
a,b=1,...,\textup{dim} \gbn \sp
j,l=0,...,n(\sv)-1 \sp
\bar{\hj}=0,...,\s_j-1\sp\bar{\hl}=0,...,\s_l-1 
\)
\reseteqn
which is a special case of the general twisted current algebra in (\ref{ghatgroup}).  This algebra consists of a product of $n(\sv)$ commuting sets of orbifold affine algebras$^{\rf{Chris}}$ $\gb_{\s_j}$ of order $\s_j$, where the factor $\gb_{\s_j}$ has orbifold affine level $\hat{k}_j=\s_jk$ and $k$ is the level of the untwisted algebra.  According to the identification (\ref{njrelation}), the quantity $\bar{\hj}=\hj-\s_j \lfloor \hj/\s_j \rfloor$ is the effective twist class of $\hat{J}_{aj}^{(\hj)}$, relative to the order $\s_j$ of $\gb_{\s_j}$.

For this case, the orbifold adjoint operation (\ref{Jdagger}) takes the form
\(
\Ltheta_{\hj aj}^{\hsp{.2} \hl bl}(\sv)=\delta_{jl} \r_a^{\ b}  \delta_{\hj+\hl,0\mod\s_j} \ \ \rightarrow \ \ \hat{J}_{aj}^{(\hj)}(m+\srac{\hj}{\s_j})^\dag
=\r_a^{\ b}\hat{J}_{bj}^{(-\hj)}(-m-\srac{\hj}{\s_j}) \label{SNdagger} \ .
\)
This is the standard adjoint operation of orbifold affine algebra,$^{\rf{Chris},\rf{us2}}$ which guarantees unitarity of the twisted affine Hilbert space given unitarity of the untwisted affine Hilbert space.  As noted above, unitarity of the permutation orbifolds $A(S_N)/S_N$ then follows from unitarity of the CFT $A(S_N)$.

\subsection{The stress tensor of sector $\sv$}
The stress tensors of the $S_N$-invariant CFT's $A(S_N)$ are
\group{untwistsec}
\(
T=L^{aJ,bL}\sum_{J,L=0}^{N-1}:J_{aJ}J_{bL}:\sp
L^{aJ,bL}=\l^{ab} \delta^{JL}+l^{ab} \label{lform}
\)
\(
\l^{ab}=\l^{ba}\sp l^{ab}=l^{ba} \sp c=2N k\e_{ab}(\l^{ab}+l^{ab})
\)
\reseteqn
where the inverse inertia tensors $\l^{ab}$ and $l^{ab}$ satisfy the reduced Virasoro master equation 
\alpheqn
\(
\l^{ab}=2N\l^{ac}\e_{cd}\l^{db}-\l^{cd}(\l^{ef}+2l^{ef})f_{ce}^{\ \ a}f_{df}^{\ \ b}-(\l^{cd}+l^{cd})f_{ce}^{\ \ f}f_{df}^{\ \ (a}\l^{b)e}
\)
\(
l^{ab}=4kl^{ac}\e_{cd}(\l^{db}+l^{db})+2k(N-2)l^{ac}\e_{cd}\l^{db}-l^{cd}l^{ef}f_{ce}^{\ \ a}f_{df}^{\ \ b}-(\l^{cd}+l^{cd})f_{ce}^{\ \ f}f_{df}^{\ \ (a}l^{b)e}.
\)
\reseteqn
This is a consistent set of $\textup{dim} \gbn (\textup{dim} \gbn +1)$ quadratic equations for the same number of unknowns and so the generically-expected number of inequivalent solutions at each level is 
\(
M(\gbn,N)=2^{(\textup{\scriptsize{dim}} \sgbn)^2} \ .
\)
Remarkably, $M(\gbn,N)$ is independent of $N$.

Then the  stress tensor of sector $\sv$ of $A(S_N)/S_N$
\group{Tsv}
\(
\hat{T}_{\sv}=\sum_{j,l=0}^{n(\sv)-1}\ \sum_{\hj=0}^{\sgcd\hsp{-.03}(\s_j,\s_l)-1} \lr^{\hj aj ;-\hj, bl}(\sv):\jh_{aj}^{(\frac{\hj\s_j}{\sgcd\hsp{-.03}(\s_j,\s_l)})}\jh_{bl}^{(-\frac{\hj\s_l}{\sgcd\hsp{-.03}(\s_j,\s_l)})}: \label{gcdform}
\)
\(
\lr \rightarrow \lr^{\hj aj;-\hj, bl}(\sv)=\l^{ab} \frac{\delta^{jl}}{\s_j}+ l^{ab} \delta_{\hj,0} \label{larrow}
\)
\reseteqn
follows as a special case of the general Virasoro construction (\ref{twistTrecapgroup}).  Here $z=$gcd($x,y$) is the greatest integer such that $x/z$ and $y/z$ are also integers.  Combining (\ref{gcdform}) and (\ref{larrow}), we find a more transparent form of these stress tensors
\group{SNrecapgroup}
\(
\hat{T}_{\sv}=\l^{ab}\sum_{j=0}^{n(\sv)-1}\frac{1}{\s_j}\sum_{\hj=0}^{\s_j-1}:\jh_{aj}^{(\hj)}\jh_{bj}^{(-\hj)}:+l^{ab}\sum_{j,l=0}^{n(\sv)-1}:\jh_{aj}^{(0)}\jh_{bl}^{(0)}:
\)
\[
L_{\sv}(m)=\sum_{p\in \sz} \{ \l^{ab}\sum_{j=0}^{n(\sv)-1}\frac{1}{\s_j}\sum_{\hj=0}^{\s_j-1}:\jh_{aj}^{(\hj)}(p+\srac{\hj}{\s_j})\jh_{bj}^{(-\hj)}(m-p-\srac{\hj}{\s_j}):
\]
\(
+l^{ab}\sum_{j,l=0}^{n(\sv)-1}:\jh_{aj}^{(0)}(p)\jh_{bl}^{(0)}(m-p): \}
\)
\(
\hat{c}(\sv)=c=2N k\e_{ab}(\l^{ab}+l^{ab})
\)
\(
\hat{\Delta}_0(\sv)=\frac{k\e_{ab}\l^{ab}}{12}\sum_{j=0}^{n(\sv)-1}(\s_j-\frac{1}{\s_j})=\frac{k\e_{ab}\l^{ab}}{12}(N-\sum_{j=0}^{n(\sv)-1}\frac{1}{\s_j})
\ . \label{cw}
\)
\reseteqn
\\
The ground state conformal weight of sector $\sv$ is given in (\ref{cw}).

The result (\ref{cw}) for the ground state conformal weights of $A(S_N)/S_N$ is in agreement with our earlier result (\ref{genpermcw}).  To see this, start with (\ref{genpermcw}) and follow the steps  
\ba
\hat{\Delta}_0(\s)&=&k\e_{ab}\l^{ab}\sum_r\frac{\bar{n}(r)}{2\r(\s)}(1-\frac{\bar{n}(r)}{\r(\s)})\dim[\bar{n}(r)] = \ \ k\e_{ab}\l^{ab}\sum_{r,\mu(r)}\frac{\bar{n}(r)}{2\r(\s)}(1-\frac{\bar{n}(r)}{\r(\s)})\nonumber \\
& = & k\e_{ab}\l^{ab}\sum_{j=0}^{n(\sv)-1}\sum_{\hj=0}^{\s_j-1}\frac{\hj}{2\s_j}(1-\frac{\hj}{\s_j}) = \ \ \frac{k\e_{ab}\l^{ab}}{12}\sum_{j=0}^{n(\sv)-1}(\s_j-\frac{1}{\s_j}) \ .
\ea
Here we have used Eq.~(\ref{njrelation}) in the fundamental range (where $\bar{\hj}=\hj$) and the identities
\(
\sum_{\mu(r)}\ \ =\ \dim[\bar{n}(r)] \sp \sum_{r,\mu(r)}=\sum_{j=0}^{n(\sv)-1}\sum_{\hj=0}^{\s_j-1}
\)
to convert from the notation of Sec.~\ref{cwsec} to the present notation.  As examples, the cases $A(S_3)/S_3$ and $A(S_4)/S_4$ are further discussed in App.~\ref{SNapp}. 

\newsection{The Inner-Automorphic Orbifolds $A(H(d))/H(d)$}
\label{innersec}
Our next large example is the set of all\footnote{The special case of inner-automorphic WZW orbifolds has been considered in Refs.~\rf{KP}, \rf{Bouw1}, \rf{FH}, \rf{Bouw2}, \rf{KT} and \rf{Birke}, and inner-automorphic coset orbifolds were also considered in Ref.~\rf{FH}.} inner-automorphic orbifolds $A(H(d))/H(d)$, where $A(H(d))$ is any inner automorphic invariant CFT.  The story of these orbifolds is particularly interesting, not least because of their overlap with the orbifolds $\talieH$.  Indeed, we will argue that this overlap contains almost all the inner-automorphic orbifolds which can be equivalently described by stress-tensor spectral flow$^{\rf{BH},\rf{FH},\rf{vme},\rf{rev}}$ whereas the generic inner-automorphic orbifold apparently can not be described in this way. 

\subsection{Inner Automorphisms of simple $g$}
\label{autoinnersec}
We begin in the Cartan-Weyl basis of the general affine algebra on simple $g$
\group{affalg}
\(
G_{ab}=k \eta_{ab} \sp \eta_{ab}=\stheta_a^{\ b}= \pmatrix{\delta_{AB}& 0 \cr  0& \delta_{\a+\beta,0}}
\)
\(
H_{A}(z) H_{B}(w) = \frac{k \delta_{AB}}{(z-w)^2} + \reg
\sp
H_{A}(z) E_{\a }(w) =\frac{\a_A E_{\a}(w)}{z-w} + \reg
\)
\(
E_{\alpha}(z) E_{\beta}(w)
=\left\{\begin{array}{lll}
\frac{N_\gamma(\alpha,\beta)E_{\gamma}(w)}{z-w} + \reg && \textrm{if}\  \alpha+\beta=\gamma\\
\frac{k}{(z-w)^2} +\frac{\alpha\cdot H (w)}{z-w}+  \reg && \textrm{if}\ \alpha+\beta=0\\
\reg && \textrm{otherwise}\\
\end{array}\right.
\)
\(
a=(A,\a) \sp A,B=1,...,{\textrm{rank}} g  \sp \alpha,\beta,\gamma\in\Delta(g) \ .
\)
\reseteqn
In this basis, the action of the general group $H(d)$ of inner automorphisms has the form
\group{innerautoactiongroup}
\(
H(d)\subset \textup{Lie} \hsp{.025} G \subset Aut(g)
\)
\(
H_A(z)\p=H_A(z) \sp
E_\a(z)\p=e^{2 \pi i \sigma \a \cdot d}E_\a(z) \label{HEprime}
\)
\(
\s=0,...,\r(1)-1 \sp N_c=\r(1)
\)
\reseteqn
where $\textup{Lie} \hsp{.025} G$ is (the action in the adjoint of) the Lie group whose algebra is $\textup{Lie} \hsp{.025} g$ and $\r(1)=\r(\s=1)$ is the order of the $\s=1$ element of $H(d)$.  The general form\footnote{Further discussion of the allowed vectors $d$ is given in Ref.~\rf{Bouw1}.  More general forms of the vector $d$ describe automorphism groups of infinite order.  Our results below are well defined in this case, leading at least formally to orbifolds with an infinite number of sectors $\s=0,1,...\infty$.} of the vector $d$ 
\alpheqn
\(
d=\frac{2}{N}\sum_{i=1}^{\textup{\scriptsize{rank}} g} \frac{q_i \l_i}{\a_i^2} \sp N\in \z^+ \sp  q_i \in \z^+  \label{innerd}
\)
\(
\a_i \cdot \l_j=\frac{\a_i^2}{2} \delta_{ij} \sp \textup{gcd}(N, q_1,...,q_{\textup{\scriptsize{rank}} g} )=1
\)
\reseteqn
gives $\r(1)=N$.  Here $\{ \a_i \}$ and $\{ \l_i \}$ are the simple roots and weights of $g$, and $z=\textup{gcd}(\{ x_n \} )$ is the greatest integer such that $(x_n/z) \in \z^+$ for all n. Note that $d$ is inversely proportional to the length of the highest root. For $\s \geq 2$ one finds that $\r(\s)=\r(1)/\textup{gcd}(\s,\r(1))$, which holds for all cyclic groups.   As an example, the grade automorphism is discussed in App.~\ref{gradeapp}.  

We also need the $H$-invariance condition (\ref{Hinv}) for the inverse inertia tensor in this case:
\group{Hinvinnergroup}
\(
L^{A \a}(1-e^{2 \pi i \sigma \a \cdot d})=0 \sp
L^{\a \beta}(1-e^{2 \pi i \sigma (\a+\beta) \cdot d})=0 \sp
\textup{ no restriction on } L^{AB} \label{innerLrestrictions}
\)
\(
\s=0,...,\r(1)-1 \ .
\)
\reseteqn
This gives the stress tensors of the inner automorphic invariant CFT's $A(H(d))$ 
\group{innerTgroup}
\(
T=L^{ab}:J_a J_b: \sp L^{ab}=L^{ba}
\)
\(
L^{A\a}=0 \textup{  unless } \a \cdot d \in \z \sp L^{\a \beta}=0 \textup{  unless } (\a + \beta) \cdot d \in \z \label{innerLsolutions}
\)
\reseteqn
where (\ref{innerLsolutions}) is the solution of (\ref{innerLrestrictions}).

As a simple example, the affine-Sugawara construction$^{\rf{BH},\rf{Halp}-\rf{Segal},\rf{rev}}$ on $g$ 
\(
T_g=L^{ab}_g:J_a J_b: \sp
L_g^{AB}=\frac{\delta^{AB}}{2k+Q_\psi} \sp L_g^{\a \beta}=\frac{\delta_{\a+\beta,0}}{2k+Q_\psi}
\sp \sum_\a \a_A \a_B = Q_\psi \delta_{AB} \label{sumroots}
\)
describes an $H(d)$-invariant CFT for any $H(d)$.  When we have in mind a particular $H(d)$, we say that the affine-Sugawara construction describes the $H(d)$-invariant CFT $A_g(H(d))$. 

\subsection{Inner-automorphically twisted currents}
\label{innercurrentssec}
The action of the automorphism group $H(d)$ in (\ref{innerautoactiongroup}) is already diagonal, so the eigenvalues, spectral indices $n(r)$ and twist classes $\bar{n}(r)$ read
\alpheqn
\(
E_{n(r)}=e^{-\frac{2 \pi i n(r)}{\r(\s)}}=e^{2 \pi i \sigma \a \cdot d} \sp \s=0,...,\r(1)-1
\)
\(
n(r) \rightarrow \ \ n_\a= -\r(\s)\sigma \alpha \cdot d, \hsp{.15} n_A=0  \label{innern}
\)
\(
\bar{n}_\a = -\r(\s)(\sigma \alpha \cdot d + \lfloor - \sigma \alpha \cdot d \rfloor ) \sp \bar{n}_A=0 \label{nforinners} 
\)
\reseteqn
where (\ref{nforinners}) is obtained from (\ref{nnbar}).  By the same token, we may choose
\(
U(\s)=U^\dagger(\s)=1 \sp {\foot{\chi(\s)}}=1 \ \ \rightarrow \ \ \j_a(\s)=J_a \sp a=1,...,\textup{dim}g
\)
so that the eigencurrents $\j$ are the untwisted currents $J$.  It follows that the twisted tensors of the general inner-automorphic orbifold $A(H(d))/H(d)$ are identical to the tensors of the untwisted sector
\(
\fr_{ab}^{\ \ c}(\s)= f_{ab}^{\ \ c} \sp \g_{ab}(\s)= G_{ab} \sp \lr^{ab}(\s)= L^{ab} \sp \Ltheta_a^{\ b}(\s)= \r_a^{\ b} \ . \label{innertwistedtensors}
\)
Then, the $\lr$-selection rule (\ref{leel}) is nothing but the $H$-invariance of $L$ in (\ref{Hinvinnergroup}), and the $\g$- and $\fr$-selection rules in (\ref{dualg}) and (\ref{dualf}) are automatically satisfied by the metric and structure constants in the Cartan-Weyl basis.

Then Eq.~(\ref{fulltwistedopegroup}) gives the inner-automorphically twisted current system of sector $\s$ 
\group{twistaffOPE}
\(
\hat{H}_{A}(z) \hat{H}_{B}(w) = \frac{k \delta_{AB}}{(z-w)^2} + \reg 
\sp \hat{H}_{A}(z) \hat{E}_{\a }(w) =\frac{\a_A \hat{E}_{\a}(w)}{z-w} + \reg \label{twistinnerHHOPE}
\)
\getletter{twistinnerEEOPE}
\vspace{-.2in}
\(
\hat{E}_{\alpha}(z) \hat{E}_{\beta}(w)
=\left\{\begin{array}{lll}
\frac{N_\gamma(\alpha,\beta) \hat{E}_{\gamma}(w)}{z-w} + \reg && \textrm{if}\  \alpha+\beta=\gamma\\
\frac{k}{(z-w)^2} +\frac{\alpha\cdot \hat{H} (w)}{z-w}+  \reg && \textrm{if}\ \alpha+\beta=0\\
\reg && \textrm{otherwise}\\
\end{array}\right.
\)
\(
\hat{E}_\a(z e^{2 \pi i})=e^{2 \pi i \sigma \a \cdot d} \hat{E}_\a(z)
\sp \hat{H}_A(z e^{2 \pi i}) = \hat{H}_A (z) \label{innermonodromies}
\)
\(
A,B=1,...,{\textrm{rank}}g\sp
\alpha,\beta,\gamma\in\Delta(g) 
\)
\reseteqn
of each orbifold $A(H(d))/H(d)$.  Here we have suppressed our usual labeling by the spectral indices of the twisted currents, but the monodromies (\ref{innermonodromies}) are recorded in the modes 
\(
\hat{H}_A(z)=\sum_{m \in \sz} \hat{H}_A(m) z^{-m-1} \sp 
\hat{E}_\a (z)=\sum_{m \in \sz} \hat{E}_\a (m-\sigma \alpha \cdot d) z^{-(m - \sigma \alpha \cdot d)-1} \label{innermodeexpansion} \ .
\)
These expansions lead to the twisted current algebra $\gb(\s)=\gb(H(d)\subset Aut(g);\s)$
\group{twistaffalg}
\(
[\hat{H}_{A}(m), \hat{H}_{B}(n)] = km \delta_{AB} \delta_{m+n,0}
\sp [\hat{H}_{A}(m), \hat{E}_{\a }(n-\s \a \cdot d)] =\a_A \hat{E}_{\a}(m+n-\s \a \cdot d)
\)
\(
[\hat{E}_{\alpha}(m-\s \a \cdot d), \hat{E}_{\beta}(n-\s \beta \cdot d)]
=\left\{\begin{array}{lll}
N_\gamma(\alpha,\beta) \hat{E}_{\gamma}(m+n-\s \gamma \cdot d)  && \textrm{if}\  \alpha+\beta=\gamma\\
\alpha\cdot \hat{H} (m+n)+k(m-\s \a \cdot d)\delta_{m+n,0} && \textrm{if}\ \alpha+\beta=0\\
0 && \textrm{otherwise}\\
\end{array}\right.
\)
\(
A,B=1,...,{\textrm{rank}}g \sp \alpha,\beta,\gamma\in\Delta(g) \sp \s=0,...,\r(1)-1
\)
\reseteqn
which is a special case of the general twisted current algebra (\ref{ghatgroup}).  As expected, this algebra  is a sector-dependent set of inner-automorphically twisted$^{\rf{KP},\rf{GO},\rf{FH},\rf{Lerche}}$ affine Lie algebras, and we note that, in this case, the integral affine subalgebra $\gb^{(0)}(\s)$ of each sector is at least the affine Cartan subalgebra.  

The adjoint of the twisted currents
\(
\hat{H}_A(m)^\dagger=\hat{H}_A(-m)
\sp \hat{E}_\a(m-\sigma \alpha \cdot d)^\dagger=\hat{E}_{-\a}(-m+\sigma \alpha \cdot d) 
\)
follows from the orbifold adjoint operation (\ref{Jdagger}), using $\Ltheta(\s)=\r$ in this case, and we know from the orbifold induction procedure for inner-automorphic twists$^{\rf{KP},\rf{GO},\rf{FH},\rf{Lerche}}$ that this adjoint guarantees unitarity of the orbifolds $A(H(d))/H(d)$ when the CFT $A(H(d))$ is unitary.

In this case, it is convenient for our discussion below to choose $M \p$  ordering (mode ordering with respect to $m\in \z$) defined in App.~\ref{NOapp}. This gives the exact operator products
\group{exactinnerOP}
\(
\hat{H}_{A}(z) \hat{H}_{B}(w) = \frac{k \delta_{AB}}{(z-w)^2} + :\hat{H}_A(z) \hat{H}_B (w):_{M \p}
\)
\(
\hat{H}_{A}(z) \hat{E}_{\a }(w) =\frac{\a_A \hat{E}_{\a}(w)}{z-w} + :\hat{H}_A(z) \hat{E}_\a (w):_{M \p}
\)
\(
\hat{E}_{\alpha}(z) \hat{E}_{\beta}(w) 
=
\ :\hat{E}_\a (z) \hat{E}_\beta (w):_{M \p}+
\left\{\begin{array}{lll}
(\frac{z}{w})^{\s \alpha \cdot d} \frac{N_\gamma(\alpha,\beta) \hat{E}_{\gamma}(w)}{z-w}  && \textrm{if}\  \alpha+\beta=\gamma\\
(\frac{z}{w})^{\s \alpha \cdot d}[\frac{k}{(z-w)^2}-\frac{k\s \alpha \cdot d}{w(z-w)} +\frac{\alpha\cdot \hat{H} (w)}{z-w}] && \textrm{if}\ \alpha+\beta=0\\
0 && \textrm{otherwise}\\
\end{array}\right.
\)
\(
A,B=1,...,{\textrm{rank}}g\sp
\alpha,\beta,\gamma\in\Delta(g) 
\)
\reseteqn
and then the relations
\group{exactinnerOPNO}
\(
:\hat{H}_{A}(z) \hat{H}_{B}(z): \ = \ :\hat{H}_A(z) \hat{H}_B (z):_{M \p}
\sp :\hat{H}_{A}(z) \hat{E}_{\a }(z): \ = \ :\hat{H}_A(z) \hat{E}_\a (w):_{M \p}
\)
\(
:\hat{E}_{\alpha}(z) \hat{E}_{\beta}(z): 
\ = \ :\hat{E}_\a (z) \hat{E}_\beta (z):_{M \p}+
\left\{\begin{array}{lll}
\frac{\s \alpha \cdot d}{z} N_\gamma(\alpha,\beta) \hat{E}_{\gamma}(z) && \textrm{if}\  \alpha+\beta=\gamma\\
\frac{\s \alpha \cdot d}{z} \alpha \cdot \hat{H}(z) -\frac{k \s \alpha\cdot d(\s \alpha\cdot d +1)}{2z^2}  && \textrm{if}\ \alpha+\beta=0 \\
0 && \textrm{otherwise} \\
\end{array}\right. \label{innerOPEMpNOing}
\)
\reseteqn
express the OPE normal ordered products in terms of $M\p$ ordering.

\subsection{The Virasoro generators of $A(H(d))/H(d)$}
\label{innerVirsec}
For each inner-automorphic orbifold $A(H(d))/H(d)$, the stress tensor of twisted sector $\s$ 
\(
\hat{T}_\s=L^{ab} : \hat{J}_a \hat{J}_b: \hsp{.1 }  = \hsp{.1} :(L^{AB} \hat{H}_A \hat{H}_B + L^{A\a} ( \hat{H}_A \hat{E}_\a + \hat{E}_\a \hat{H}_A ) + L^{\a \beta}  \hat{E}_\a \hat{E}_\beta ): \label{generalinnertwistT}
\)
is obtained from (\ref{bestTrecap}), (\ref{innerTgroup}) and (\ref{innertwistedtensors}).  For these orbifolds, all the sector dependence of $\hat{T}_\s$ resides in the twisted currents.

With $M\p$ ordering, the corresponding Virasoro generators of sector $\s$ are
\group{twistLgroup}
\(
L_\s(m)=L_\s^q(m)+L_\s^l(m)-\delta_{m,0} \frac{k}{2} \sum_\a L^{\a,-\a} (\sigma \alpha \cdot d)^2
\)
\vspace{-.4in}
\ba
L_\s^q(m) &\equiv& \sum_{p\in \sz} \{ \sum_{A,B} L^{AB}:\hat{H}_A(p) \hat{H}_B (m-p):_{M \p}  \label{quadraticterm} \\
&& \hsp{.3} + \sum_{A,\a} L^{A \a} : [ \hat{H}_A(p) \hat{E}_\a ((m-p+\s \a \cdot d)-\s \a \cdot d) \nonumber \\
&&\hsp{1.2} +\hat{E}_\a ((p+\s \a \cdot d)-\s \a \cdot d) \hat{H}_A(m-p) ] :_{M \p} \nonumber \\
&& \hsp{.3} + \sum_{\a, \beta} L^{\a \beta}:\hat{E}_\a(p-\s \a \cdot d) \hat{E}_\beta((m-p+\s(\a+\beta)\cdot d) - \sigma \beta \cdot d):_{M \p} \} \nonumber \\
L_\s^l(m)&\equiv&\sum_{\a+\beta=\gamma} L^{\a \beta} N_\gamma(\a,\beta) (\sigma \a \cdot d )\hat{E}_\gamma ((m+\s \gamma \cdot d)-\s \gamma \cdot d) \label{innerlinearterm} \\
&& \ + \sum_\a L^{\a,-\a}(\s \alpha \cdot d) \alpha \cdot \hat{H}(m)  \nonumber 
\ea
\(
L^{A\a}=0 \textup{  unless } \a \cdot d \in \z \sp L^{\a \beta}=0 \textup{  unless } (\a + \beta) \cdot d \in \z \sp \s=0,...,\r(1)-1 \label{innertLsolutions}
\)
\reseteqn
where $q$ and $l$ label the terms quadratic and linear in the twisted current modes.  The result (\ref{twistLgroup}) follows from (\ref{innermodeexpansion}), (\ref{exactinnerOPNO}) and (\ref{generalinnertwistT}), or as a special case of the general $M\p$-ordered orbifold Virasoro generators in (\ref{LmodeMp}).  In this result, we have written the arguments of the twisted root operators in the form
\(
\hat{E}_\a((\textup{integer}) -\s \a \cdot d) \label{argument}
\)
to exhibit their proper modeing.  To see that the quantities in the inner parentheses are integers, one must use\footnote{For example, consider the term proportional to $\hat{E}_\a \hat{E}_\beta$ in Eq.~(\ref{quadraticterm}).  The quantity in the inner parentheses of $\hat{E}_\beta$ is an integer because the solution of the $L^{\alpha \beta}$ selection rule allows the term to contribute only when $\s(\a+\beta)\cdot d \in \z$.  This term can also be written simply as $L^{\a \beta} \hat{E}_\a(p-\s \a \cdot d) \hat{E}_\beta(m-p+\s \a \cdot d)$.  Since this term can contribute only when $\hat{E}_\beta$ is in the twist class of $\hat{E}_{-\a}$, the simple expression is actually in the form $\hat{J}_{n(r)} \hat{J}_{-n(r)}$ of the general result (\ref{LmodeMp}).  Similarly, all the other terms in (\ref{twistLgroup}) can be put in the form of (\ref{LmodeMp}). } the solutions (\ref{innertLsolutions}) of the $L$-selection rules.

As a simple example of (\ref{twistLgroup}), we use the affine-Sugawara construction (\ref{sumroots}) to obtain the Virasoro generators of the general inner-automorphic WZW orbifold:
\group{innerWZWgroup}
\(
\frac{A_g(H(d))}{H(d)} \sp \s=0,...,\r(1)-1
\)
\ba
(2k+Q_\psi)L_\s^{\sgb (\s)} (m)&=& \sum_{p\in \sz} \{ :\sum_{\a}\hat{E}_\a(p-\s \a \cdot d) \hat{E}_{-\a}(m-p+\s \a \cdot d):_{M \p}  \\
&& \hsp{.4} +\sum_{A} :\hat{H}_A(p) \hat{H}_A (m-p):_{M \p} \} + Q_\psi \{\s  d \cdot \hat{H}(m)-\delta_{m,0} \frac{k}{2} \s^2 d^2 \} \nonumber 
\ea
\(
[L_\s^{\sgb (\s)} (m),\hat{H}_A(n)]=-n\hat{H}_A(m+n) \label{innerLHcommutator}
\)
\getletter{innerLEcommutatorlett}
\vspace{-.2in}
\(
[L_\s^{\sgb (\s)} (m),\hat{E}_\a(n-\s \a \cdot d)]=-(n-\s \a \cdot d) \hat{E}_\a(m+n-\s \a \cdot d) \ .
\)
\reseteqn
This result (including the fact that $\hat{E}$, $\hat{H}$ are twisted $(1,0)$ operators) is a special case of the result given for the general WZW orbifold in Eq.~(\ref{OASgroup}).  In sector $\s$, the orbifold affine-Sugawara construction $L_\s^{\sgb (\s)}$ is equivalent (with $\s d\rightarrow d$) to the inner-automorphically twisted affine-Sugawara construction of Refs.~\rf{KP} and \rf{FH}.  

Other special cases included in the result (\ref{twistLgroup}) are the Virasoro generators of the general inner-automorphic coset orbifold 
\(
\frac{\frac{g}{h}(H(d))}{H(d)} \subset \frac{A(H(d))}{H(d)} \ .
\)
These orbifolds were discussed at the level of stress-tensor spectral flow in Ref.~\rf{FH}. 

\subsection{Action on the untwisted affine vacuum}

In this and the following subsection, we study the action of the orbifold Virasoro generators (\ref{twistLgroup}) on a particular state $|0 \rangle$ which satisfies 
\(
(\hat{H}_A(m) - \delta_{m,0} \sigma k d_A)|0 \rangle=\hat{E}_\alpha(m-\sigma \alpha \cdot d)|0 \rangle =0 
\textup{ when } m \geq 0 \ .
\)
According to the orbifold induction procedure$^{\rf{KP},\rf{GO},\rf{FH},\rf{Lerche}}$ for inner-automorphically twisted affine Lie algebras, the state $|0 \rangle$ is the untwisted affine vacuum (see Eq.~(\ref{HEvac})).  (Except for small $\{ \s \a \cdot d \}$, the untwisted affine vacuum $|0\rangle$ is not the true ground state $|0\rangle_\s$ of twisted sector $\s$, but, so far as we know, identification of the true ground state is an unsolved problem.)  As emphasized in Ref.~\rf{FH}, the untwisted affine vacuum $|0\rangle$ is a twisted affine highest weight state only so long as $\s \a \cdot d > -1$ for all $\a\in \Delta (g)$.  One may compute 
\(
\langle 0 | L_\s(m) |0 \rangle =\delta_{m,0} \ \{ k^2\sum_{A,B}L^{AB}(\sigma d_A)(\sigma d_B) +\frac{k}{2}\sum_\a L^{\a,-\a}(\sigma \alpha \cdot d)^2 \}
\)
for all $\{ \s \a \cdot d \}$, but, as also emphasized in Ref.~\rf{FH},  $|0 \rangle$ is not Virasoro primary in general unless $\s \a \cdot d > -1$ for all $\a\in \Delta (g)$.

This situation can presumably be avoided by computing on twisted affine primary states or, as we will study here, by choosing \textit{only} $L^{AB}$ and $L^{\a,-\a}$ not equal to zero in (\ref{twistLgroup}).  In this case, we find that $|0\rangle$ is Virasoro primary
\vspace{.1in}
\group{goodinnerLorbgroup}
\(
\hsp{-3} L_\s(m) = \sum_{p \in \sz}  \sum_{A,B} L^{AB}:\hat{H}_A(p) \hat{H}_B (m-p):_{M\p}
\)
\[
\hsp{.5} + \sum_{\a} L^{\a,- \a} \{ \sum_{p \in \sz} :\hat{E}_\a(p-\s \a \cdot d) \hat{E}_{-\a}(m-p+\s \a \cdot d):_{M\p} + \s \a \cdot d(\a \cdot \hat{H}(m)- \delta_{m,0} \s \a \cdot d) \}  \label{noprobL} 
\]
\(
L_\sigma(m \geq 0) |0 \rangle=\delta_{m,0} \hat{\Delta}_0(L,d;\s) |0 \rangle \sp \s=1,...,\r(1)-1
\)
\(
\hat{\Delta}_0(L,d;\s)=k^2\sum_{A,B}L^{AB}(\sigma d_A)(\sigma d_B) +\frac{k}{2}\sum_\a L^{\a,-\a}(\sigma \alpha \cdot d)^2 \label{generalinnercw} 
\)
\reseteqn
without restriction on $\{ \s \a \cdot d \}$.  Recall from (\ref{innerLsolutions}) that there is no $H$-invariance restriction on $L^{AB}$ or $L^{\a,-\a}$. 

\subsection{Connection with spectral flow}
\label{flowsec}
When there is an orbifold induction procedure, one may rewrite orbifold Virasoro generators in terms of untwisted current modes\footnote{The inverse of this unconventional step$^{\rf{FH}}$ was introduced by Freericks and Halpern, who studied the twisted current formulation of inner-automorphic WZW orbifolds starting from the spectral flow discussed below.} $J_a(m)$ which satisfy (\ref{untwistJalg}).  Such procedures are known for inner-automorphically twisted affine Lie algebra$^{\rf{KP},\rf{GO},\rf{FH},\rf{Lerche}}$ and orbifold affine algebra$^{\rf{Chris}}$
\alpheqn
\(
\hat{H}_A(m) = H_A(m) + \delta_{m,0} k \sigma d_A
\sp \hat{E}_\a (m-\sigma \alpha \cdot d)=E_\a (m)  \label{innerinduction}
\)
\(
\hat{J}_{aj}^{(r)}(m+\frac{r}{\r(\s)})=J_{aj}(\r(\s) m+r) \sp r=0,...,\r(\s)-1 \label{zlinduction}
\)
\reseteqn
as well as the doubly-twisted affine algebras$^{\rf{us1},\rf{us2}}$ which combine (\ref{innerinduction}) and (\ref{zlinduction}).  

As discussed in App.~\ref{modesunbalancedapp}, rewriting orbifold stress tensors in terms of untwisted currents leads to generically exotic and unfamiliar forms of the Virasoro generators $L_\s(m)$.  In these forms one sees a generic \textit{mode imbalance}, in which the modes of the untwisted currents do not sum to the integer $m$.

We rewrite here only the special case $L^{AB},L^{\a,-\a} \neq 0$ of the inner-automorphic orbifolds in (\ref{goodinnerLorbgroup}), which avoids this mode imbalance phenomenon.  In this case one finds
\group{FHuntwistedJgroup}
\(
L_\s(m)=L(m)+D(L,d;\s)\cdot H(m)+\delta_{m,0} \hat{\Delta}_0(L,d;\s) \sp \s=0,...,\r(1)-1 \label{Lorbifoldflow}
\)
\(
L(m) \equiv \sum_{p \in \sz} \{ \sum_{A,B} L^{AB}:H_A(p) H_B (m-p):
+ \sum_{\a} L^{\a,- \a}:E_\a(p) E_{-\a}(m-p): \} \label{usualL} 
\)
\(
D(L,d;\s)^A \equiv \s \sum_B d_B M(L)^{BA}  \sp M(L)^{AB} =  2k L^{AB} + \sum_\a L^{\a,-\a} \alpha^A \alpha^B  \label{Ddefa}
\)
\(
(L_\sigma(m \geq 0) - \delta_{m,0} \hat{\Delta}_0(L,d;\s)) |0 \rangle=0 \sp  \hat{\Delta}_0(L,d;\s) =\frac{k\s^2}{2} \sum_{AB} d_A d_B M(L)^{AB} \label{generalinnercwa} 
\)
\reseteqn
where the matrix $M(L)^{AB}$ is $\delta^{AC} M(L)_C^{\ \ B}$ and $M(L)_A^{\ \ B}$ is the Cartan block of $M(L)_a^{\ b}$ in (\ref{MNdef}).  In (\ref{generalinnercwa}), the conformal weight $\hat{\Delta}_0(L,d;\s)$ of the untwisted affine vacuum $|0 \rangle$ is the same as that given in (\ref{generalinnercw}).  All the current modes in (\ref{FHuntwistedJgroup}) are untwisted and we have written the operator $L(m)$ in (\ref{usualL}) as an OPE normal ordered product
\alpheqn
\(
:J_a(m)J_b(n): \ = \ :J_a(m)J_b(n):_{M\p} \ = \theta(m\geq 0)J_b(n)J_a(m)+\theta(m\leq 0)J_a(m)J_b(n)
\)
\(
H_A(m \geq 0) |0\rangle=E_\alpha(m \geq 0)|0 \rangle =L(m\geq -1)|0 \rangle=0 \label{HEvac}
\)
\reseteqn
although OPE normal ordering is the same as $M\p$ ordering for untwisted currents.  

The form (\ref{FHuntwistedJgroup}) (but not the more general case in (\ref{unbalancedgroup})) provides an opportunity to check some results of the orbifold program against known results in affine-Virasoro theory.  In particular, we want to compare the result (\ref{FHuntwistedJgroup}) to the \textit{general $c$-fixed conformal deformation}$^{\rf{BH},\rf{FH},\rf{vme},\rf{rev}}$ of the general affine-Virasoro construction:
\group{cfixedgroup}
\(
L(m;D)=L^{ab} \sum_{p \in \sz} :J_a(p) J_b(m-p): + D^a J_a (m) + \Delta_0(D) \delta_{m,0}
 \label{spectralflow}
\)
\vspace{-.2in}
\(
L^{ab}= 2 L^{ac} G_{cd} L^{db} - L^{cd} L^{ef} f_{ce}^{\ \ a}f_{df}^{\ \ b}-L^{cd} f_{cd}^{\ \ f}f_{df}^{\ \ (a}L^{b) e} \sp c(D)=c=2G_{ab}L^{ab}
\)
\(
D^b M(L)_b^{\ a}=D^a \sp M(L)_a^{\ b}=2 G_{ac} L^{cb}+f_{ad}^{\ \ e}L^{dc}f_{ce}^{\ \ b}  \label{Ddef}
\)
\(
(L(m\geq0;D) - \delta_{m,0} \Delta_0(D)) |0\rangle =0
\sp  \Delta_0(D) =\frac{1}{2} G_{ab}D^a D^b \ . \label{cfixedcw}
\)
\reseteqn
Here the first term in (\ref{spectralflow}) is the undeformed affine-Virasoro construction with central charge $c$, and $\Delta_0(D)$ is the conformal weight of the untwisted affine vacuum $|0\rangle$ under the deformed Virasoro $L(m;D)$ as a function of the deformation $D$.  The matrix $M(L)$ in (\ref{Ddef}) is the same matrix defined in (\ref{MNdef}), and the eigenvector condition $DM(L)=D$ constrains the allowed values of the deformation $D$.  The eigenvector condition is equivalent to the requirement that $D\cdot J$ is a $(1,0)$ operator of the undeformed Virasoro.  Because the scale of $D$ is not fixed by the eigenvector condition, the $c$-fixed conformal deformations are also called \textit{stress-tensor spectral flow}$^{\rf{BH},\rf{FH},\rf{vme},\rf{rev}}$ which is equivalent$^{\rf{FH}}$ to inner-automorphic twisting or spectral flow of the underlying currents.$^{\rf{KP},\rf{GO},\rf{FH},\rf{Lerche}}$  The first example of stress-tensor spectral flow was given by Bardakci and Halpern in Ref.~\rf{BH}.

To begin this comparison, we note first that the orbifold result (\ref{FHuntwistedJgroup}) \textit{must} be in the spectral flow (\ref{cfixedgroup}), with the identifications 
\group{identificationgroup}
\(
L_{\s=0}(m)=L(m)=L(m;D=0) \sp L^{ab}=\{ L^{AB}, L^{\a,-\a} \}  \sp \hat{c}(\s)=c(D)=c \label{Lnames}
\)
\(
D^A=D(L,d;\s)^A \sp D^\a=0 \sp \Delta_0(D)=\hat{\Delta}_0 (L,d;\s)
\)
\reseteqn
because the first term $L(m)$ in (\ref{Lorbifoldflow}) is precisely the Virasoro generator $L_{\s=0}(m)$ we started with in the untwisted sector of each orbifold.  For the affine-Sugawara construction on $g$ in (\ref{sumroots}), the two systems (\ref{FHuntwistedJgroup}) and (\ref{cfixedgroup}) are indeed equivalent with
\group{ASflowgroup}
\(
L_\s^{\sgb (\s)} (m) = L_g(m;D_g) = \frac{\eta^{ab}}{2k+Q_\psi}\sum_{p\in \z}:J_a(p) J_b(m-p): + D_g^A J_A(m)+\Delta_0(D_g) \delta_{m,0}
\)
\(
M(L_g)_a^{\ b}=\delta_a^{\ b} \sp D_g^A=D(L_g,d;\s)=\s d^A \sp \Delta_0(D_g)=\hat{\Delta}_0(L_g,d;\s)=\frac{k}{2} D_g^2  
\)
\(
\hat{c}_g(\s)=c_g(D)=c_g \sp a=1,...,\textup{dim}g \sp A=1,...,\textup{rank}g \sp \s=0,...,\r(1)-1
\)
\reseteqn
and these Virasoro generators are also the same as those in (\ref{innerWZWgroup}).  For arbitrary $D_g$ the result (\ref{ASflowgroup}) is the canonical example of stress-tensor spectral flow first studied by Freericks and Halpern in Ref.~\rf{FH}, and applied more recently in Refs.~\rf{KT} and \rf{Oog}. As emphasized in Ref.~\rf{FH}, the orbifold sectors in (\ref{ASflowgroup}) are special points of the spectral flow.

But the more general identification given in (\ref{identificationgroup}) is surprising because the ``deformation'' $D(L,d;\s)$ in (\ref{Ddefa}) is a function of the inverse inertia tensor $L^{ab}$ and it is not obvious that the conformal weight $\hat{\Delta}_0(L,d;\s)$ in (\ref{generalinnercwa}) of the untwisted affine vacuum agrees with the spectral flow form $\Delta_0(D)$ in (\ref{cfixedcw}).

The key to understanding this identification is an unsuspected\footnote{Motivation for a Lie $h$ invariance of the orbifold Virasoro generators $L_{\s=0}(m)=L(m;D=0)$ comes from the spectral-flow side of the identification: Lie $h$-invariant CFT's with a non-trivial global component $h_1$ (see (\ref{01untwistedcommute})) are the only known CFT's with (1,0) operators.  Conversely, for any Lie $h$-invariant CFT, arbitrary deformation by the (1,0) operators of $h_1$ automatically solve the spectral flow system (\ref{cfixedgroup}).} Lie $h$ invariance in the Virasoro generators $L_{\s=0}(m)$ of the untwisted sectors of these orbifolds!  

To see this Lie invariance, one must verify the identity
\(
\delta_AL^{ab} \equiv L^{c (a}f_{c A}^{\ \ \ b)}=-i N(L)_A^{\hsp{.1} ab}=0 \sp A=1,...,\textup{rank}g \label{anotherLieh}
\)
which follows (when only $L^{AB}$ and $L^{\a,-\a}$ are non-zero) from the form of the structure constants in the Cartan-Weyl basis.  The identity (\ref{anotherLieh}) tells us (see Subsec.~\ref{Liehsec}) that the subset $L^{AB}$, $L^{\a,-\a}\neq 0$ of inverse inertia tensors in $A(H(d))$ also has a Lie symmetry 
\(
\textup{Lie} \hsp{.025} h = \textup{Cartan} \hsp{.025}g
\)
in addition to the inner-automorphic invariance $H(d)$.  

In the nomenclature of Subsec.~\ref{Hliehcftsec}, we have shown that $L_{\s=0}(m)=L(m)$ in (\ref{Lnames}) describes a large set of \textit{doubly-invariant} or ($H$ and Lie $h$)-invariant CFT's which we will call
\(
A(\textup{Cartan} \hsp{.025} g(H(d))) \subset A(H(d)) \sp A(\textup{Cartan} \hsp{.025} g(H(d))) \subset \alie \ .
\)
It follows that the orbifold Virasoro generators $L_\s(m)$ in (\ref{goodinnerLorbgroup}) or (\ref{FHuntwistedJgroup}) describe the orbifolds
\(
\frac{A(\textup{Cartan} \hsp{.025} g(H(d)))}{H(d)} \subset \frac{A(H(d))}{H(d)} \sp \frac{A(\textup{Cartan} \hsp{.025} g(H(d)))}{H(d)} \subset \alieHH
\)
by $H(d)$ of the doubly-invariant CFT's $A(\textup{Cartan} \hsp{.025} g(H(d)))$.

Returning to the untwisted sectors, the Cartan invariance (\ref{anotherLieh}) implies that 
\group{anotherliehgroup}
\(
M(L)_A^{\hsp{.1} C} M(L)_C^{\hsp{.1} B}= M(L)_A^{\hsp{.1} B} \sp M(L)_A^{\ \a}=0  \label{Liehagain}
\)
\getletter{anotherlielettb}
\vspace{-.3in}
\(
[L(m),H_A(n)]=-n M(L)_A^{\ B}H_B(m+n) \label{LHCartancommutator}
\)
\(
[L(m),D^A(L,d;\s)  H_A(n)]=-n D(L,d;\s)^A  H_A(m+n) \label{DH10}
\)
\reseteqn
at least for unitary CFT's,$^{\rf{Giveon},\rf{Lieh}}$ where (\ref{Liehagain},\ref{anotherlielettb}) are special cases of the general properties of Lie $h$-invariant CFT's in (\ref{hblockMgroup}).  The relation (\ref{DH10}), which follows from (\ref{Ddefa}) and (\ref{Liehagain},\ref{anotherlielettb}), says that $D\cdot H$ is a $(1,0)$ operator of the undeformed Virasoro.  As noted above, this means that $D^A(L,d;\s)$ solves the eigenvector condition for $D$ in (\ref{Ddef}).  More explicitly, the eigenvector condition in (\ref{Ddef}) takes the form
\(
\s \sum_B d^B \ \{ M(L)_B^{\ A} - \sum_C M(L)_B^{\ C} M(L)_C^{\ A} \}=0 \label{DforD}
\)
upon substitution of $D(L,d;\s)$ for $D$.  Then, using $M^2=M$ in (\ref{Liehagain}), we see that the eigevector condition is satisfied identically for arbitrary $\s$ and $d$.  

Finally, the equality of the conformal weights
\(
\Delta_0(D=D(L,d;\s))  =  \frac{1}{2}kD^2(L,d;\s) =  \hat{\Delta}_0(L,d;\s)
\)
is established by using the definition of $D(L,d;\s)$ in (\ref{Ddefa}) and $M^2=M$ in (\ref{Liehagain}).  This completes our check that, at least for unitary $A(\textup{Cartan} \hsp{.025} g(H(d)))$, the orbifold systems (\ref{FHuntwistedJgroup}) are points in the general spectral flow (\ref{cfixedgroup}).

We close this section with a number of remarks:

\bu \ Conditions on the vector $d$.  The vector $d$ is determined in the orbifolds by the choice of $H(d)$, but, as we saw in (\ref{DforD}), the vector $d$ is not determined by the eigenvector condition (\ref{Ddef}) for the spectral flow.  This is of course the phenomenon described by Freericks and Halpern in their original study$^{\rf{FH}}$ of inner-automorphic orbifolds as special points of stress-tensor spectral flow.

\bu \ Unitarity revisited.  In fact, the Lie $h$ relations (\ref{anotherliehgroup}) -- and hence the conclusions above for the orbifolds -- can be checked directly, without using unitarity.  The Lie $h$ relations $N(L)_A^{\ \ bc}=0$ and $M_A^{\ \a}=0$ (which are necessary for (\ref{LHCartancommutator})) follow immediately from $L^{AB},L^{\a,-\a}\neq 0$.  To check the last Lie $h$ relation $M^2=M$ in (\ref{Liehagain}) we need the reduced Virasoro master equation of the doubly-invariant CFT's $A(\textup{Cartan} \hsp{.025} g(H(d)))$
\group{VMEsubansatzgroup}
\(
L^{AB}=2k \sum_D L^{AD}L^{DB} - \sum_\a (L^{\a,-\a})^2 \a^A \a^B + \sum_{\a,D} L^{\a,-\a} \a^D L^{D (A} \a^{B)} 
\)
\(
L^{\a,-\a}=2(k+\a^2)(L^{\a,-\a})^2 + \sum_{\beta+\gamma=\a} (2 L^{\a,-\a}-L^{\gamma,-\gamma})L^{\beta,-\beta} N_\a^2(\beta,\gamma) 
\)
\(
c=2k(\sum_A L^{AA}+\sum_\a L^{\a,-\a}) 
\)
\reseteqn
where the structure constants $N_\gamma(\a,\beta)$ are defined in (\ref{affalg}).  The inverse inertia tensors of $A(\textup{Cartan} \hsp{.025} g(H(d)))$ are controlled entirely by this system because there are no $H$-invariance restrictions on $L^{AB}$, $L^{\a,-\a} \neq 0$.  The reduced Virasoro master equation (\ref{VMEsubansatzgroup}) is a consistent subansatz of the Virasoro master equation, with the generically-expected number of inequivalent solutions at each level 
\(
N(g)=2^{n(g)}=2^{\frac{1}{2}((\textup{\scriptsize{rank}} g)^2 + \textup{\scriptsize{dim}} g)} \label{numsolutionsVME}
\)
where $n(g)$ is the number of equations and unknowns in the system.  

Using the reduced master equation (\ref{VMEsubansatzgroup}) to eliminate terms linear in $L$, we find that the relation $M^2=M$ in (\ref{Liehagain}) can be reduced to the consistency relation
\[
\sum_{\a,\beta}\a^A \beta^BL^{\a,-\a} L^{\beta,-\beta} (\a \cdot \beta) = \sum_\a \a^A \a^B \{ 2 \a^2 (L^{\a,-\a})^2 +\sum_{\beta+\gamma=\a} 2 L^{\a,-\a} L^{\beta,-\beta} N_\a^2(\beta,\gamma) 
\]
\(
\hsp{2.5} - \sum_{\beta+\gamma=\a} L^{\gamma,-\gamma} L^{\beta,-\beta} N_\a^2 (\beta,\gamma) \} 
\)
and this relation is in fact an identity because 
\alpheqn
\(
N_\gamma(\a,\beta)=N_{-\beta}(-\gamma,\a)=-N_{-\gamma}(-\a,-\beta)
\) 
\(
\a \cdot \beta = (\delta_{-\a+\beta,0}-\delta_{\a+\beta,0})\a^2 + \delta_{-\a+\beta,\gamma\p} N_{\gamma\p}^2 (-\a,\beta)-\delta_{\a+\beta,\gamma} N_\gamma^2 (\a,\beta) \label{alphabeta}
\)
\reseteqn
where (\ref{alphabeta}) follows from the Jacobi identity of Lie $g$.  This completes the check that the Lie $h$ relations (\ref{anotherliehgroup}) are true independent of unitarity, and we conjecture that the general Lie $h$ relations in (\ref{hblockMgroup}) are similarly true independent of unitarity.

\bu \ $K$-conjugation in spectral flow.  We note that orbifold $K$-conjugation
\group{innerKconjgroup}
\(
L_\s(m;D(L,d;\s))+\tilde{L}_\s(m;D(\tilde{L},d;\s))=L_\s^{\sgb (\s)} (m;D(L_g,d;\s))
\)
\(
L+\tilde{L}=L_g \sp c+ \tilde{c}=c_g \sp [L_\s(m;D(L,d;\s)),\tilde{L}_\s(n;D(\tilde{L},d;\s))]=0
\)
\reseteqn
holds for the special points of the stress-tensor spectral flow which describe the orbifolds $A(\textup{Cartan} \hsp{.025} g(H(d)))/H(d)$.  Here the modes of $L_\s^{\sgb (\s)}$ are given in Eq.~(\ref{ASflowgroup}), and (\ref{innerKconjgroup}) is a special case of the general orbifold $K$-conjugation in (\ref{genK}).  Eq.~(\ref{innerKconjgroup}) is the first observation of $K$-conjugation in stress-tensor spectral flow, and $K$-conjugation in the general spectral flow (\ref{cfixedgroup}) deserves further study.

\bu \ Twisted $h$ currents.  We return for a moment to the twisted current formulation of the orbifolds $A(\textup{Cartan} \hsp{.025} g(H(d)))/H(d)$.  Another consequence of the Lie$\hsp{.025}h=$Cartan$\hsp{.025} g$ invariance of $A(\textup{Cartan} \hsp{.025} g(H(d)))$ is the orbifold statement 
\(
[L_\s (m),\hat{H}_A(n)]=-n M(L)_A^{\ \ B} \hat{H}_B(m+n) \sp A=1,...,\textup{rank}g \sp \s=\r(1)-1 \label{innerLHhatcommutator}
\)
where $L_\s(m)$ is given in (\ref{goodinnerLorbgroup}) and $\hat{H}_A(m)$ are the Cartan modes of sector $\s$.  In a left eigenbasis of $M(L)$, Eq.~(\ref{innerLHhatcommutator}) is a special case of the more general result (\ref{TJOPELiehgroup}).

\bu \ Beyond spectral flow.  We have seen that $A(\textup{Cartan} \hsp{.025} g(H(d)))/H(d)$ is a large subset of inner-automorphic orbifolds which can also be described by stress-tensor spectral flow.  We emphasize however that Eqs.~(\ref{twistLgroup}) and (\ref{unbalancedgroup}) contain the Virasoro generators of a much larger class of inner-automorphic orbifolds which apparently (due to the mode-imbalance phenomenon discussed in App.~\ref{modesunbalancedapp}) can not be described by spectral flow.

\bu \ SL(2,R) WZW models.  We finally note that the stress tensors proposed for SL(2,R) WZW models in Ref.~\rf{Oog} are special cases (with $\s \a \cdot d \in \z$ for $g=\ $SL(2,R)) of the inner-automorphic WZW orbifold stress tensors (\ref{innerWZWgroup}) or (\ref{ASflowgroup}).

\bigskip

\noindent
{\bf Acknowledgements} 

We thank P. Bouwknegt, R. Dijkgraaf, M. Gaberdiel, A. Giveon, H. Ooguri, N. Nekrasov, N. Reshetikin and K. Skenderis for helpful discussions.  We also thank J. Evslin for his participation in the early stages of this paper, and C. Schweigert for reading the manuscript.  

J. E. W. was supported by the Department of Education, GAANN.  M. B. H. was supported in part by the Director, Office of Science, Office of High Energy and Nuclear Physics, Division of High Energy Physics, of the U.S. Department of Energy under Contract DE-AC03-76SF00098 and in part by the National Science Foundation under grant PHY95-14797.

\appendices

\app{selectionapp}{Incorporation of the Selection Rules}
In this appendix we incorporate the solutions of various selection rules to obtain reduced forms of relations in the text.

For example, we may combine the solution (\ref{fwithdelta}) of the $\fr$-selection rule and the Jacobi identity (\ref{jacobi}) for $\fr$ to obtain the reduced form of the Jacobi identity
\[
\hsp{-.75} {\textstyle {\Big {\bf \sum}} _{\mu \p}} \{  \fr_{n(r) \mu; n(s) \nu}^{\ \ \ \ \ \ \ \ \ \ \ n(r)+n(s), \mu \p}(\s) \fr_{n(t) \delta;  n(r)+n(s), \mu \p}^{\ \ \ \ \ \ \ \ \ \ \ \ \ \ \ \ \ \ n(r)+n(s)+n(t), \gamma}(\s)
\] 
\[
\hsp{-.3} + \fr_{n(t) \delta; n(r) \mu}^{\ \ \ \ \ \ \ \ \ \ \hsp{.04} n(r)+n(t),\mu \p}(\s) \fr_{n(s) \nu; n(r)+n(t), \mu \p}^{\ \ \ \ \ \ \ \ \ \ \ \ \ \ \ \ \ \hsp{.05} n(r)+n(s)+n(t), \gamma}(\s)
\]
\(
+ \fr_{n(s) \nu; n(t) \delta}^{\ \ \ \ \ \ \ \ \ \ \hsp{.03} n(s)+n(t), \mu \p}(\s) \fr_{n(r) \mu; n(s)+n(t),  \mu \p}^{\ \ \ \ \ \ \ \ \ \ \ \ \ \ \ \ \ \hsp{.05} n(r)+n(s)+n(t), \gamma}(\s) \} =0 \label{sJacobi}
\)
in which no spectral indices are summed.  Similarly the solution of the $\fr$-selection rule and (\ref{frantisymm}) imply the relation
\(
\fr_{n(r) \mu;n(s) \nu; -(n(r)+n(s)), \delta}(\s) = - \fr_{n(r) \mu; -(n(r)+n(s)), \delta; n(s) \nu}  \label{sfanti} 
\)
among the reduced, twisted totally-antisymmetric structure constants of sector $\s$.  The relations (\ref{sJacobi}) and (\ref{sfanti}) are found to guarantee the Jacobi identity of the general twisted current algebra $\gb (\s)$ in Eq.~(\ref{ghatgroup}).

Using the solution (\ref{Lthetawithdelta}) of the selection rule for the orbifold conjugation matrix $\Ltheta$, we obtain the reduced form of (\ref{invphi})
\(
{\textstyle {\Big {\bf \sum}}_\delta} \ \Ltheta_{n(r) \mu}^{\hsp{.3} -n(r), \delta}(\s) \Ltheta_{-n(r), \delta}^{\ \ \ \ \ \ \ \ n(r) \nu}(\s)^* ={\textstyle {\Big {\bf \sum}}_\delta}  \ \Ltheta_{n(r) \mu}^{\hsp{.3} -n(r) , \delta}(\s)^* \Ltheta_{-n(r), \delta}^{\hsp{.4} n(r) \nu}(\s) =\delta_\mu^{\ \nu} \label{partinvphi}
\)
\noindent which is needed to verify Eq.~(\ref{hatJdoubledagger}).

Similarly, the reduced forms of (\ref{grstar}), (\ref{frstar}) and (\ref{lrstar})
\alpheqn
\(
\g_{n(r) \mu;  -n(r), \nu}(\s)^* = \sum_{\mu\p,\nu\p} \Ltheta_{n(r) \mu}^{\hsp{.3} -n(r), \mu '}(\s) \Ltheta_{-n(r), \nu}^{\ \ \ \ \ \ \ \ n(r) \nu '}(\s) \g_{-n(r), \mu '; n(r)  \nu '}(\s) \label{sgrstar}
\)
\(
\fr_{n(r) \mu; -n(r), \nu}^{\ \ \ \ \ \ \ \ \ \ \ \ \ \ 0 \delta}(\s)^*= \sum_{\mu\p,\nu\p, \delta\p} \Ltheta_{n(r) \mu}^{\hsp{.3} -n(r), \mu '}(\s) \Ltheta_{-n(r), \nu}^{\ \ \ \ \ \ \ \  n(r) \nu '}(\s) \fr_{-n(r), \mu ' ; n(r) \nu '}^{\ \ \ \ \ \ \ \ \ \ \ \ \ \ \  0 \delta '}(\s)  \Ltheta_{0 \delta '}^{\ \ \ 0 \delta}(\s)^*  \label{sfrstar}
\)
\(
\lr^{n(r) \mu; -n(r), \nu}(\s)^* = \sum_{\mu\p,\nu\p} \lr^{-n(r), \mu \p;n(r) \nu \p}(\s) \Ltheta_{-n(r), \mu \p}^{\hsp{.45} n(r) \mu}(\s)^* \Ltheta_{n(r) \nu \p}^{\hsp{.3} -n(r), \nu}(\s)^* \label{slrstar} 
\)
are needed to verify (\ref{twistLdagger}) and (\ref{cwreal}), and to check the consistency of (\ref{consistentJdagger}). 

\app{projectorsapp}{The Projectors of \tahh}
The projectors $\pr(\bar{n}(r);\s)$ onto the $\bar{n}(r)$ subspaces of sector $\s$,
\namegroup{projectorappgroup}
\alpheqn
\vspace{-.15in}
\(
\pr(\bar{n}(r);\s)_a^{\ b} \equiv {\textstyle {\Big {\bf \sum}}_{\mu}} \ \uds_a^{\ n(r) \mu} U(\s)_{n(r) \mu}^{\ \ \ \ \ \ b} \sp \pr(\bar{n}(r);\s)_a^{\ a}=\textup{dim}[\bar{n}(r)]
\)
\(
\pr(\bar{n}(r) \pm \r(\s);\s)= \pr(\bar{n}(r);\s) 
\sp \pr(-\bar{n}(r);\s)= \pr(\r(\s)-\bar{n}(r);\s) \label{projperiod}
\)
\(
\pr(\bar{n}(r);\s) \ \pr(\bar{n}(s);\s) =\delta_{n(r)}^{\hsp{.2} n(s)} \pr(\bar{n}(r);\s)
\sp {\textstyle {\Big {\bf \sum}}_{r}} \ \pr(\bar{n}(r);\s)=1 \label{projectorpropertyapp}
\)
\(
\w(h_\s)={\textstyle {\Big {\bf \sum}} _{r}} \ E_{n(r)}(\s) \pr(\bar{n}(r);\s) 
\sp [\w(h_\s), \pr(\bar{n}(r);\s)]=0 \label{wprc} 
\)
\(
\pr(\bar{n}(r);\s)_a^{\ c} \pr(\bar{n}(s);\s)_b^{\ d} G_{cd} (1-E_{n(r)} E_{n(s)})=0 \label{projectorselctiongroup}
\)
\agetletter{psglett}
\vspace{-.05in}
\(
\pr(\bar{n}(r);\s)_a^{\ d} \pr(\bar{n}(s);\s)_b^{\ e} f_{de}^{\ \ f} \pr(\bar{n}(t);\s)_f^{\ c}(1-E_{n(r)} E_{n(s)}E_{n(t)}^*)=0
\)
\(
L^{ab}  \pr(\bar{n}(r);\s)_a^{\ c} \pr(-\bar{n}(r);\s)_b^{\ d} G_{cd}=L^{ac} G_{cb} \pr(\bar{n}(r);\s)_a^{\ b} \label{prforcw}
\)
\reseteqn
are defined for all orbifolds $A(H)/H$.  The identity in (\ref{prforcw}) can be proven by using duality transformations to write out $\sum_{r,s,\mu,\nu} \ f(n(r)) \lr^{n(r)\mu;n(s)\nu}\g_{n(r)\mu;n(s)\nu}$ in two ways, both with and without the selection rules, for arbitrary periodic $f$.  These projectors were encountered in Sec.~\ref{cwsec} and will play a central role in App.~\ref{constrainedbasisapp}.

\app{constrainedbasisapp}{A Constrained Basis for Twisted Currents}
We consider another basis for the twisted currents of \tahh
\alpheqn
\(
\hat{J}_a^{n(r)}(m+\srac{n(r)}{\r(\s)}) \equiv {\textstyle {\Big {\bf \sum}} _{\mu}} \ {\foot{\chi(\s)_{n(r) \mu}^{-1}}} U^\dagger(\s)_a^{\ n(r) \mu} \hat{J}_{n(r) \mu}(m+\srac{n(r)}{\r(\s)})
\)
\(
\hat{J}_a^{n(r)}(m+\srac{n(r)}{\r(\s)})^\dagger=\stheta_a^{\ b} \hat{J}_b^{-n(r)}(-m-\srac{n(r)}{\r(\s)})
\)
\(
\pr(\bar{n}(s);\s)_a^{\ b} \hat{J}_b^{n(r)}(m+\srac{n(r)}{\r(\s)}) = \hat{J}_a^{n(s)}(m+\srac{n(s)}{\r(\s)}) \ \delta_{n(s)}^{\hsp{.2} n(r)} \label{PJ}
\)
\(
a=1,...,\textup{dim}g \sp \bar{n}(r)\in \{ 0,...,\r(\s)-1 \}
\)
\reseteqn
where $\{ \pr(\bar{n}(r);\s) \}$ is the set of projectors in App.~\ref{projectorsapp}.  These twisted currents carry the original Lie algebra index $a$, but the basis is overcomplete with the constraints (\ref{PJ}).

In this basis, the twisted current algebra and the orbifold stress tensors take the form
\alpheqn
\(
\hat{T}_\s(z)={\textstyle {\Big {\bf \sum}}_{r}} \ L^{ab}:\hat{J}_a^{\ n(r)}(z) \hat{J}_b^{\ -n(r)}(z):
\)
\(
\hat{J}_a^{n(r)}(z)=\sum_{m\in \sz} \hat{J}_a^{n(r)}(m+\srac{n(r)}{\r(\s)}) z^{-(m+\frac{n(r)}{\r(\s)})-1} \sp \hat{J}_a^{n(r)}(ze^{2 \pi i})=E_{n(r)}(\s) \hat{J}_a^{n(r)}(z)
\)
\[
[ \hat{J}_a^{n(r)}(m+\srac{n(r)}{\r(\s)}),\hat{J}_b^{n(s)}(n+\srac{n(s)}{\r(\s)})] =\pr(\bar{n}(r);\s)_a^{\ c} \pr(\bar{n}(s);\s)_b^{\ d} \{  (m+\srac{n(r)}{\r(\s)}) G_{cd} \delta_{m+n+\frac{n(r)+n(s)}{\r(\s)},0}
\]
\(
\hsp{1.5} + i f_{cd}^{\ \ e}  \hat{J}_e^{n(r)+n(s)}(m+n+\srac{n(r)+n(s)}{\r(\s)}) \} \label{constrainedJalg}
\)
\reseteqn
where $L^{ab}$ is the $H$-invariant inverse inertia tensor of the untwisted sector of the orbifold.  The Jacobi identity for (\ref{constrainedJalg}) is verified with the selection rules (\ref{projectorselctiongroup},\ref{psglett}).

\app{nnbarapp}{Conversion from $n(r)$ to $\bar{n}(r)$}
In Eq.~(\ref{exactJJ}), various integers $n(r)$ are converted into their corresponding twist class $\bar{n}(r)$.  To see how this happens, follow the steps
\alpheqn
\(
\hsp{-1.5} \frac{1}{z} (\frac{w}{z})^{\frac{n(r)}{\r(\s)}} \sum_{m \in \sz} \theta(m+\srac{n(r)}{\r(\s)} \geq 0) (\frac{w}{z})^m 
\)
\vspace{-.25in}
\ba
&=&\frac{1}{z} (\frac{w}{z})^{\frac{\bar{n}(r)}{\r(\s)}} \sum_{m\p \in \sz} \theta(m\p+\srac{\bar{n}(r)}{\r(\s)} \geq 0) \ (\frac{w}{z})^{m\p} \sp 0\leq \frac{\bar{n}(r)}{\r(\s)}<1 \label{nnbarm} \\
&=&\frac{1}{z} (\frac{w}{z})^{\frac{\bar{n}(r)}{\r(\s)}} \sum_{m\p=0}^{\infty} (\frac{w}{z})^{m\p}=(\frac{w}{z})^{\frac{\bar{n}(r)}{\r(\s)}} \frac{1}{z-w} \label{nnbarf}
\ea
\reseteqn
where we have used (\ref{nnbar}) and the change of variable $m\p=m+\lfloor n(r)/\r(\s) \rfloor$ to obtain (\ref{nnbarm}).

\app{TJapp}{Direct Computation of the $\hat{T} \hat{J}$ OPE's}
Orbifold operator products of the general currents $\hat{J}_{n(r)\mu}$ can be obtained in one step by the prescription 
\(
r a \rightarrow n(r) \mu \sp \hat{J}_a^{(r)} \rightarrow \hat{J}_{n(r)\mu}
\)
from the formulas of Appendix A of Ref.~\rf{us1}.  For the operator product $\hat{T}_\s \hsp{.025} \hat{J}(\s)$, one finds the same OPE's as in Eq.~(\ref{hatsTJOPE}), but the forms of the twisted tensors $\mc(\s)$ and $\nc(\s)$ which result from this computation are:
\namegroup{mcncappgroup}
\alpheqn
\vspace{-.2in}
\(
\mc_{n(r) \mu}^{\hsp{.3} n(s) \nu}(\s) = \lr^{n(t) \delta; n(u) \epsilon}(\s) \mc_{n(t) \delta;n(u) \epsilon;n(r) \mu}^{\hsp{.9} n(s) \nu }(\s)
\)
\(
\nc_{n(r) \mu}^{\hsp{.3} n(s) \nu; n(t) \delta}(\s) = \lr^{n(u) \epsilon ; n(v) \gamma} (\s) \nc_{n(u) \epsilon ; n(v) \gamma; n(r) \mu}^{\hsp{.9} n(s) \nu; n(t) \delta}(\s)
\)
\vspace{-.2in}
\ba
\mc_{n(r) \mu;n(s) \nu;n(t) \delta}^{\hsp{.9} n(u) \epsilon }(\s) & = & \delta_{(\hsp{.02} n(r) \mu}^{\hsp{.38} n(u) \epsilon} \g_{n(s) \nu \hsp{.02} ); n(t) \delta} (\s) \nonumber \\ 
&& + \frac{1}{2} \fr_{n(v) \gamma; (\hsp{.02} n(r) \mu}^{\hsp{.75} n(u) \epsilon}(\s) \fr_{n(s) \nu \hsp{.02} ) ; n(t) \delta}^{\hsp{.75} n(v) \gamma} (\s)
\ea
\(
\nc_{n(r) \mu; n(s) \nu; n(t) \delta}^{\hsp{.9} n(u) \epsilon; n(v) \gamma}(\s)
=\frac{i}{2} \delta_{(\hsp{.02}  n(r) \mu}^{\hsp{.38} (\hsp{.02} n(u) \epsilon} \fr_{n(s) \nu \hsp{.02} ) ; n(t) \delta}^{\hsp{.68} n(v) \gamma \hsp{.02} )}(\s) \ .
\)
\reseteqn
As expected, these results can be obtained by the duality algorithm (\ref{fdalgorithm}) from the untwisted $TJ$ OPE's in (\ref{MNgroup}) and the standard relations among the untwisted tensors$^{\rf{vme},\rf{Yam},\rf{Giveon},\rf{Lieh},\rf{rev}}$
\alpheqn
\(
M(L)_a^{\ b}=L^{cd}M_{cda}^{\ \ \ b} \sp N(L)_a^{\ bc}=L^{de}N_{dea}^{\ \ \ \ bc}
\)
\(
M_{abc}^{\ \ \ d}=\delta_{(a}^{\ d}G_{b) c}+\frac{1}{2} f_{e (a}^{\ \ \ d}f_{b) c}^{\ \ \ e} \sp
N_{abc}^{\ \ \ de}=\frac{1}{2}\delta_{(a}^{\ (d}f_{b)c}^{\ \ \ e)} \ .
\)
\reseteqn
Using the duality transformations (\ref{twistg},\ref{twistflett}) and (\ref{lr}) for $\g,\fr$ and $\lr$, one finds that the forms of $\mc(\s)$ and $\nc(\s)$ in (\ref{mcncappgroup}) are equivalent to the duality transformations for $\mc(\s)$ and $\nc(\s)$ in (\ref{mcdef},\ref{ncdeflett}).

\app{SNapp}{$A(S_3)/S_3$ and $A(S_4)/S_4$}
\bu $A(S_3)/S_3$\\ 
We begin this appendix by checking the results of Sec.~\ref{SNsec} against the results for the permutation orbifolds $A(S_3)/S_{3}$ given earlier in Ref.~\rf{us2}.  The permutation orbifolds $A(S_3)/S_3$ have three sectors named by the partitions $\sv=\{1,1,1\}$, $\{2,1\}$ and $\{3\}$.  The first of these is the untwisted sector (see (\ref{untwistsec})), where the ambient algebra $g$ consists of three copies of an untwisted ($\s_0=\s_1=\s_2=1$) affine algebra at level $\hat{k}_0=\hat{k}_1=\hat{k}_2=k$.  

The twisted sector $\sv=\{3\}$ has an order $\s_0=3$ orbifold affine algebra$^{\rf{Chris}}$ whose currents $\jh_{a,j=0}^{(\hj)}$, $\hj=0,1,2$ have orbifold affine level $\hat{k}_0=\s_0k=3k$.  The stress tensor, central charge and ground state conformal weight of this sector are
\namegroup{z3sec}
\alpheqn
\(
\hat{T}_{\{3\}} \=\frac{\l^{ab}}{3}:(\jh_{a0}^{(0)}\jh_{b0}^{(0)}\+\jh_{a0}^{(1)}\jh_{b0}^{(-1)}\+\jh_{a0}^{(2)}\jh_{b0}^{(-2)}):\hsp{.04}+\hsp{.0666}l^{ab}:\jh_{a0}^{(0)}\jh_{b0}^{(0)}: \label{z3T3}
\)
\(
\hat{c}(\{3\})=c=6 k\e_{ab}(\l^{ab}+l^{ab}) \sp \hat{\Delta}_0(\{3\})=\frac{2k\e_{ab}\l^{ab}}{9}.
\)
\reseteqn
With the identification $ \hat{J}_a^{(\hj)}\equiv\jh_{a,j=0}^{(\hj)}$  and the symmetry of the bilinears given in Ref.~\rf{us1}, the result (\ref{z3sec})  is recognized as the stress tensor and ground state conformal weight of the sector $\w_1$ given in Ref.~\rf{us2}.  

Finally, the twisted sector $\sv=\{2,1\}$ has an order $\s_0=2$ orbifold affine algebra whose currents $\jh_{a,j=0}^{(\hj)}$, $\hj=0,1$ have orbifold affine level $\hat{k}_0=\s_0 k=2k$, and a commuting order $\s_1=1$ untwisted affine algebra with currents $\jh_{a,j=1}^{(0)}$ at level $\hat{k}_1=k$.  The stress tensor, central charge and ground state conformal weight of this sector are
\alpheqn
\ba
\hat{T}_{\{2,1\}} &=&\frac{\l^{ab}}{2}:(\jh_{a0}^{(0)}\jh_{b0}^{(0)}+\jh_{a0}^{(1)}\jh_{b0}^{(-1)}):+\hsp{.04}\l^{ab}:\jh_{a1}^{(0)}\jh_{b1}^{(0)}:\nonumber\\&+&l^{ab}:(\jh_{a0}^{(0)}\jh_{b0}^{(0)}+\jh_{a0}^{(0)}\jh_{b1}^{(0)}+\jh_{a1}^{(0)}\jh_{b0}^{(0)}+\jh_{a1}^{(0)}\jh_{b1}^{(0)}):
\ea
\(
\hat{c}(\{2,1\})=c=6 k\e_{ab}(\l^{ab}+l^{ab}) \sp \hat{\Delta}_0(\{2,1\})=\frac{k\e_{ab}\l^{ab}}{8}.
\)
\reseteqn
With the identifications $\jh_a^{(\hj)}\equiv\jh_{a,j=0}^{(\hj)}$, $J_a\equiv\jh_{a,j=1}^{(0)}$ this is recognized as the stress tensor and ground state conformal weight of sector $\w_2$ in Ref.~\rf{us2}.  Agreement is obtained in this case because $\w_2$ and our $\w(\sv=\{2,1\})$ are in the same conjugacy class of $S_3$.

\noindent
\bu$A(S_4)/S_4$\\
We turn next to the permutation orbifolds $A(S_4)/S_4$, which have five sectors named by the partitions $\sv$ given in Eq.~(\ref{s4}).  Consider first the sector $\sv=\{3,1\}$ with the identifications
\(
\jh_a^{(\hj)}\equiv\jh_{a,j=0}^{(\hj)}\sp\hj=0,1,2;\hsp{.5}
J_a\equiv\jh_{a,j=1}^{(0)}
\)
where $\hat{J}$ is an orbifold affine algebra of order three and $J$ is a commuting integral affine algebra.  In this notation, we obtain
\alpheqn
\ba
\hat{T}_{\{3,1\}} &=&\frac{\l^{ab}}{3}:(\jh_{a}^{(0)}\jh_{b}^{(0)}+\jh_{a}^{(1)}\jh_{b}^{(-1)}+\jh_{a}^{(2)}\jh_{b}^{(-2)}):+\l^{ab}:J_aJ_b:\nonumber\\&+&l^{ab}:(\jh_{a}^{(0)}\jh_{b}^{(0)}+\jh_{a}^{(0)}J_b+J_a\jh_{b}^{(0)}+J_aJ_b):
\ea
\(
\hat{c}(\{3,1\})=c=8 k\e_{ab}(\l^{ab}+l^{ab}) \sp \hat{\Delta}_0(\{3,1\})=\frac{2k\e_{ab}\l^{ab}}{9}
\)
\reseteqn
from Eq.~(\ref{SNrecapgroup}).  The remaining ground state conformal weights of $A(S_4)/S_4$ are easily computed
\(
\hat{\Delta}_0(\{4\},\{2,2\},\{2,1,1\})=k\eta_{ab}\l^{ab} (\frac{5}{16}, \frac{1}{4}, \frac{1}{8})
\)
from the partitions $\sv$ in Eq.~(\ref{s4}).

\app{gradeapp}{The Grade Automorphism Group}
The grade (inner) automorphism group is defined by 
\alpheqn
\(
d=\frac{1}{h_g} \sum_{i=1}^{r} \frac{\l_i}{\a_i^2} \sp \a \cdot d = \frac{1}{2 h_g}G(\a) \sp \forall \ \a\in \Delta(g) \label{graded}
\)
\(
G(\a) = \sum_{i=0}^r n_i (\a)  \hsp{.3} \textup{when } \a= \sum_{i=0}^r n_i (\a) \a_i
\)
\(
N_c=\r(\s=1)=2h_g \sp \s=0,...,2h_g-1 \ . \label{gradesectors}
\)
\reseteqn
Here $r=\textup{rank} g$, $h_g$ is the Coxeter number of $g$ and $G(\a)$ is the grade of $\a\in \Delta(g)$.  The order $\r(1)$ of the $\s=1$ automorphism given in (\ref{gradesectors}) is computed from (\ref{order}) and the fact that $G(\psi_g)=h_g-1$, where $\psi_g$ is the highest root of $g$.  Using (\ref{nnbar}), one may also compute the twist classes of the twisted currents $\hat{J}$ of sector $\s=1$
\(
\bar{n}_\a (\s=1)
=\left\{\begin{array}{lll}
2h_g-G(\a) \sp \a > 0 \\
-G(\a), \hsp{.625} \a < 0 
\end{array}\right. 
\)
and the other sectors may be similarly analyzed.

Invariance under the grade automorphism group restricts the inverse inertia tensors of the untwisted sectors to the form $L^{AB}$, $L^{\a,-\a} \neq 0$, so the stress tensors of the orbifolds by the grade automorphism group are included in Eqs.~(\ref{goodinnerLorbgroup}), (\ref{FHuntwistedJgroup}) and (\ref{cfixedgroup}).

\app{NOapp}{Another Mode Ordering}
$M\p$ ordering is mode ordering by the integer part $m$ of the mode number ($m+\frac{n(r)}{\r(\s)}$),
\[
:\hat{J}_{n(r) \mu} (m+\srac{n(r)}{\r(\s)}) \hat{J}_{n(s) \mu} (n+\srac{n(s)}{\r(\s)}):_{M\p} \equiv \theta(m\geq 0) \hat{J}_{n(s) \mu} (n+\srac{n(s)}{\r(\s)}) \hat{J}_{n(r) \mu} (m+\srac{n(r)}{\r(\s)})
\]
\(
+ \theta(m <0) \hat{J}_{n(r) \mu} (m+\srac{n(r)}{\r(\s)}) \hat{J}_{n(s) \mu} (n+\srac{n(s)}{\r(\s)}) \ .
\)
With $M\p$ ordering, we find in place of Eqs.~(\ref{exactJJ}), (\ref{twiststress}) and (\ref{Lmodeb}):
\namegroup{MpNOappgroup}
\alpheqn
\[
\hsp{-1} \hat{J}_{n(r) \mu}(z) \hat{J}_{n(s) \nu}(w)=(\frac{w}{z})^{\frac{n(r)}{\r(\s)}} \{ [\frac{1}{(z-w)^2} + \frac{n(r)/\r(\s)}{w(z-w)}] \g_{n(r) \mu;n(s) \nu}(\s)
\]
\(
\hsp{1} +\frac{i \fr_{n(r) \mu;n(s) \nu}^{\hsp{.6} n(r)+n(s), \delta}(\s) \hat{J}_{n(r)+n(s), \delta}(w)}{z-w} \} + :\hat{J}_{n(r) \mu}(z) \hat{J}_{n(s) \nu}(w):_{M\p} \label{JJMp}
\)
\vspace{-.1in}
\ba
\hat{T}_\s(z)&=& \sum_{r,\mu,\nu} \lr^{n(r) \mu;-n(r), \nu}(\s) \{ :\hat{J}_{n(r) \mu}(z) \hat{J}_{-n(r), \nu}(z):_{M\p} - \frac{i n(r)}{\r(\s)} \fr_{n(r) \mu;-n(r) \nu}^{\hsp{.7} 0 \delta}(\s) \frac{\hat{J}_{0 \delta}(z)}{z} \nonumber \\
&&\hsp{1.3} + \frac{1}{z^2}   \frac{n(r)}{2 \r(\s)}(1-\frac{n(r)}{\r(\s)}) \g_{n(r) \mu;-n(r), \nu}(\s) \} \label{TMp}
\ea
\[
\hsp{-.5} L_\s (m) = \sum_{r,\mu,\nu} \lr^{n(r) \mu; -n(r), \nu}(\s) \ \{ \sum_{p \in \sz} :\hat{J}_{n(r) \mu}(p+\srac{n(r)}{\r(\s)}) \hat{J}_{-n(r), \nu}(m-p-\srac{n(r)}{\r(\s)}):_{M\p} 
\]
\agetletter{LmodeMplett}
\(
\hsp{1} -i\frac{n(r)}{\r(\s)} \fr_{n(r) \mu; -n(r), \nu}^{\ \ \ \ \ \ \ \ \ \ \ \ \ \ 0 \delta}(\s) \hat{J}_{0 \delta}(m) 
 + \delta_{m,0} \frac{n(r)}{2 \r(\s)}(1-\frac{n(r)}{\r(\s)}) \g_{n(r) \mu;-n(r), \nu}(\s) \}  \label{LmodeMp} \ .
\)
\reseteqn
Curiously, the $M\p$ results (\ref{MpNOappgroup}) can be obtained by the map $M\rightarrow M\p,$ $ \bar{n}(r) \rightarrow n(r)$ from their $M$-counterparts in (\ref{exactJJ}), (\ref{twiststress}) and (\ref{Lmodeb}).  The $M\p$ ordered product in (\ref{JJMp}) is not periodic under $n(r)\rightarrow n(r)+\r(\s)$, but the change is compensated by the $\g$ term so that the operator product on the left is periodic.  Similarly, the total summands of (\ref{TMp},\ref{LmodeMplett}) are periodic, so that each $n(r)$ can be replaced by $\bar{n}(r)$.

Because of its relation to the untwisted affine vacuum $|0\rangle$, $M\p$ ordering is used to discuss the general inner-automorphic orbifold in Sec.~\ref{innersec}.

\app{modesunbalancedapp}{The Mode Imbalance Phenomenon}

We begin this appendix by using the orbifold induction procedure (\ref{innerinduction}) to rewrite the stress tensors (\ref{twistLgroup}) of the inner-automorphic orbifolds in terms of untwisted currents.  The result is 
\namegroup{unbalancedgroup}
\alpheqn
\vspace{-.25in}
\(
L_\s(m)=L_\s^{q\p}(m)+L_\s^{l\p}(m)+\delta_{m,0} \hat{\Delta}_0(L,d;\s)
\)
\[
\hsp{-2.5} L_\s^{q\p}(m) \equiv \sum_{p\in \sz} \{ \sum_{A,B} L^{AB}:H_A(p) H_B (m-p):_{M \p}
\]
\[
\hsp{1.3} + \sum_{A,\a} L^{A \a} : [ H_A(p) E_\a (m-p+\s \a \cdot d)
+ E_\a (p+\s \a \cdot d) H_A(m-p) ] :_{M \p} 
\]
\(
\hsp{-.3} + \sum_{\a,\beta} L^{\a \beta}:E_\a(p) E_\beta(m-p+\s(\a+\beta)\cdot d):_{M \p} \} 
\)
\[
L_\s^{l\p}(m) \equiv \sum_{\a+\beta=\gamma} L^{\a \beta} N_\gamma(\a,\beta) (\sigma \a \cdot d) E_\gamma (m+\s \gamma \cdot d) +2 k  \sum_{A,\a} \s d_A L^{A \a} E_\a (m+\s \a \cdot d)
\]
\(
+ 2 k \sum_{A,B} \s d_A L^{AB} H_B(m) + \sum_\a L^{\a,-\a} (\s \alpha \cdot d) \alpha \cdot H(m) 
\)
\(
L^{A\a}=0 \textup{  unless } \a \cdot d \in \z \sp L^{\a \beta}=0 \textup{  unless } (\a + \beta) \cdot d \in \z  \sp \s=0,...,\r(1)-1 \label{innertLsolutionsapp}
\)
\reseteqn
where the quantity $\hat{\Delta}_0(L,d;\s)$ is given in (\ref{generalinnercw}).  In this form, the arguments (modes) of the root operators are exactly the quantities in the inner parentheses of Eq.~(\ref{twistLgroup}), so (according to the discussion around (\ref{argument})) all the modes here are integers.

In spite of their unfamiliar form, these constructions are Virasoro generators when their inverse inertia tensors satisfy the H-invariance conditions (\ref{innertLsolutionsapp}) and the Virasoro master equation (\ref{VME}).  Although the current modes were balanced (i.e. summed to $m$) when the generators were written in terms of the twisted currents (see Eq.~(\ref{twistLgroup})), we see in the form (\ref{unbalancedgroup}) a generic \textit{mode imbalance} (when $L^{A \a}$ and $L^{\a \beta}$, $\a+\beta\neq 0$ are non-zero) in which the modes of the currents of $L_\s(m)$ do not necessarily sum to $m$.  It follows that the generic inner-automorphic orbifold can not be described by the stress-tensor spectral flow (\ref{cfixedgroup}), in which the current modes sum to $m$.  (The mode imbalance phenomenon is avoided however for the special case when only $L^{A \a}$ and $L^{\a \beta}$ are non-zero, so that, as discussed in the text, these special inner-automorphic orbifolds can be described by spectral flow.)

A similar mode imbalance is found for all the stress tensors of all the permutation orbifolds when the orbifold induction procedure (\ref{zlinduction}) is used to express these stress tensors in terms of untwisted current modes.  As an example, we mention the form 
\ba
\hsp{-.5} L_\s (m) & = & \sum_{r=0}^{\r(\s)-1}[\ \sum_{j,l=0}^{\frac{\l}{\r(\s)}-1} \lr^{r a j; -r, bl}(\s) \ \sum_{p \in \sz} : J_{a j} (\r(\s)p+r) J_{bl} (\r(\s)(m-p) -r):_M   \nonumber \\
&& + \sum_{j=0}^{\frac{\l}{\r(\s)}-1} \lr^{r a j; -r, b j}(\s)  \{ \frac{-i  r}{\r(\s)} f_{ab}^{\ \ c} J_{cj} (\r(\s) m)  + \delta_{m,0}  \frac{r}{2 \r(\s)}(1-\frac{r}{\r(\s)}) k \r(\s) \eta_{ab} \} ] \label{zlLapp}
\ea
obtained from (\ref{zlL}) for the sectors of the orbifold \taz.  

\app{dlapp}{The OVME and \tadd}
In Ref.~\rf{us2} it was shown that every solution of the orbifold Virasoro master equation$^{\rf{us1}}$ (OVME) at order $\l$ is a sector of a permutation orbifold of type $A(\dl)/\zl$.  Moreover, it was asserted in Ref.~\rf{us2} that the sectors described by the OVME also occur in the permutation orbifolds $A(\dl)/\dl$. 

To see this we begin with (the inverse of) the first ($\s=1$) duality transformation,$^{\rf{us2}}$ which constructs a particular $\dl$-invariant CFT described by $L$ from any particular solution $\lr_{OVME}=\lr$ of the OVME at order $\l$:
\namegroup{firstzldgroup}
\vspace{-.2in}
\alpheqn
\(
L^{ab}_{I-J}= \l \sum_{r=0}^{\l-1} \lr_r^{ab} U(1)_{r;I} U(1)_{-r;J}
\sp \lr^{ab}_r=\lr^{ba}_{-r}=\lr^{ba}_r \ \rightarrow \ L_K^{ab}=L^{ba}_{-K}=L^{ba}_K \label{ovmesymmetry}
\)
\(
U^\dagger(1)_{I;r}=\frac{1}{\sqrt{\l}} e^{-\frac{2 \pi i r I}{\l}}  \sp \r(1)=\l \sp  I,J,r=0,...,\l-1 \ .
\)
\reseteqn
The solution $\lr$ of the OVME was identified in Ref.~\rf{us2} as a sector of the orbifold $A(\dl)/\zl$ because of the $\dl$ symmetry of $L$ in (\ref{ovmesymmetry}) and the fact that $U^\dagger(1)$ satisfies 
\(
\w(h_1)_{I}^{\ J} U^\dagger(1)_{J;r}=U^\dagger(1)_{I;r} e^{-\frac{2 \pi ir}{\l}}, \hsp{.2}  \w(h_1)_{I}^{\ J}= \delta_{I+1, J \mod \l}, \hsp{.2} h_1 \in \zl \label{zlwgroup}
\)
which is the $\zl$-eigenvalue problem at $\s=1$. 

Consider next the permutation orbifold $A(\dl)/\dl$, where the untwisted sector is described by the same $\dl$-symmetric $L$.  The $2 \l$ elements of the group $\dl$ are 
\(
\{ h \}=\{ r^\s, s r^\s ; \hsp{.1} \s=0,...,\l-1  \} \sp  r^\l=s^2=1 \sp rs=sr^{-1} 
\)
and $\dl$ acts by permuting the currents according to  
\alpheqn
\(
J_{aI}^{\ \ \p} = \w(h)_{IJ} J_{aJ} \sp  \forall \ h \in \dl
\)
\(
\w(r^\s)_{I}^{\ J}= \delta_{I+\s, J \mod \l}, \hsp{.15} r^\s \in \zl \subset \dl; \hsp{.2} \w(s r^\s)_{I}^{\ J}= \delta_{I+\s, -J \mod \l}. \label{dlzlsector1}
\)
%\(
%\w(r^{\s=1})=\w(h_{\s=1}) \sp h_1 \in \zl \label{dlzlsector1} \ .
%\)
\reseteqn 
The action of the element $\w(r^{\s=1}) \in \dl$ in (\ref{dlzlsector1}) is the same as the  action of the element $\w(h_{\s=1}) \in \zl$ in (\ref{zlwgroup}).  Then, the first duality transformation into the first twisted sector of $A(\dl)/\dl$
\(
\lr^{ab}_r=\frac{1}{\l} \sum_{I,J=0}^{\l-1}L_{I-J}^{ab} U^\dagger (1)_{I;r}
U^\dagger (1)_{J;-r}
\sp  \lr^{ab}_r=\lr^{ba}_{-r}=\lr^{ba}_r
\ \leftarrow \ L_K^{ab}=L^{ba}_{-K}=L^{ba}_K \label{firstdualitydlgroup}
\)
gives the original solution $\lr$ of the OVME.  This establishes the assertion of Ref.~\rf{us2}.

\app{outersapp}{The Setup for Outer-automorphic Orbifolds}
In our notation, the outer automorphism$^{\rf{book},\rf{GO}}$ groups of simple $g$ are described by 
\alpheqn
\vspace{-.1in}
\(
\a_i\p=\tau(\a_i) \ \rightarrow \ \a\p =\sum_{i=1}^{\textup{\scriptsize{rank}}g} n_i \tau(\a_i) \ : \hsp{.2}  \w_\a^{\ \b}=\xi_\a \delta_{\tau(\a),\beta} \sp \w_A^{\ B} = \sum_{i=1}^{\textup{\scriptsize{rank}} g}\l_{iA} \tau(\a_i)^B 
\)
\(
\xi_{\a_i}=1 \sp \xi_{\a}\xi_{-\a}=1 \sp \xi_{\a}\xi_{\beta}\xi_{\gamma}^{-1} N_{\tau(\gamma)}(\tau(\a),\tau(\beta))=N_{\gamma}(\a,\beta)
\)
\reseteqn
where $\a, \beta, \gamma \in \Delta(g)$, $\{ \l_{i} \}$ are the fundamental weights of $g$ and we have taken $\a^2=2$.

For $SU(3)$, the outer automorphism group is a $\z_2$ 
\(
\r(\s=1)=2 \sp \tau(\a_{1,2})=\a_{2,1} \sp \xi_{\pm(\a_1+\a_2)}=-1 
\)
and solution of the $H$-eigenvalue problem gives eight twisted currents $\hat{J}=\{ \hat{H}_{n(r)=0,1},$ $\hat{E}^{\pm \a_i}_{n(r)=0,1},$ $\hat{E}_{n(r)=1}^{\pm(\a_1+\a_2)} \}$ which, as expected, satisfy the outer-automorphically twisted affine Lie algebra $A_2^{(2)}$.  The inverse inertia tensors of the outer automorphic invariant CFT's on $SU(3)$ satisfy the $H$-invariance conditions
\alpheqn
\(
L^{\a_1\a_1}=L^{\a_2\a_2} \sp L^{\a_1,-\a_2}=L^{\a_2,-\a_1}
\)
\(
L^{\a_1+\a_2,\pm\a_1}=-L^{\a_1+\a_2,\pm \a_2} \sp L^{-(\a_1+\a_2),\pm \a_1}=-L^{-(\a_1+\a_2),\pm \a_2}
\)
\reseteqn
and corresponding conditions on $L^{AB}, L^{A\a}$.  The reader is encouraged to work out the outer-automorphic orbifolds in further detail.

\pagebreak


\begin{thebibliography}{99}
{\small 
\bibitem {o1} \vspace{-10pt} M. B. Halpern and C. B. Thorn, \emph{Phys. Rev.} {\textbf{D4}}, 3084 (1971). \label{refTh}
\bibitem {o3} \vspace{-10pt} E. Corrigan and D. B. Fairlie, \emph{Nucl. Phys.} {\textbf{B91}}, 527 (1975). \label{refCo}
\bibitem {o4} \vspace{-10pt} J. Lepowsky and R. Wilson, \emph{Commun. Math. Phys.} {\textbf{62}}, 43 (1978). \label{refLep}
\bibitem {o5} \vspace{-10pt} I. B. Frenkel, J. Lepowsky and A. Meurman, \emph{Proc. Natl. Acad. Sci. U.S.A.} {\textbf{81}}, 3256 (1984). \label{refFr}
\bibitem {o6} \vspace{-10pt} L. J. Dixon, J. A. Harvey, C. Vafa and E. Witten, \emph{Nucl. Phys.} {\textbf{B261}}, 678 (1985). \label{refDixO1}
\bibitem {o7} \vspace{-10pt} L. J. Dixon, J. A. Harvey, C. Vafa and E. Witten, \emph{Nucl. Phys.} {\textbf{B274}}, 285 (1986). \label{refDixO2}
\bibitem {o8} \vspace{-10pt} L. J. Dixon, D. H. Friedan, E. J. Martinec and S. H. Shenker, \emph{Nucl. Phys.} {\textbf{B282}}, 13 (1987). \label{refDixC}
\bibitem {o9} \vspace{-10pt} S. Hamidi and C. Vafa, \emph{Nucl. Phys.} {\textbf{B279}}, 465 (1987). \label{refHam}
\bibitem {o10} \vspace{-10pt} L. J. Dixon, P. Ginsparg and J. A. Harvey, \emph{Commun. Math. Phys.} {\textbf{119}}, 221 (1988). \label{refDix}
\bibitem {o11} \vspace{-10pt} R. Dijkgraaf, C. Vafa, E. Verlinde and H. Verlinde, \emph{Commun. Math. Phys.} {\textbf{123}}, 485 (1989). \label{refDVVV}
\bibitem {o12} \vspace{-10pt} A. Klemm and M. G. Schmidt, \emph{Phys. Lett.} {\textbf{B245}}, 53 (1990). \label{refKS}
\bibitem {o13} \vspace{-10pt} J. Fuchs, A. Klemm and M. G. Schmidt, \emph{Ann. Phys.} {\textbf{214}}, 221 (1992). \label{refFuchs}
\bibitem h \vspace{-10pt} R. Dijkgraaf, G. Moore, E. Verlinde and H. Verlinde, \emph{Comm. Math. Phys.} {\textbf{185}}, 197 (1997). \label{refDMVV}
\bibitem i \vspace{-10pt} V. G. Kac and I. T. Todorov, \emph{Commun. Math. Phys.} {\textbf{190}} 57 (1997). \label{refKT}
\bibitem a \vspace{-10pt} L. Borisov, M. B. Halpern and C. Schweigert, \emph{Int. J. Mod. Phys.} {\textbf{A13}}, 125 (1998).\label{refChris}
\bibitem {o14} \vspace{-10pt} P. Bantay, \emph{Phys. Lett.} {\textbf{B419}}, 175 (1998). \label{refBan1}
\bibitem 9 \vspace{-10pt} K. Barron, C. Dong and G. Mason, ``Twisted sectors for tensor product vertex operator algebras associated to permutation groups,'' math.QA/9803118. \label{refDong}
\bibitem {o15}  \vspace{-10pt} L. Birke, J. Fuchs and C. Schweigert, ``Symmetry Breaking Boundary Conditions and WZW Orbifolds,'' ETH-TH/99-11, hep-th/9905038. \label{refBirke}
\bibitem {us2} \vspace{-10pt} J. de Boer, J. Evslin, M. B. Halpern and J. E. Wang, \emph{Int. J. Mod. Phys.} {\textbf{A15}}, 1297 (2000). \label{refus2}
\bibitem 6 \vspace{-10pt} P. Bantay, ``Permutation Orbifolds,'' hep-th/9910079. \label{refBan2}
\bibitem {us3} \vspace{-10pt} J. Evslin, M. B. Halpern and J. E. Wang, ``Cyclic Coset Orbifolds,'' UCB-PTH-99/53, hep-th/9912084, to appear in \emph{Int. J. Mod. Phys. A}. \label{refus3}
\bibitem {us1} \vspace{-10pt} J. Evslin, M. B. Halpern and J. E. Wang, \emph{Int. J. Mod. Phys.} {\textbf{A14}}, 4985 (1999). \label{refus1}
\bibitem j \vspace{-10pt} M. B. Halpern, E. B. Kiritsis and N. A. Obers, \emph{Int. J. Mod. Phys.} {\textbf{A7}} [Suppl. 1A], 339 (1992). \label{refLieh}
\bibitem k \vspace{-25pt} M. B. Halpern and N. Sochen, \emph{Int. J. Mod. Phys} {\textbf{A10}}, 1181 (1995). \label{refSochen}
\bibitem d \vspace{-10pt} M. B. Halpern, E. Kiritis, N. A. Obers and K. Clubok, ``Irrational Conformal Field Theory,'' \emph{Physics Reports} {\textbf{265}} No. 1\&2 (1996) 1. \label{refrev}
\bibitem b \vspace{-10pt} M. B. Halpern and E. Kiritsis, \emph{Mod. Phys. Lett.} {\textbf{A4}}, 1373 (1989); Erratum, ibid. {\textbf{A4}}, 1797 (1989). \label{refvme}
\bibitem c \vspace{-10pt} A. Yu. Morozov, A. M. Perelomov, A. A. Rosly, M. A. Shifman and A. V. Turbiner, \emph{Int. J. Mod. Phys.} {\textbf{A5}}, 803 (1990). \label{refruss}
\bibitem 5 \vspace{-10pt} J. de Boer and M. B. Halpern, \emph{Int. J. Mod. Phys.} {\textbf{A13}}, 4487 (1998). \label{refJan}
\bibitem g \vspace{-10pt} K. Bardakci and M. B. Halpern, \emph{Phys. Rev. } {\textbf{D3}}, 2493 (1971). \label{refBH}
\bibitem t \vspace{-10pt} J.K. Freericks and M. B. Halpern, \emph{Ann. Phys.} {\textbf{188}}, 258 (1988); Erratum, ibid. \textbf{190}, 212 (1989). \label{refFH}
\bibitem e \vspace{-20pt} V. G. Kac, \emph{Funct. Anal. App.} {\textbf{1}}, 328 (1967). \label{refKac}
\bibitem f \vspace{-10pt} R. V. Moody, \emph{Bull. Am. Math. Soc.} {\textbf{73}}, 217 (1967). \label{refMoody}
\bibitem w \vspace{-10pt} M. B. Halpern, E. Kiritsis, N. A. Obers, M. Porrati and J. P. Yamron, \emph{Int. J. Mod. Phys.} {\textbf{A5}}, 2275 (1990). \label{ref5ofus}
\bibitem r \vspace{-10pt} M. B. Halpern and N. A. Obers, \emph{Commun. Math. Phys.} {\textbf{138}}, 63 (1991). \label{refObers}
\bibitem 3 \vspace{-10pt} F. A. Bais, P. Bouwknegt, K. Schoutens and M. Surridge, \emph{Nucl. Phys.} {\textbf{B304}}, 348 (1988).\label{refBouw}
\bibitem l \vspace{-10pt} M. B. Halpern, \emph{Phys. Rev.} {\textbf{D4}}, 2398 (1971). \label{refHalp}
\bibitem o \vspace{-10pt} R. Dashen and Y. Frishman, \emph{Phys. Rev.} {\textbf{D11}}, 278 (1975). \label{refDash}
\bibitem p \vspace{-10pt} V. G. Knizhnik and A. B. Zamolodchikov, \emph{Nucl. Phys.} {\textbf{B247}}, 83 (1984). \label{refKZ}
\bibitem q \vspace{-10pt} G. Segal, unpublished. \label{refSegal}
\bibitem 4 \vspace{-10pt} V. G. Kac and D. H. Peterson, \emph{Adv. Math.} {\textbf{53}}, 125 (1984). \label{refKP}
\bibitem s \vspace{-10pt} P. Goddard and D. Olive, \emph{Int. J. Mod. Phys.} {\textbf{A1}}, 303 (1986). \label{refGO}
\bibitem u \vspace{-10pt} W. Lerche, C. Vafa and N. P. Warner, \emph{Nucl. Phys.} {\textbf{B324}}, 427 (1989). \label{refLerche}
\bibitem v \vspace{-10pt} V. G. Kac, \emph{Infinite-dimensional Lie Algebras}, third edition (Cambridge University Press, Cambridge 1990). \label{refbook}
\bibitem 7 \vspace{-10pt} M. B. Halpern and J. Yamron, \emph{Nucl. Phys.} {\textbf{B332}}, 411 (1990). \label{refYam}
\bibitem 1 \vspace{-10pt} A. Giveon, M. B. Halpern, E. B. Kiritsis and N. A. Obers, \emph{Nucl. Phys.} {\textbf{B357}}, 655 (1991). \label{refGiveon}
\bibitem m \vspace{-10pt} P. Goddard, A. Kent and D. Olive, \emph{Phys. Lett.} {\textbf{B152}}, 88 (1985). \label{refGKO}
\bibitem n  \vspace{-10pt} E. Kiritsis, \emph{Mod. Phys. Lett.} {\textbf{A4}}, 437 (1989). \label{refKir}
\bibitem x  \vspace{-10pt} P. Bouwknegt, \emph{J. Math Phys.} {\textbf{30}}, 571 (1989). \label{refBouw1}
\bibitem y  \vspace{-10pt} P. Bouwknegt and A. Ceresole, \emph{Phys. Lett.} {\textbf{B238}}, 224 (1990). \label{refBouw2}
\bibitem 8 \vspace{-10pt} J. Maldacena and H. Ooguri, ``Strings in AdS$_3$ and SL(2,R) WZW Model: I,'' hep-th/0001053. \label{refOog}
}

\end{thebibliography}
\end{document}